\documentclass[10pt,preprint]{aastex}

\pdfoutput=1

\usepackage{url}
\usepackage{grffile}
\newcounter{column_number}
\newcommand{\numberthecolumn}{\colhead{(\arabic{column_number})}\stepcounter{column_number}}

\newcommand{\anchorfoot}[2] {\anchor{#1}{#2}\footnote{\url{#1}}}

\newcommand{\hii}{H{\scriptsize II} }
\newcommand{\etacar}{$\eta$~Car }
\newcommand{\Chandra}{{\em Chandra} }

\shorttitle{CCCP Intro Paper}
\shortauthors{Townsley et al.} 
\begin{document}
% http://authors.iop.org/atom/help.nsf/LookupJournalSpecific/WebAuthorGuidelines~AJ

\title{AN INTRODUCTION TO THE {\em CHANDRA} CARINA COMPLEX PROJECT}

\author{ Leisa K. Townsley\altaffilmark{*}\altaffilmark{1}, 
% The Leads
Patrick S. Broos\altaffilmark{1}, Michael F. Corcoran\altaffilmark{2}, Eric D. Feigelson\altaffilmark{1}, Marc Gagn{\'e}\altaffilmark{3}, Thierry Montmerle\altaffilmark{4}, M. S. Oey\altaffilmark{5}, Nathan Smith\altaffilmark{6}, 
% The Locals
Gordon P.\ Garmire\altaffilmark{1}, Konstantin V.\ Getman\altaffilmark{1}, Matthew S.\ Povich\altaffilmark{1,7}, 
% The First Authors
Nancy Remage Evans\altaffilmark{8}, Ya{\"e}l Naz{\'e}\altaffilmark{9}, E. R. Parkin\altaffilmark{10,}\altaffilmark{11}, Thomas Preibisch\altaffilmark{12}, Junfeng Wang\altaffilmark{8}, Scott J. Wolk\altaffilmark{8},
% The Co-authors
You-Hua Chu\altaffilmark{13}, David H. Cohen\altaffilmark{14}, Robert A. Gruendl\altaffilmark{13}, Kenji Hamaguchi\altaffilmark{15}, Robert R. King\altaffilmark{16}, Mordecai-Mark Mac Low\altaffilmark{17}, Mark J. McCaughrean\altaffilmark{18,16}, Anthony F. J. Moffat\altaffilmark{19}, L. M. Oskinova\altaffilmark{20}, Julian M. Pittard\altaffilmark{11}, Keivan G. Stassun\altaffilmark{21}, Asif ud-Doula\altaffilmark{22}, Nolan R. Walborn\altaffilmark{23}, Wayne L. Waldron\altaffilmark{24},
% The Remaining Team (those willing at least) 
Ed Churchwell\altaffilmark{25}, J. S. Nichols\altaffilmark{8}, Stanley P. Owocki\altaffilmark{26}, N. S. Schulz\altaffilmark{27}
 }

\altaffiltext{*}{townsley@astro.psu.edu} 

\altaffiltext{1} {Department of Astronomy \& Astrophysics, 525 Davey Laboratory, Pennsylvania State University, University Park, PA 16802, USA}

\altaffiltext{2} {CRESST and X-ray Astrophysics Laboratory NASA/GSFC, Greenbelt, MD 20771, USA; Universities Space Research Association, 10211 Wincopin Circle, Suite 500, Columbia, MD 21044, USA}

\altaffiltext{3} {Department of Geology and Astronomy, West Chester University, West Chester, PA 19383, USA} 

\altaffiltext{4} {Institut d'Astrophysique de Paris, 98bis, Bd Arago, 75014 Paris, France}

\altaffiltext{5} {Department of Astronomy, University of Michigan, 830 Dennison Building, Ann Arbor, MI   48109-1042, USA}

\altaffiltext{6} {Steward Observatory, University of Arizona, 933 North Cherry Avenue, Tucson, AZ 85721, USA}

\altaffiltext{7} {NSF Astronomy and Astrophysics Postdoctoral Fellow}

\altaffiltext{8} {Harvard-Smithsonian Center for Astrophysics, 60 Garden Street, Cambridge, MA 02138, USA}

\altaffiltext{9} {GAPHE, D{\'e}partement AGO, Universit{\'e} de Li{\`e}ge, All{\'e}e du 6 Ao{\^u}t 17, Bat. B5C, B4000-Li{\`e}ge, Belgium}

\altaffiltext{10} {Institut d'Astrophysique et de G\'{e}ophysique, Universit\'{e} de Li\`{e}ge, 17, All\'{e}e du 6 Ao\^{u}t, B5c, B-4000 Sart Tilman, Belgium}

\altaffiltext{11} {School of Physics and Astronomy, The University of Leeds, Woodhouse Lane, Leeds LS2 9JT, UK}

\altaffiltext{12} {Universit\"ats-Sternwarte, Ludwig-Maximilians-Universit\"at, Scheinerstr.~1, 81679 M\"unchen, Germany}

\altaffiltext{13}{Department of Astronomy, University of Illinois at Urbana-Champaign, 1002 West Green Street, Urbana, IL 61801, USA}

\altaffiltext{14}{Department of Physics and Astronomy, Swarthmore College, 500 College Ave., Swarthmore, PA 19081, USA}

\altaffiltext{15} {CRESST and X-ray Astrophysics Laboratory NASA/GSFC, Greenbelt, MD 20771, USA; Department of Physics, University of Maryland, Baltimore County, 1000 Hilltop Circle, Baltimore, MD 21250, USA}

\altaffiltext{16} {Astrophysics Group, College of Engineering, Mathematics, and Physical Sciences, University of Exeter, Exeter EX4 4QL, UK}

\altaffiltext{17} {Division of Physical Sciences, American Museum of Natural History, 79th St.\ at CPW, NY, NY, 10024-5192, USA}

\altaffiltext{18} {European Space Agency, Research \& Scientific Support Department, ESTEC, Postbus 299, 2200 AG Noordwijk, The Netherlands}

\altaffiltext{19} {D{\'e}partement de physique, Universit{\'e} de Montr{\'e}al and Centre de Recherche en Astrophysique du Qu{\'e}bec, C.P. 6128, Succ. C-V, Montr{\'e}al, QC, H3C 3J7, Canada}

\altaffiltext{20} {Institute for Physics and Astronomy, University of Potsdam, 14476 Potsdam, Germany}

\altaffiltext{21} {Department of Physics \& Astronomy, Vanderbilt University, VU Station B 1807, Nashville, TN 37235, USA; Department of Physics, Fisk University, 1000 17th Ave. N., Nashville, TN 37208, USA}

\altaffiltext{22} {Penn State Worthington Scranton, 120 Ridge View Drive, Dunmore, PA 18512, USA}

\altaffiltext{23}{Space Telescope Science Institute, Baltimore, MD 21218, USA}

\altaffiltext{24}{Eureka Scientific, Inc., 2452 Delmer Street, Suite 100, Oakland, CA 94602-3017, USA}

\altaffiltext{25}{Department of Astronomy, University of Wisconsin-Madison, 475 N. Charter Street, Madison, WI 53706, USA}

\altaffiltext{26}{Bartol Research Institute, Department of Physics \& Astronomy, University of Delaware, Newark, DE 19716, USA}

\altaffiltext{27}{Kavli Institute for Astrophysics and Space Research, Massachusetts Institute of Technology, Cambridge, MA 02139, USA}

%\email{townsley@astro.psu.edu}

\begin{abstract}
The Great Nebula in Carina provides an exceptional view into the violent massive star formation and feedback that typifies giant \hii regions and starburst galaxies.  We have mapped the Carina star-forming complex in X-rays, using archival \Chandra data and a mosaic of 20 new 60-ks pointings using the {\em Chandra X-ray Observatory}'s Advanced CCD Imaging Spectrometer, as a testbed for understanding recent and ongoing star formation and to probe Carina's regions of bright diffuse X-ray emission.  This study has yielded a catalog of properties of $>$14,000 X-ray point sources; $>$9800 of them have multiwavelength counterparts.  Using {\em Chandra}'s unsurpassed X-ray spatial resolution, we have separated these point sources from the extensive, spatially-complex diffuse emission that pervades the region; X-ray properties of this diffuse emission suggest that it traces feedback from Carina's massive stars.  In this introductory paper, we motivate the survey design, describe the \Chandra observations, and present some simple results, providing a foundation for the 15 papers that follow in this {\em Special Issue} and that present detailed catalogs, methods, and science results.
\end{abstract}

% http://authors.iop.org/atom/help.nsf/LookupJournalSpecific/WebSubjectKeywords~ApJ?OpenDocument&journalid=ApJ

%============================================================================= 
% Enclose first mention of each cluster in ''object'' macro. 
% http://www.journals.uchicago.edu/AAS/objects/objectlinking.aas.html

\section{INTRODUCTION \label{sec:intro}}

\subsection{Overview}

\begin{figure}[htb] 
\begin{center}
\includegraphics[width=0.5\textwidth]{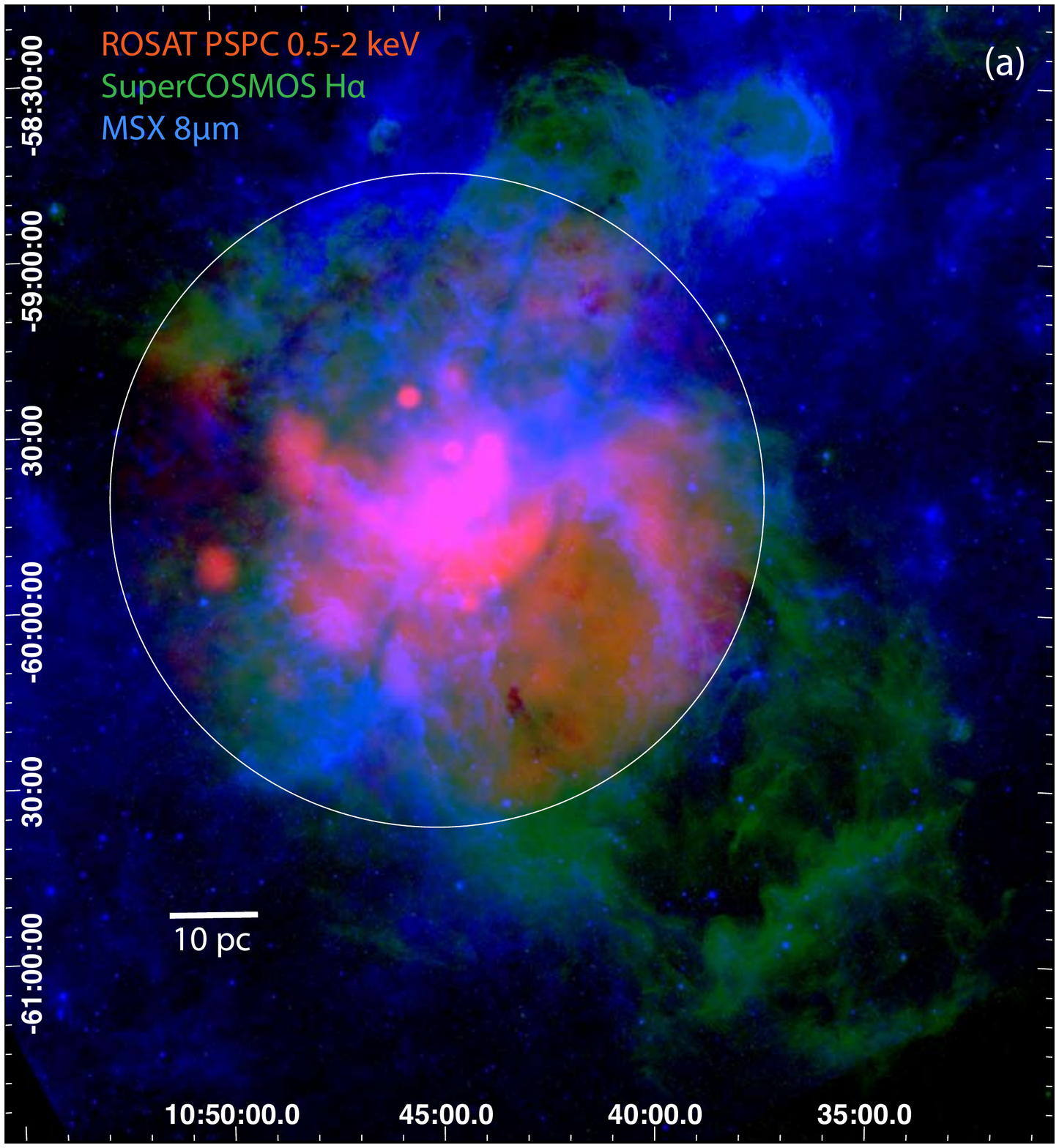}
\includegraphics[width=0.45\textwidth]{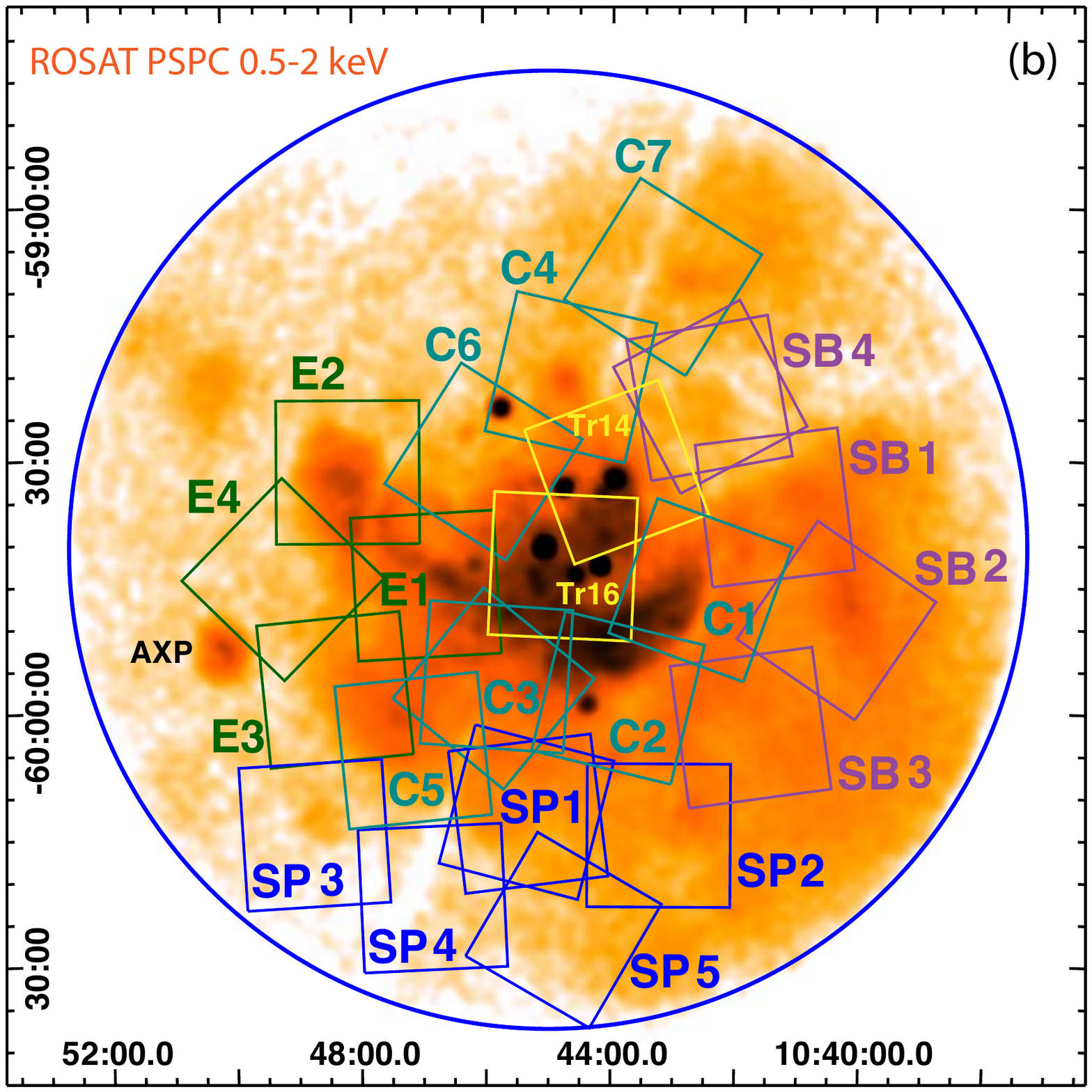}
\caption{
The Great Nebula in Carina (NGC~3372).  Here and throughout the paper, image coordinates are celestial J2000 unless otherwise noted and the mapping of image colors to multiwavelength datasets is denoted by a legend on the image and/or by the figure caption.
(a) A large-scale, 3$^{\circ}$ view showing the {\em MSX} 8.28~$\mu$m image (blue) overlaid with SuperCOSMOS H$\alpha$ \citep{Parker05} (green) and {\em ROSAT} PSPC hard band (0.5--2.0~keV) data (red) scaled to highlight diffuse emission; soft X-rays clearly fill the lower superbubble lobe.  A bar of length $\sim$10~pc is shown at lower left.
(b) The 22 CCCP $17\arcmin \times 17\arcmin$ ACIS-I pointings overlaid on the {\em ROSAT} hard-band image of Carina and coded by region: ÊC = Cluster of Clusters, E = Carina East, SP = South Pillars, SB = Superbubble (see Section~\ref{sec:motivation} for details).  The massive stellar clusters Tr14 and Tr16 were observed previously by ACIS-I; those archival data were included in the CCCP.  ``AXP'' marks the anomalous X-ray pulsar 1E1048.1-5937. 
In both panels, the large circle indicates the extent of the {\em ROSAT} data.  At D=2.3~kpc, 10$\arcmin \simeq 6.7$~pc, or 1~pc $\simeq 1.49\arcmin$.
} 
\normalsize 
\label{fig:intro}
\end{center}
\end{figure}

The Great Nebula in Carina is a massive star-forming complex located in the Sagittarius-Carina spiral arm, at a distance of $\sim$2.3~kpc \citep{Smith06a}.  Its ensemble of young stellar clusters exhibits ages $<$1~Myr \citep[Trumpler~14,][]{Sana10} to $\sim$6~Myr \citep[Trumpler~15,][]{Feinstein80}.  Altogether, it contains one of the richest concentrations of massive stars in the nearby Galaxy, with $>$65 O stars, 3 Wolf-Rayet (WR) stars, and the luminous blue variable \object[eta Car]{$\eta$~Carinae} \citep{Smith06b, Corcoran04}.  The Carina complex is likely to contain hundreds of intermediate-mass stars and several tens of thousands of low-mass pre-Main Sequence (pre-MS) stars, based on the known O-star population and assuming a normal initial mass function.  Infrared (IR) and visual images of Carina are dominated by highly-structured arcs, filaments, pillars, and shells on 0.1--10~pc scales; these reveal the interfaces between cavities and surrounding material and demonstrate how the OB stars are shredding their natal environment.  The remaining molecular material is scattered throughout the complex in dusty clouds with a wide range of sizes \citep{Preibisch11a}.  The ionizing photons, winds, and perhaps supernovae from Carina's OB stars are also fueling a young bipolar superbubble \citep{Smith00}.  Mid-IR data (Figure~\ref{fig:intro}a) show a $\sim$1$^\circ$ closed upper loop of emission, where the superbubble is confined by the dense medium near the Galactic Plane, and a more open lower cavity filled with hot plasma seen by {\em Einstein} \citep{Seward82} and {\em ROSAT} \citep{Corcoran95b}.  \citet{Smith08} recently reviewed star formation in the Carina complex; we refer interested readers to that work for a comprehensive overview of the complex and its literature.  This paper describes a new X-ray survey of Carina using the {\em Chandra X-ray Observatory}, the \Chandra Carina Complex Project (CCCP), a mosaic of 22 pointings with the Advanced CCD Imaging Spectrometer (ACIS) imaging CCD array \citep[ACIS-I,][]{Garmire03}; see Figure~\ref{fig:intro}b.  This paper is the first of 16 papers on the CCCP presented together as a {\em Special Issue}.

%-------------------------------------------------------------------------
%-------------------------------------------------------------------------

Concurrent star formation is underway across the Carina complex, especially in the eroding South Pillars.  \citet{Smith10a} use H$\alpha$ imaging with {\em HST} to find a wide array of Herbig-Haro objects, dark globules, proplyds, and microjets in the Carina Nebula; these are all signs of ongoing star formation.  Mid-IR images with {\em MSX} showed sites of likely embedded recent star formation in the South Pillars and other sites across the Nebula \citep{Smith08}.  These have been elucidated with new {\em Spitzer} data; \citet{Smith10b} catalog $>$900 candidate young stars that are emerging from the South Pillars as winds and radiation from Carina's massive stars destroy the surrounding molecular material.

The history of supernova activity in Carina is disputed.  We may be seeing the complex just before its first supernova \citep{Smith08} so its environment has not been altered by such explosions, however an {\em ISO} study found a broad emission feature at 22$\mu$m reminiscent of newly-made dust seen in the Cas~A supernova remnant \citep{Chan00}, implying that Carina has also seen supernova activity.  Recent X-ray evidence also suggests \citep{Townsley09a, Hamaguchi07a, Ezoe08} that at least one supernova might have already occurred, driving Carina's infant superbubble and enriching its interstellar medium (ISM).  The recent X-ray discovery of a $\sim$$10^6$ year old neutron star in the Carina complex \citep{Hamaguchi09, Pires09} provides strong evidence for past supernovae in Carina, as well as the suggestion that a slightly older, 5--10~Myr population of young stars pervades the complex.  Several other authors have reported evidence for a 5--10~Myr old stellar population superposed on Carina's young, dense clusters and/or signs that star formation has been ongoing throughout the past several million years \citep[see the introduction of][and references therein]{Ascenso07}.

Carina's rich stellar clusters, ionization fronts, star-forming dust pillars, and emerging superbubbles constitute one of the closest analogs of a Giant Extragalactic \hii Region (GEHR).  \citet{Walborn97} showed that there are multiple generations of massive stellar clusters in 30 Doradus, the Local Group's most powerful GEHR; the Carina complex might represent a miniaturized version of 30 Doradus -- a smaller sample in both space and time of the processes at work in a GEHR.  Carina's ``cluster of clusters'' even more closely resembles NGC~604, a GEHR in M33 (the second-largest GEHR in the Local Group) that \citet{Maiz04} describe as a ``Scaled OB Association,'' an extended structure made up of multiple adjacent massive star clusters and their bubbles and superbubbles.

\subsection{Past X-ray Surveys}
 
The \object{Carina Nebula} has long been a favorite target of X-ray telescopes, most often because \etacar is a bright, variable X-ray source \citep{Corcoran95a}.  An unusual X-ray nebula surrounds \etacar and the Homunculus; it was detected by {\em Einstein} \citep{Seward79}, better imaged by {\em ROSAT} \citep{Corcoran95b, Weis01}, and studied in detail with \Chandra \citep{Seward01, Weis04, Corcoran04, Corcoran07}.  Since \etacar is completely saturated in the \Chandra observations to be described here, we will not discuss it further; the reader is referred to several recent studies and reviews to learn more about this amazing luminous blue variable system and its unparalleled X-ray emission \citep[e.g.,][]{Davidson97,Pittard02,Hamaguchi07b,Henley08,Okazaki08,Parkin09,Parkin11a}.  Here we will focus on wide-field X-ray imaging surveys of the Carina Nebula; papers on individual X-ray point sources in Carina (usually massive stars) will be described later.
 
The first X-ray mission capable of imaging multiple point sources and diffuse emission in the Carina field was {\em Einstein}; it showed that several of the most massive stars in Carina were bright X-ray emitters and that the field was pervaded by diffuse emission that could be due to stellar winds or past supernova activity in the complex \citep{Seward79}.  A later {\em Einstein} paper \citep{Seward82} used the Imaging Proportional Counter (with $\sim 1\arcmin$ resolution) to establish the broad extent of the diffuse X-ray emission ($\sim$2800 square arcminutes) and its total luminosity ($2.09 \times 10^{35}$~erg~s$^{-1}$ assuming D=2.6~kpc) and the High Resolution Imager (with $\sim$$3\arcsec$ resolution) to catalog $\sim$25 X-ray point sources associated with massive stars, including all three Wolf-Rayet stars and $\eta$~Car.  Seward and Chlebowski found a strong luminosity correlation for Carina's massive stars ($L_{X} \simeq 2 \times 10^{-7} L_{Bol}$) and established that their X-ray emission is harder than the diffuse emission.  They fit X-ray spectra of the diffuse emission with a thermal plasma model and found a typical extinction of $N_{H} \sim 3 \times 10^{21}$~atoms~cm$^{-2}$ and plasma temperature of $kT \sim 0.7$~keV (8 million degrees Kelvin).  They note that X-ray absorption by the dense nebula requires that the diffuse X-ray emission lie on the near side of the Carina complex and that the energetics of the region support the idea that powerful winds from Carina's massive stars fuel the diffuse X-ray emission.  At about the same time, an {\em IUE} study of interstellar absorption lines \citep{Walborn82} showed that high-velocity gas coincides with the regions of bright X-ray emission seen by {\em Einstein}.
 
The {\em ROSAT} data (Figure~\ref{fig:intro}) also show bright point sources suffused by diffuse X-ray emission.  \citet{Corcoran95b} performed two-temperature thermal plasma spectral fits on the Position-Sensitive Proportional Counter (PSPC) detections of 5 bright massive stars in Carina, finding lower-temperature components of 0.2--0.8~keV and higher-temperature components of 1.4--4.0~keV.  These fits yielded absorption-corrected luminosities roughly consistent with the {\em Einstein} results \citep{Seward82}; PSPC luminosity estimates for $\sim$20 massive stars confirmed the {\em Einstein} $L_{X} \simeq 2 \times 10^{-7} L_{Bol}$ correlation for massive stars.  \citet{Corcoran95b} also fit the {\em ROSAT}/PSPC diffuse emission data, dividing the Nebula into 7 separate regions.  Except for the region near $\eta$~Car, all of these single-temperature spectral fits yielded soft thermal plasmas, with $0.1 < kT < 0.4$~keV.

{\em XMM-Newton} observed a $\sim 30\arcmin$ field centered on \etacar early in its mission, obtaining a 44-ks exposure in the very broad energy range 0.3--12~keV \citep{Albacete03}.  This dataset covered the young cluster Tr16 and part of the nearby young cluster Tr14; Albacete-Colombo et al.\ tabulated 80 point source detections, including 24 known massive stars.  They performed spectral fits on the 14 brightest sources (all known massive stars) and found that most of these sources were best-fit with sub-solar metallicities.  Many of them required a two-temperature thermal plasma model, with lower-temperature components in the range 0.1--0.8~keV and higher-temperature components in the range 0.9--6.5~keV.  The lower-temperature components showed a correlation of $L_{X} \simeq 6 \times 10^{-7} L_{Bol}$ (consistent with the {\em Einstein} and {\em ROSAT} results given the broader bandpass of {\em XMM}) but the higher-temperature components showed no correlation and likely arise from a different physical mechanism:  magnetic fields \citep{Babel97}, wind collisions in binaries \citep[e.g.,][]{Stevens92,Pittard09}, or non-thermal emission from inverse Compton scattering of UV photons \citep{Chen91}.

\citet{Antokhin08} combined the {\em XMM} \etacar pointings described above with 3 later {\em XMM} observations of WR~25, for a total exposure of $\sim$160~ks.  They restricted their study to the energy range 0.4--10~keV and found 235 point sources, mostly massive stars and pre-MS stars in the Carina complex.  They performed spectral fits on 24 massive stars with results similar to \citet{Albacete03}:  many sources required multiple-temperature thermal plasmas with a cooler component below 1~keV and a hotter component above 1~keV.  These authors also note that the X-ray properties of binary and single massive stars are often more similar than simple theories would predict, due to many confounding effects in the physics of the interactions of massive binary systems.  Extreme X-ray properties, on the other hand, may indicate binarity or strong magnetic fields in massive stars.

Although the {\em Suzaku} ({\em Astro-E2}) telescope has poor spatial resolution ($\sim$$2\arcmin$) compared to {\em XMM} ($\sim$$10\arcsec$) or \Chandra ($\sim$$1\arcsec$), it has excellent spectral resolution, thus it has proved useful in studying Carina's diffuse X-ray emission.  \citet{Hamaguchi07a} analyzed 60~ks of {\em Suzaku} data centered on $\eta$~Car, carefully accounting for point source emission by comparing to {\em XMM} and \Chandra data.  They found a dominant thermal plasma temperature for the diffuse emission of 0.6~keV with spatial variations in spectral lines indicating elemental abundance variations across the nebula, with much higher Si and Fe abundances south of \etacar than north of it.  They postulate that these abundance variations could be caused by one or more cavity supernovae exploding inside the Carina complex, or they could be caused by dust creation and destruction associated with this young star-forming region.  \citet{Ezoe08} performed a similar analysis on a 77-ks {\em Suzaku} observation of the eastern arm of diffuse emission seen by {\em Einstein} and {\em ROSAT} (Figure~\ref{fig:intro}); plasma temperatures there are consistent with the results of Hamaguchi et al.  These authors conclude that the eastern arm of diffuse emission has the same origin as that in the center of the nebula and that either stellar winds or young cavity supernovae could supply sufficient energy to generate the X-ray plasma.

Very early in the \Chandra mission, $\sim$22~ks of data were obtained for a $17\arcmin \times 17\arcmin$ field centered on $\eta$~Car, using ACIS-I.  These early mission check-out data were obtained at a non-standard focal plane temperature when ongoing radiation damage was altering the charge transfer efficiency of the CCDs, so they are not well-calibrated.  Despite these difficulties, \citet{Evans03} identified 151 point sources in a 9.6~ks subset of these data, including 23 massive stars (in addition to $\eta$~Car).  They note that intermediate-mass stars (late-B and A spectral types) have X-ray luminosities similar to the lower-mass pre-MS stars seen in their sample, thus their X-ray emission may be due to lower-mass companions.  X-ray spectra for 13 massive stars and 3 pre-MS from this sample were analyzed by \citet{Evans04} using the full 22-ks dataset; results for the massive stars were similar to the {\em XMM} studies, often requiring two-temperature thermal plasma fits with a soft and a hard component.  The lower-mass stars were systematically harder than the massive stars.  This same early \Chandra dataset was reanalyzed by \citet{Sanchawala07a}, resulting in a net exposure of $\sim$18~ks and the detection of 454 point sources.  All known O stars in the field were detected in this analysis of the {\em Chandra}/ACIS-I data but only 10\% of the known early B stars were found.  These authors use the \Chandra source list with near-IR photometry to propose a list of 16 candidate OB stars, 180 candidate pre-MS stars, and a small clump of X-ray sources that may be an obscured, physically-related group.  By scaling the X-ray luminosity function (XLF) of this population to that in the Orion Nebula Cluster \citep{Getman05}, Sanchawala et al.\ estimate the total population of Tr16 to be 1000--1300 stars.

In 2004, {\em Chandra}/ACIS-I observed Tr14 for 57~ks (PI Townsley).  Using custom analysis methods developed for the crowded starburst cluster R136 in 30~Doradus \citep{Townsley06a,Townsley06b}, this observation yielded over 1600 X-ray point sources, evidence for spatial variations in the spectra of the diffuse emission (including Ne and Fe enhancements), and comparisons between early-O star spectra indicating that some O3 stars show only soft thermal plasma components while others require both soft and hard components \citep{Townsley05, Feigelson07, Townsley09a, Townsley09b}.  An independent analysis of this dataset with more standard techniques yielded only $\sim$500 X-ray sources \citep{Sanchawala07b}, demonstrating that standard \Chandra data analysis methods are inadequate for crowded fields pervaded by diffuse X-ray emission.

Finally, in 2006 {\em Chandra}/ACIS-I re-observed Tr16 for 88~ks (PI S.\ Murray).  These data were analyzed by \citet{Albacete08}, resulting in a published list of 1035 X-ray sources; they did not analyze the diffuse emission in the field.  Albacete-Colombo et al.\ performed spectral fitting on all sources with $\geq$20 net counts using a one-temperature thermal plasma model, finding a median aborption of $A_{V} \sim 3.3$~mag and a median temperature of $kT \sim 2$~keV.  Variable sources showed harder spectra, presumably due to coronal heating in magnetic-reconnection flares; this behavior is typical of pre-MS stars \citep[e.g.,][]{Favata05}.  The 28 massive stars in the field typically were fit with two-component thermal plasmas; a soft plasma was always needed and for many massive stars a hard component was also required, as other studies have shown.  These authors find that the Orion Nebula Cluster (ONC) shows systematically brighter X-ray luminosities for 1.5--2.5~M$_{\odot}$ pre-MS stars \citep{Preibisch05a} than Tr16 exhibits and assert that this is due to the age difference between these clusters (assuming an age of 1~Myr for the ONC and an age of 3~Myr for Tr16).

\subsection{The Present Survey}

The \Chandra Carina Complex Project (CCCP) was selected in 2007 as a {\em Chandra X-ray Observatory} Cycle 9 Very Large Project (VLP).  Observations were carried out over 9 months in 2008.  Its primary goal is to catalog and characterize the X-ray-emitting point sources and diffuse emission in the Carina star-forming complex using the highest spatial resolution currently available, to advance the community's understanding of the high-energy processes at work in massive star-forming complexes.  The CCCP involves over 50 scientists organized into seven science groups, each with a group leader reporting to the project Principal Investigator (Townsley).  These groups and their lead scientists are:  Data Products (Patrick Broos), Massive Stars (Marc Gagn{\'e} and Michael Corcoran), Revealed Stellar Populations (Eric Feigelson), Obscured Stellar Populations (Thierry Montmerle), Diffuse Emission (Leisa Townsley), Multiwavelength Studies (Nathan Smith), and Global Synthesis (Sally Oey).  

In this brief introduction to the CCCP, first we present our motivation for the survey and an orientation to the observations (Section~\ref{sec:motivation}).  Next the \Chandra observations are tabulated and a few important data analysis steps are noted (Section~\ref{sec:observations}).  An orientation to the survey is given in Section~\ref{sec:overview}, using a variety of full-field mosaic images.  Then we present some early findings (Section~\ref{sec:results}), organized by relevance to each CCCP science group.  Detailed descriptions of CCCP data analysis, data products, and extensive science results will be presented in the suite of papers that follow in this {\em Special Issue}, led by many of the CCCP participating scientists.

%=============================================================================

\section{SURVEY DESIGN AND MOTIVATION \label{sec:motivation}}

The CCCP is a 1.42 square degree \Chandra X-ray survey of the Great Nebula in Carina.  This 1.2~Ms mosaic was obtained with the Imaging Array of the ACIS camera (ACIS-I) \citep{Garmire03} and consists of 20 new, spatially contiguous pointings combined with archival {\em Chandra}/ACIS-I data to provide a comprehensive view of thousands of young stars and hot gas that are contributing to the X-ray emission from this complex of massive star-forming regions.  For the purposes of this paper, we assume that all X-ray sources in Carina are at the distance of the famous luminous blue variable $\eta$~Car, established by \citet{Smith06a} to be 2.3~kpc using kinematic measurements of the Homunculus.  At this distance, our survey samples an irregular region in the center of the Carina Nebula roughly 50~pc across.  In reality, Carina's constituent massive stellar clusters may span a substantial range of distances, but a spatially-dependent reddening law makes it difficult to establish those distances accurately \citep{Smith08}.  

The ``backbone'' of our \Chandra survey maps the highest concentration of O stars in the region, following a ridge of young clusters that defines the Carina star-forming complex.  The locations of our other ACIS pointings are governed by this backbone of cluster observations:  cluster aimpoints were placed to achieve the highest spatial resolution at the crowded cluster centers, then other pointings filled out the field, overlapping enough to minimize holes in the survey coverage (observation roll angles could not be specified in advance).  Pointings extend far enough to the east and west to capture the brightest parts of the unresolved emission seen in the {\em ROSAT} data, but the survey is too small to capture the full extent of the {\em ROSAT} emission, especially in the lower, more open superbubble lobe.  Several pointings south of the ridge of young clusters sample the South Pillars region and cover much of the {\em Spitzer} South Pillars Survey \citep{Smith10b}. 

Figure~\ref{fig:intro}b shows a smoothed {\em ROSAT}/PSPC image of the Great Nebula in Carina, with the CCCP $17\arcmin \times 17\arcmin$ ACIS-I observations outlined with boxes. ÊThe large blue circle roughly outlines the {\em ROSAT} field of view. ÊMultiple ACIS-I boxes overlaid on a single pointing denote the different roll angles used in different observations (catalogued as uniquely-numbered ``ObsID's''); details of the CCCP observations are given in Section~\ref{sec:observations}. ÊMost pointings consist of multiple ObsID's, but often the same roll angle was used.  Yellow labeled pointings in the center identify data that came from the \Chandra archive: Ê``Tr14'' is the 2004 observation of Trumpler~14 and ``Tr16'' is the 2006 observation of Trumpler~16.  The anomalous X-ray pulsar 1E1048.1-5937 lies just off the eastern edge of the CCCP field of view; it is seen as a bright point source in the {\em ROSAT} image.  It is believed to be an unrelated background object \citep{Gaensler05}.  

To motivate the design of the CCCP, we conceptually divided the ACIS mosaic into four major regions based on their morphology:  the main Cluster of Clusters, the star-forming South Pillars, the complex mix of pointlike and diffuse emission in Carina East, and the soft-X-ray-filled Superbubble.  These are defined in Figure~\ref{fig:intro}b.  Each of these regions highlights different long-term science goals that we have for the CCCP data.

{\bf The Cluster of Clusters}\\
In Carina's clustered massive star-forming regions we will study OB stars and their winds, mass segregation, pre-MS star populations, and, through X-ray selection of IR sources, the evolution of protostellar disks, which are affected both by hard radiation and hot gas in the cluster environment and by flares from the host pre-MS stars. \Chandra has discovered that some \hii regions are filled with diffuse X-ray plasma (T$\sim$$10^7$~K, $L_X$$\sim$$10^{32-33}$~erg~s$^{-1}$) from OB wind shocks, pushing the $10^4$~K H$\alpha$-emitting plasma due to UV ionization into a thin Str{\" o}mgren shell \citep[e.g.,][]{Townsley03}.  The X-ray morphology of this shocked plasma in Carina's massive star-forming regions will provide a detailed diagnostic of the injection of kinetic energy into the ISM.

{\bf Carina East}\\
This part of the Nebula shows complex diffuse X-ray emission that may be associated with recent supernovae or with OB wind shocks interacting with dense portions of the ISM \citep{Ezoe08}.  The \Chandra mosaic allows us to determine abundances and temperatures throughout this region, largely uncontaminated by point source emission that is unresolved by other X-ray observatories.  These pointings and others around the periphery of the ``cluster of clusters'' may sample distributed star formation that occurred spontaneously in the Carina complex, independent of the massive clusters that dominate the region today, or that trace an older (5--10~Myr) population of young stars that has drifted far enough from its formation site that its clusters are no longer distinguishable.  How do pre-MS stars distribute themselves around more massive stars in these distributed samples?  Are protoplanetary disks more or less prevalent in these less hostile environments?  These issues can be addressed by our X-ray observations alone and by studying IR properties of X-ray-selected samples.

{\bf The South Pillars}\\
This has been described as the center of ongoing star formation in Carina, influenced by \etacar and other massive stars in Tr16 and Tr14 \citep{Smith08,Smith10b}.  Here O star winds and radiation are eroding pillars; we can study how that material in turn affects the hot plasma.  \Chandra observations penetrate deeply into obscuring molecular clouds, so even in the South Pillars we can study the most recent generation of stars just now forming; the {\em Spitzer} survey of this region finds over 900 young stellar objects there \citep{Smith10b}.  More broadly, the CCCP mosaic covers all regions of possible embedded star formation throughout the complex seen by {\em MSX} \citep{Smith08}.

{\bf The Superbubble}\\
The CCCP samples the upper closed loop, the lower more open shell, and the dust-enshrouded ``waist'' of Carina's proto-superbubble \citep{Smith00}.  These pointings also survey the wider stellar content of the Nebula, due either to distributed star formation or to the dispersal of past generations of massive clusters as described above.  In GEHRs such as 30~Doradus, multiple ``cavity supernovae'' produce soft X-rays filling $\sim$30--100~pc bubbles with $L_X$$\sim$$10^{35-36}$~erg~s$^{-1}$ plasma while showing no classical radio or visual supernova signatures \citep{Chu90, Wang91, Townsley06a}.  Carina's brightest diffuse X-ray regions may also be supernova-powered; the overabundance of Fe suggested by our Tr14 data \citep{Townsley05, Townsley09a} and by recent {\em Suzaku} observations \citep{Hamaguchi07a}, as well as high-velocity expanding structures \citep{Walborn07}, imply that one or more supernovae has already occurred in Carina's infant superbubble.  Theory predicts that hot gas should quickly fill the superbubble cavities and show up as center-filled soft X-rays \citep{Basu99}; if this occurred in Carina, the hot plasma seen throughout our mosaic should be the same.  Differences in intrinsic surface brightness or temperature should define distinct structures such as cavity supernova remnants or wind shocks.  From {\em Einstein} data, Carina's global X-ray properties are $L_X$$\sim$$2 \times 10^{35}$~erg~s$^{-1}$, kT$\sim$0.7~keV \citep{Seward82}.  These values are similar to the diffuse X-ray emission seen in supernova-brightened superbubbles in 30~Doradus, thus our mosaic will elucidate the energetics of that canonical GEHR.

%=============================================================================

\section{{\em CHANDRA} OBSERVATIONS AND DATA ANALYSIS \label{sec:observations}}

As noted above, the CCCP covers $\sim$1.42 square degrees (5112 square arcminutes), or $\sim$2300~pc$^2$ at an assumed distance of 2.3~kpc \citep{Smith06a}, using 20 new ACIS-I pointings; the nominal VLP survey exposure is 60~ks.  Generally we only include data from the ACIS-I array; although ACIS-S array CCDs S2 and S3 were operational for all observations, the \Chandra point spread function (PSF) is highly degraded at large off-axis angles, so most sources on S2 and S3 become hopelessly crowded and dominated by background.  The exceptions to this practice were a few bright sources in Carina, typically important massive stars; for these bright sources, background and events from faint nearby sources become negligible, while the added time samples and exposure yield valuable information, so special ACIS-S extractions were performed.

Data from three earlier ACIS-I observations were obtained from the \Chandra archive to augment the VLP survey:  ObsID 6402, an 88-ks observation of Tr16 (PI S.\ Murray, part of the HRC Instrument Team Guaranteed Time program from Cycle 7), ObsID 4495, a 58-ks observation of Tr14 (Cycle 5, PI L.\ Townsley), and ObsID 6578, a 10-ks snapshot of the Treasure Chest Cluster (PI G.\ Garmire, part of the ACIS Instrument Team Guaranteed Time program from Cycle 7).  Two of the new pointings were contributed to the VLP survey by G.\ Garmire as part of the ACIS Team Guaranteed Time for Cycle 9.  The \Chandra archive contains several transmission gratings observations of \etacar and the O stars HD~93250 and HD~93129 but, due to limited resources and complications that arise when combining data from front-illuminated and back-illuminated CCDs, we omit all gratings observations from CCCP analysis.

Table~\ref{tbl:obslog} gives the observing log for the project.  The VLP data were acquired over 9 months, from 12 February 2008 through 15 October 2008; the nominal survey exposure was often accumulated via 2 or 3 observations per pointing due to spacecraft thermal constraints.  The final dataset includes \dataset[ADS/Sa.CXO#DefSet/CCCP]{38 ObsID's}.  All data were obtained in the ``Very Faint'' ACIS observing mode; this mode telemeters $5 \times 5$-pixel event islands and is useful for suppressing background\footnote{\url{http://cxc.harvard.edu/cal/Acis/Cal_prods/vfbkgrnd/index.html}}.  

%-----------------------------------------------------------------------------
\begin{deluxetable}{lrccrccrl}
\centering 
%\rotate
\tabletypesize{\tiny} \tablewidth{0pt}
\tablecolumns{9}
\tablecaption{ Log of \Chandra Observations 
 \label{tbl:obslog}}
\tablehead{
\colhead{Target} & 
\colhead{ObsID} & 
\colhead{Sequence} & 
\colhead{Start Time} & 
\colhead{Exposure} & 
\multicolumn{2}{c}{Aimpoint} & 
\colhead{Roll Angle} & 
\colhead{Notes} \\
\cline{6-7}
\colhead{} & 
\colhead{} & 
\colhead{} &
\colhead{(UT)} & 
\colhead{(s)} & 
\colhead{$\alpha_{\rm J2000}$} & 
\colhead{$\delta_{\rm J2000}$} & 
\colhead{(\arcdeg)} & 
\colhead{} 
}
\startdata
% Cluster of Clusters
Clusters 1 & 9481 & 900818 & 2008 Aug 10 10:48 & 44583 & 10:42:39.04 & -59:46:47.5 & 200 & ACIS GTO\tablenotemark{a} \\
Clusters 1 & 9894 & 900818 & 2008 Aug 09 13:22 & 14408 & 10:42:39.04 & -59:46:47.5 & 200 & ACIS GTO 
\medskip \\
Clusters 2 & 9482 & 900819 & 2008 Aug 18 09:21 & 56512 & 10:43:55.67 & -59:59:40.0 & 193 & ACIS GTO, \object{QZ Car}, \object{WR~24}
\medskip \\
Clusters 3 & 9483 & 900820 & 2008 Aug 28 23:10 & 59385 & 10:45:50.10 & -59:57:14.6 & 184 & \object{Treasure Chest} \\
Clusters 3 & 6578 & 400483 & 2006 Apr 16 01:26 &  9829 & 10:45:54.59 & -59:57:13.9 & 309 & ACIS GTO, Treasure Chest
\medskip \\
Clusters 4 & 9484 & 900821 & 2008 Aug 19 06:35 & 55322 & 10:44:40.62 & -59:21:39.5 & 193 & \object{Trumpler 15}, \object{HD 93403}
\medskip \\
Clusters 5 & 9485 & 900822 & 2008 Sep 08 12:20 & 57389 & 10:47:08.69 & -60:05:55.8 & 174 & \object{Bochum 11}
\medskip \\
Clusters 6 & 9486 & 900823 & 2008 Jul 30 20:28 & 38678 & 10:46:04.21 & -59:31:31.0 & 212 & HD 93403 \\
Clusters 6 & 9891 & 900823 & 2008 Aug 01 14:56 & 20657 & 10:46:04.21 & -59:31:31.0 & 212 & HD 93403
\medskip \\
Clusters 7 & 9487 & 900824 & 2008 Jul 28 05:58 & 38675 & 10:43:17.24 & -59:09:33.4 & 212 & \object{Bochum 10} \\
Clusters 7 & 9890 & 900824 & 2008 Jul 29 22:38 & 19914 & 10:43:17.24 & -59:09:33.3 & 212 &         Bochum 10
\medskip \\
Trumpler 14 & 4495 & 200264 & 2004 Sep 21 17:29 & 56634 & 10:43:55.91 & -59:32:41.5 & 160 & \object{Trumpler 14}, \object{HD 93250}
\medskip \\
Trumpler 16 & 6402 & 200379 & 2006 Aug 30 18:40 & 86905 & 10:44:47.93 & -59:43:54.2 & 183 & HRC GTO, \object{Trumpler 16}, \object{WR~25}
\bigskip \\
%
% Carina East
East 1 & 9488 & 900825 & 2008 Sep 05 21:24 & 59389 & 10:46:55.97 & -59:46:13.2 & 177 & \object[EHG7]{neutron star}
\medskip \\
East 2 & 9489 & 900826 & 2008 Sep 03 06:16 & 58403 & 10:48:08.74 & -59:32:39.5 & 180 & the eastern ``hook''
\medskip \\
East 3 & 9490 & 900827 & 2008 Sep 09 19:29 & 55894 & 10:48:22.20 & -59:58:28.4 & 174 &
\medskip \\
East 4 & 9491 & 900828 & 2008 Oct 14 06:14 & 50021 & 10:49:06.02 & -59:45:10.2 & 134 &  \\
East 4 & 10784 & 900828 & 2008 Oct 15 17:13 & 9529 & 10:49:06.01 & -59:45:10.0 & 134 &
\bigskip \\
%
% South Pillars
South Pillars 1 & 9492 & 900829 & 2008 Feb 12 15:56 & 18803 & 10:45:18.41 & -60:11:49.0 & 15 &  \\
South Pillars 1 & 9816 & 900829 & 2008 Feb 15 06:23 & 20767 & 10:45:18.42 & -60:11:49.0 & 15 &  \\
South Pillars 1 & 9817 & 900829 & 2008 Mar 07 10:46 & 19927 & 10:45:18.95 & -60:11:59.4 & 353 &
\medskip \\
South Pillars 2 & 9493 & 900830 & 2008 Feb 25 20:30 & 19788 & 10:43:13.43 & -60:14:30.1 & 0 & WR~24 \\
South Pillars 2 & 9830 & 900830 & 2008 Feb 28 12:29 & 19785 & 10:43:13.42 & -60:14:30.1 & 0 & WR~24 \\
South Pillars 2 & 9831 & 900830 & 2008 Mar 01 02:38 & 14261 & 10:43:13.42 & -60:14:30.2 & 0 & WR~24
\medskip \\
South Pillars 3 & 9508 & 900840 & 2008 May 23 02:40 & 35582 & 10:48:48.41 & -60:14:55.9 & 266 &  \\
South Pillars 3 & 9857 & 900840 & 2008 May 22 05:19 & 19813 & 10:48:48.43 & -60:14:56.0 & 266 & 
\medskip \\
South Pillars 4 & 9494 & 900831 & 2008 May 20 11:56 & 36568 & 10:46:55.98 & -60:22:29.8 & 267 & cluster of galaxies \\
South Pillars 4 & 9856 & 900831 & 2008 May 21 17:00 & 18868 & 10:46:55.98 & -60:22:29.8 & 267 & cluster of galaxies
\medskip \\
South Pillars 5 & 9495 & 900832 & 2008 Apr 24 00:58 & 30963 & 10:44:49.30 & -60:26:00.1 & 300 &  \\
South Pillars 5 & 9849 & 900832 & 2008 Apr 26 16:32 & 28480 & 10:44:49.31 & -60:26:00.2 & 300 & 
\bigskip \\
%
% Superbubble
Superbubble 1 & 9496 & 900833 & 2008 May 27 04:58 & 26723 & 10:41:32.49 & -59:36:03.4 & 263 & \object{WR~22} \\
Superbubble 1 & 9860 & 900833 & 2008 May 28 10:21 & 32126 & 10:41:32.50 & -59:36:03.4 & 263 &         WR~22
\medskip \\
Superbubble 2 & 9497 & 900834 & 2008 Jul 22 22:14 & 31866 & 10:40:31.80 & -59:49:53.5 & 214 &   \\
Superbubble 2 & 9889 & 900834 & 2008 Jul 25 13:49 & 26814 & 10:40:31.80 & -59:49:53.5 & 214 &  
\medskip \\
Superbubble 3 & 9498 & 900835 & 2008 May 24 17:32 & 31671 & 10:41:52.74 & -60:02:12.4 & 262 &   \\
Superbubble 3 & 9859 & 900835 & 2008 May 31 00:55 & 28162 & 10:41:52.74 & -60:02:12.4 & 262 &  
\medskip \\
Superbubble 4 & 9499 & 900836 & 2008 May 30 00:35 & 31443 & 10:42:34.60 & -59:23:13.7 & 260 &   \\
Superbubble 4 & 9861 & 900836 & 2008 Jun 21 05:19 & 28183 & 10:42:33.38 & -59:23:13.4 & 242 &   \\
\enddata
\tablenotetext{a}{ GTO = Guaranteed Time Observation}
\tablecomments{  Exposure times are the net usable times after various filtering steps are applied in the data reduction process. The aimpoints and roll angles are obtained from the satellite aspect solution before astrometric correction is applied.  Units of right ascension are hours, minutes, and seconds; units of declination are degrees, arcminutes, and arcseconds.}
\end{deluxetable}
%-----------------------------------------------------------------------------

Custom data analysis was performed at Penn State; a detailed account of many of these analysis steps was presented by \citet{Broos10}.  To summarize briefly, we perform our own ``Level 1 to Level 2'' ACIS event processing \citep{Townsley03}, removing the event position randomization added by the standard analysis pipeline (this sharpens the system PSF), applying our own custom bad pixel map (to retain CCD columns that are acceptable for our science goals but that are removed by the standard bad pixel map), and making various other filtering decisions that differ from the pipeline processing \citep{Broos11a}.  

To detect point sources, we apply {\em wavdetect} \citep{Freeman02} at a range of spatial scales and in several energy bands \citep{Broos10}, but {\em wavdetect} often misses sources in crowded regions or in regions with spatially-varying backgrounds.  To try to improve our detection of such sources, we perform maximum likelihood image reconstruction and find potential sources in these reconstructed images \citep{Townsley06b}.  These reconstruction sources are merged with the {\em wavdetect} sources to remove duplicates, then this list of potential sources is input to {\it ACIS Extract} \citep{Broos03}, our publicly-available custom IDL software that was specifically designed to perform event extraction, source validation, and many other analysis steps on ACIS data that consist of multiple non-aligned ObsID's \citep{Broos10}.  For the CCCP, {\it ACIS Extract} produced a final list of 14,369 X-ray point sources, with X-ray photometry for all sources and lightcurves and spectral fits for sources of sufficient brightness.  Again please note that \etacar was omitted from the CCCP source list due to its extreme \anchorfoot{http://cxc.harvard.edu/ciao/why/pileup_intro.html}{{photon pile-up}}, so technically the CCCP detected 14,370 point sources.  

The point sources are designated ``CXOGNC~J'' for ``{\em Chandra X-ray Observatory} Great Nebula in Carina epoch J2000'' followed by a number indicating celestial coordinates.  Each source has a working CCCP label as well, indicating the pointing in which it can be found (e.g., SB3\_25 is the 25th source found in Superbubble Pointing 3).  The CCCP X-ray point source positions and X-ray properties are tabulated in \citet{Broos11a}.  The reliability of {\it ACIS Extract} broad-band fluxes and absorption estimates is validated by \citet{Getman10}.  Automated spectral fits using simple spectral models were generated by {\it ACIS Extract} and used to inform more sophisticated spectral fits of some sources; they will be explored in more detail in future work.

{\it ACIS Extract} can also be used to extract events and perform spectral fits on diffuse emission regions.  Working with images where the point sources have been removed, we used the publicly-available {\em WVT Binning} software\footnote{\url{http://www.phy.ohiou.edu/~diehl/WVT/}} \citep{Diehl06} to tessellate the diffuse emission into regions of roughly constant signal-to-noise.  We converted these 161 tessellates into {\it SAOImage ds9} \citep{Joye03} region files and supplied this list of region files to {\it ACIS Extract} for event extraction and spectral fitting.  Details of our methods can be found in \citet{Broos10} and results of this analysis are given in \citet{Townsley11a}.

%=============================================================================

\section{SURVEY OVERVIEW \label{sec:overview}}

Here and in the following section we give some brief examples of the results emerging from the CCCP survey.  Some of the topics mentioned here are described in more detail in other {\em Special Issue} papers.  In other cases, unusual sources or circumstances that arose in the survey are described here as topics of interest.  First some full-field images are shown to orient readers to the CCCP survey and results, then some particularly unusual objects are described.

\subsection{Survey Images \label{sec:images}}

\begin{figure}[htb] 
\begin{center}  
\includegraphics{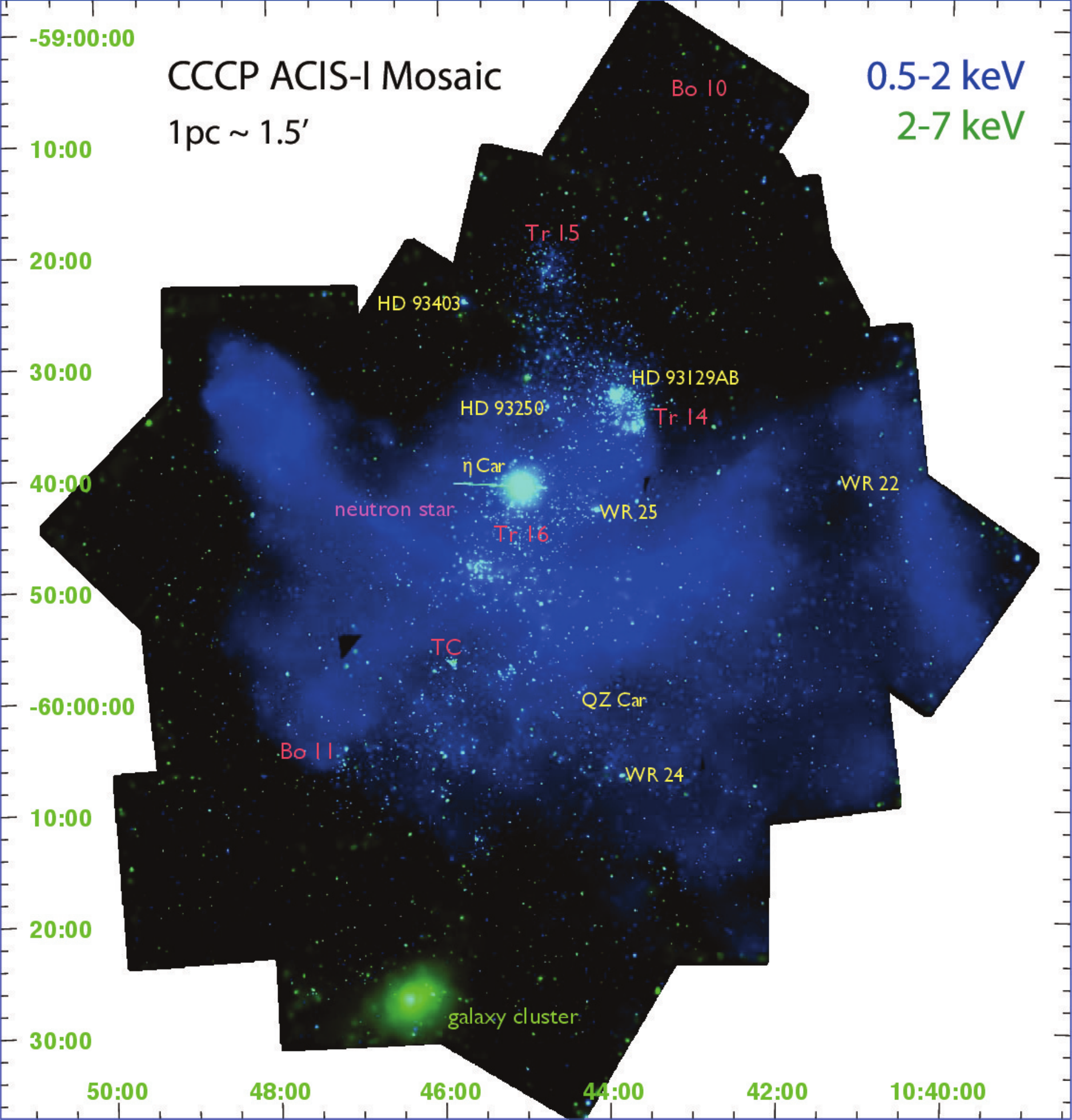}
\caption{
The \Chandra Carina Complex Project mosaic smoothed with {\it csmooth}, with soft-band (0.5--2~keV) emission shown in blue and hard-band (2--7~keV) emission shown in green.  This mosaic image should be compared to the map of ACIS-I pointings shown in Figure~\ref{fig:intro}b.  This and all other smoothed ACIS images in this paper depict the apparent surface brightness of the X-ray emission, i.e., they have been corrected for exposure and vignetting variations.  Well-known massive stars are labeled in yellow and historical clusters are shown in peach (TC = Treasure Chest).  The hard, bright diffuse structure in the south is a cluster of galaxies at a redshift of 0.1 (see Section~\ref{sec:galcluster}).
} 
\label{fig:cccpfull}
\end{center}
\end{figure}

The CCCP full survey is shown in Figure~\ref{fig:cccpfull} using a heavily-binned image smoothed with the adaptive-kernel smoothing tool {\em csmooth} \citep{Ebeling06}.  This image is meant to highlight diffuse features and orient the reader to well-known massive stars and clusters; it does not give a good rendering of the $>$14,000 X-ray point sources found in the CCCP. ÊThe nearly-horizontal line at about -59:40:00 in Figure~\ref{fig:cccpfull} is an instrumental artifact -- the readout streak from $\eta$~Car. ÊWR~25 also shows a faint readout streak.  The survey was designed to provide contiguous coverage across the Carina Nebula, but the actual observations involved slight changes in pointing centers and combinations of roll angles that left several small triangular holes in the mosaic.  These holes can be prominent in the survey images but they cause minimal loss of field coverage; none of the unexposed patches coincides with the location of any major cluster.

%-------------------------------------------------------------------------
%-------------------------------------------------------------------------

\begin{figure}[htb] 
\begin{center}  
\includegraphics[width=1.0\textwidth]{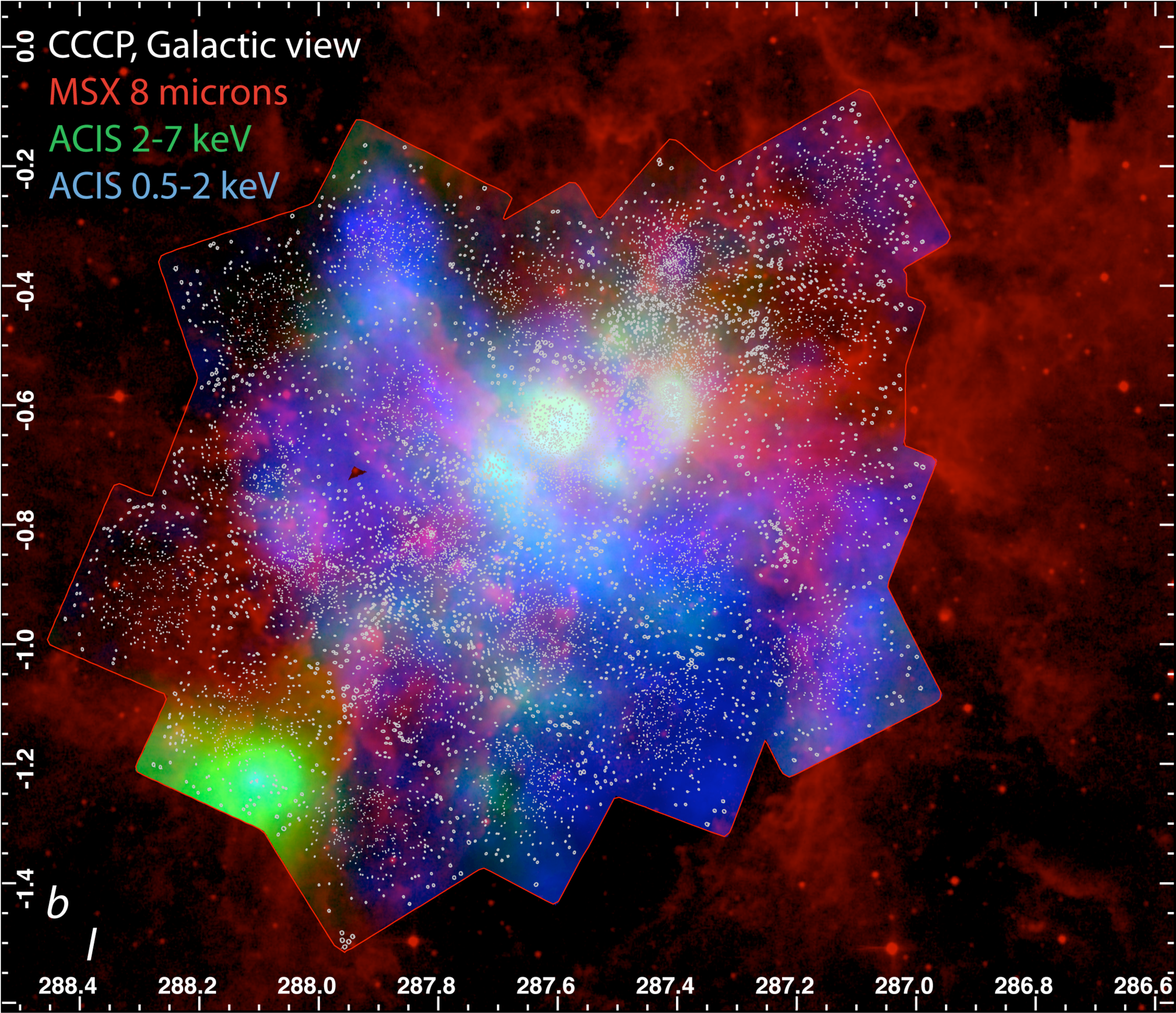}
\caption{The CCCP survey in Galactic coordinates; 10~pc $\simeq 0.25^{\circ}$.  The ACIS-I mosaic is outlined in red; comparison with Figure~\ref{fig:cccpfull} shows that the field has been rotated slightly to the right from its orientation in celestial coordinates.
Source-free adaptively-smoothed X-ray surface brightness images of the full CCCP survey region, with soft X-rays (0.5--2~keV) in blue and hard X-rays (2--7~keV) in green, are shown with the {\em MSX} 8~$\mu$m image in red.  The 14,369 ACIS point source extraction regions are overlaid (grey polygons).  The survey is centered roughly at (l,b) = (287.7$^{\circ}$,-0.8$^{\circ}$) and ends just below the Galactic midplane.  This image is extended westward of the CCCP mosaic to show the edges of the Carina bipolar superbubble outlined by PAH emission in the {\em MSX} image \citep{Smith00}.
} 
\label{fig:fluximage}
\end{center}
\end{figure}

Using a different representation of the CCCP survey, Figure~\ref{fig:fluximage} illustrates the X-ray point source population across the field by showing the event extraction region for each source used by {\it ACIS Extract}.  It is displayed in Galactic coordinates, to ease comparison with other (mainly IR) datasets and to show the orientation of the X-ray structures in relation to the Galactic Plane.  The source extraction regions are displayed on X-ray images where the point sources were excised, then the remaining counts adaptively-smoothed \citep{Townsley03,Broos10}; the {\em MSX} 8~$\mu$m image shows PAH emission and heated dust.  This technique emphasizes the diffuse X-ray emission morphology while preserving point source location information.  The {\em Chandra}/ACIS PSF is so sharp on-axis that not all source extraction regions are visible without significantly zooming this figure.  The massive stellar clusters Tr14, Trumpler~15 (Tr15), and Tr16 are easily discerned in this rendering, but the reader is warned that telescope vignetting and the rapidly-increasing size of the PSF off-axis significantly reduce {\em Chandra}'s point source detection sensitivity at off-axis angles $\geq 5\arcmin$ in each pointing, leading to an artificial spatial point source distribution that we call the ``egg crate effect.''

%-------------------------------------------------------------------------
%-------------------------------------------------------------------------

\begin{figure}[htb] 
\begin{center}
\includegraphics[width=1.0\textwidth]{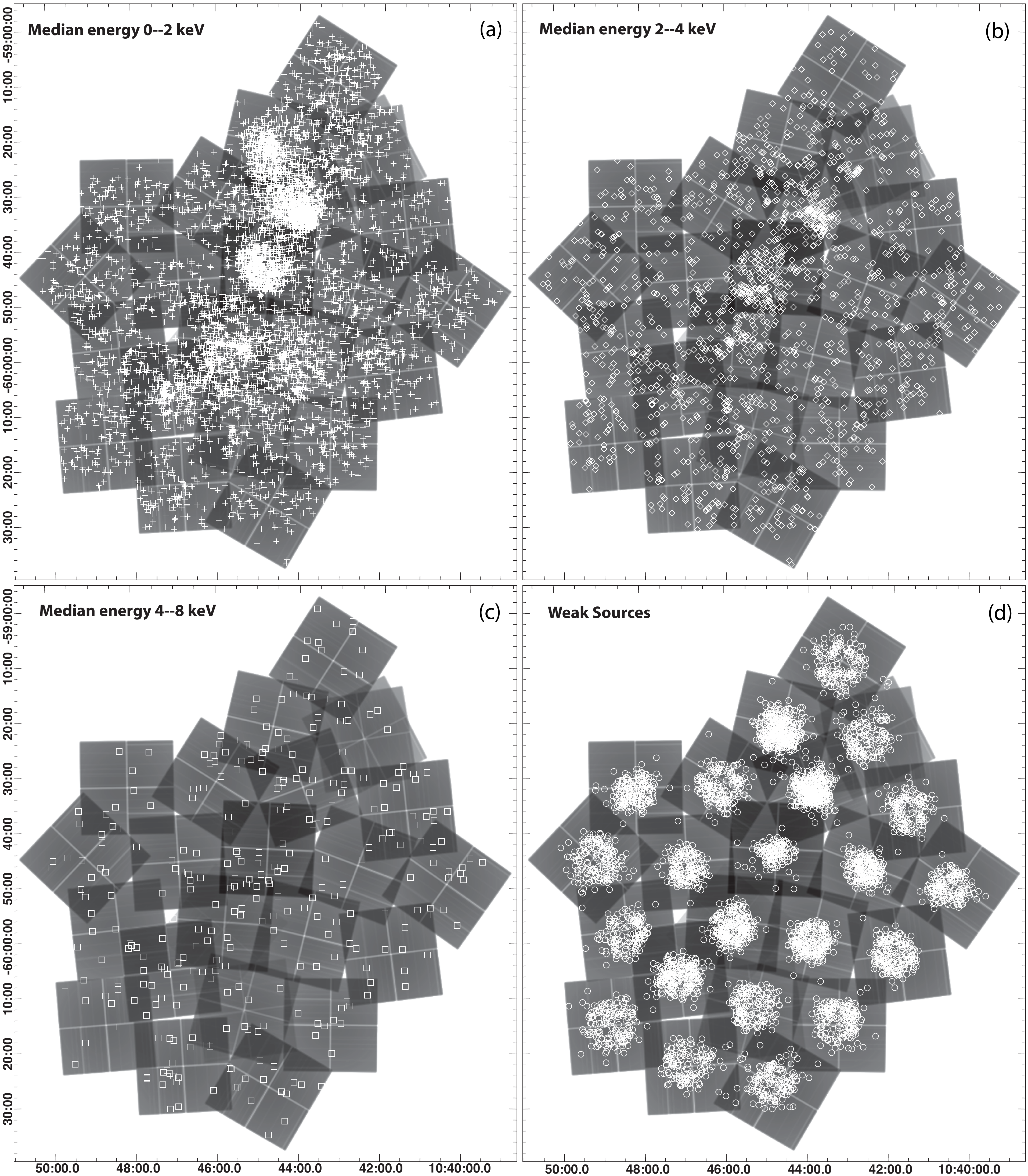}
%\mbox{
%\includegraphics[width=0.5\textwidth]{Emed_0-2_sources.pdf}
%\includegraphics[width=0.5\textwidth]{Emed_2-4_sources.pdf}}
%\mbox{
%\includegraphics[width=0.5\textwidth]{Emed_4-8_sources.pdf}
%\includegraphics[width=0.5\textwidth]{weak_sources_greyscale.pdf}
%}
\caption{Locations of X-ray point sources detected in the CCCP, subdivided first by net counts and then by median energy for the brighter sources, shown as symbols overlaid on the composite CCCP exposure map.
(a)--(c) Brighter X-ray sources (net counts $\geq$5.0), subdivided by median energy.  Weak sources (net counts $<$5.0) are not shown because their spatial distribution is strongly affected by detection sensitivity variations across the field.
(a) 7381 sources with 0~keV$< E_{med} <$2~keV; 
(b) 1696 sources with 2~keV$< E_{med} <$4~keV; 
(c) 321 sources with 4~keV$< E_{med} <$8~keV. 
(d) The 4971 weak X-ray sources (net counts $<$5.0), not subdivided by median energy, illustrating the strong detection sensitivity variations due to vignetting and increasing PSF size with off-axis angle in each ACIS pointing.  We call this spatial variation in faint source detections the ``egg crate effect'' and must account for it in large-scale spatial studies of the X-ray point source populations in Carina \citep{Feigelson11}.
} 
\normalsize 
\label{fig:sourcepops}
\end{center}
\end{figure}

Figure~\ref{fig:sourcepops} shows subsets of detected X-ray point sources displayed on the CCCP exposure map; again the full catalog of CCCP point sources and methods for deriving their X-ray properties are given in \citet{Broos11a}.  Panels (a)--(c) illustrate the spatial distribution of ``brighter'' sources, defined as those with total-band (0.5--8~keV) net counts $\geq$5.0, subdivided by median energy $E_{med}$.  By selecting only the brighter sources, we partially suppress the ``egg crate effect'' due to sensitivity variations across the field, giving a better sense of the true spatial distribution of sources.  A careful analysis of point source detection sensitivity and its variations across the CCCP field is presented by Broos et al.; Figure~\ref{fig:sourcepops} is merely a qualitative depiction of source locations and median energies and is not intended to be used for quantitative analysis or interpretation.

Carina's main clusters are most evident in the unobscured population with 0~keV$< E_{med} <$2~keV (Figure~\ref{fig:sourcepops}a); this is also the largest subsample of sources (79\% of the brighter sources).  A quantitative assessment of the clustering properties of Carina X-ray sources is performed by \citet{Feigelson11} using only the subsample of sources bright enough that their detections are complete across the entire CCCP field, based on the sensitivity and completeness analysis of \citet{Broos11a}.  

Sources with harder spectra and moderately-obscured sources account for 18\% of the sources with $\geq$5.0 net counts and appear in Figure~\ref{fig:sourcepops}b with 2~keV$< E_{med} <$4~keV; Tr14 and Tr16 are still prominent in this spatial distribution and a few small clumps of obscured sources are noticeable.  While only 3\% of the brighter sources are hard and/or highly-obscured with 4~keV$< E_{med} <$8~keV (Figure~\ref{fig:sourcepops}c), these are potentially very interesting because some of them may trace very young, embedded stars just forming in the Carina complex.  Lastly, Figure~\ref{fig:sourcepops}d completes the X-ray stellar census in Carina by showing the spatial distribution of weak sources (net counts $<$5.0) without regard for median energy.  These weak detections make up 35\% of the total X-ray point source sample in the CCCP.  Their spatial distribution is dominated by {\em Chandra}/ACIS sensitivity variations in each pointing.  Although little can be said about the X-ray properties of these very faint sources, their existence is still important to note, because their X-ray emission identifies them as part of the Carina young stellar population and their visual and IR properties can be studied.  

%-------------------------------------------------------------------------
%-------------------------------------------------------------------------

From these first impressions of the CCCP, it is clear that {\em Chandra}'s spatial resolution is critical for understanding the point source populations pervading the Carina Nebula.  This X-ray stellar sample is one of the largest ensembles of young stars born in the same molecular cloud complex catalogued to date; these source positions alone enable new studies of the clustering (and lack thereof) of Carina's young stars \citep{Feigelson11} and motivate visual and IR follow-up studies \citep[e.g.,][]{Preibisch11b, Povich11a}.  While Carina's famous massive clusters feature prominently in the X-ray sample, it is also clear that a distributed population of young stars pervades the complex; as mentioned before, this may trace an older pre-MS population suggested by earlier studies \citep[e.g.,][]{Ascenso07, DeGioia01}.  A careful assessment of the contaminating populations in this X-ray sample -- foreground stars, background stars, and extragalactic objects \citep[see Section~\ref{sec:data} below and][]{Getman11} -- gives us a useful rule-of-thumb:  we expect $<$1 contaminating source per square arcminute of the survey, evenly distributed across the field.  Thus we expect a large fraction of the brighter distributed X-ray sources to be tracing this slightly older population of young stars, with only modest contamination from unrelated X-ray sources.

These full-field images also show that the soft diffuse X-ray emission first seen by {\em Einstein} and {\em ROSAT} is truly diffuse; despite resolving out thousands of point sources cospatial with the diffuse emission, ACIS still clearly detects a bright, soft, diffuse component of highly-complex spatial structure and wide extent.  This diffuse component does not trace the spatial distribution of the point sources discovered by ACIS and it is substantially softer than the point source population; both of these facts indicate that the diffuse emission is unlikely to consist primarily of a yet-fainter, unresolved population of X-ray point sources in Carina.  Rather it likely traces hot plasma from the powerful winds of Carina's massive stars and perhaps even from past cavity supernovae that have exploded inside Carina's evolving bubble structures.  The nature of the diffuse emission will be addressed later in this paper and in other papers in this {\em Special Issue} \citep{Townsley11a,Townsley11b}.

\clearpage

%=============================================================================
\subsection{Unusual Objects \label{sec:unusual}}
\subsubsection{A New Cluster of Galaxies \label{sec:galcluster}}
%***Beware of multiple point sources that the machinery finds on top of this diffuse structure!  Check their spectra -- could pairs or multiples of them be the same (lensed) source?***

The CCCP survey uncovered a cluster of galaxies hidden (before now) by absorption in the Galactic Plane.  This object appears at the southern edge of the CCCP field, at a Galactic latitude of $\sim-1.2^{\circ}$, and is shown in Figures~\ref{fig:cccpfull} and \ref{fig:fluximage} as a bright green (hard) extended X-ray source.  In our CCCP study on diffuse emission \citep{Townsley11a} the cluster emission is mostly contained in the region named ``outside001''.  Its spectrum (Figure~5i in that study) is well-fitted by a single redshifted thermal plasma model, {\em TBabs*apec} in {\em XSPEC} \citep{Arnaud96}, with parameters  $N_{H} = 2.4 \times 10^{22}$~cm$^{-2}$, $kT = 6.3$~keV, and z = 0.10, with an intrinsic total-band (defined as 0.5--7~keV for diffuse emission) luminosity of $1.3 \times 10^{44}$~erg~s$^{-1}$.

%***Check the diffuse paper -- we should not be reporting Lx's for this tessellate unless we account for redshift and report a rough number for the cluster!***

While this cluster of galaxies is fairly bright and would be a well-known target if it didn't happen to lie very near the Galactic Plane, it is not a particularly remarkable member of its class.  Many brighter, less-absorbed examples of galaxy clusters in the Zone of Avoidance are known from {\em ROSAT} surveys \citep{Ebeling02}.  For the CCCP purpose of studying young stars and massive star feedback in Carina, this cluster of galaxies serves mainly as a background enhancement, diminishing our hard-band X-ray detection sensitivity to embedded or obscured young stars in part of the South Pillars and complicating the study of diffuse emission in that region.

\subsubsection{Isolated Neutron Star(s) \label{sec:neutronstar}}

An \object[EHG7]{isolated neutron star} in Carina was recognized recently from {\em XMM} point source catalog surveys \citep{Hamaguchi09, Pires09}.  This source (CXOGNC~J104608.71-594306.4, ACIS source label E1\_85, marked in pink in Figure~\ref{fig:cccpfull}) is a very soft X-ray emitter (with median energy 0.99~keV) and is quite bright, with 522 net counts in the CCCP data.  Its low column density and soft spectrum are evidence that it lies in the Carina complex rather than behind it along the Sagittarius-Carina spiral arm; a background soft X-ray source would have been much more highly absorbed, likely to the degree that it would become undetectable.  As mentioned above, Hamaguchi et al.\ note that this stellar remnant is compelling evidence for supernova activity in the Carina complex, likely from a population of 5--10~Myr old stars.  

We have searched for other isolated neutron star candidates in the CCCP data, using the criteria that the source must have a soft spectrum (median energy $<$1~keV) and have no known visual or IR counterpart.  We also consider only sources with $>$10 net counts, to ensure the fidelity of the median energy estimation and to guarantee a strong X-ray source detection.  Table~\ref{neutron_candidates.tbl} shows our six candidates; the known isolated neutron star (labeled E1\_85) is included for comparison.  These sources were not identified in any of the visual or IR counterpart catalogs examined by the CCCP \citep{Broos11a}; the positions of sources C1\_961 and C3\_33 were visually inspected in the deep HAWK-I  \citep{Kissler08} near-IR survey of Carina \citep{Preibisch11b} and no evidence of IR emission was seen. 
 
% See /bulk/cochise1/targets/carina/data/neutron_star_search.txt

%-----------------------------------------------------------------------------
\begin{deluxetable}{llllcrr}
\tablecaption{Candidate Neutron Stars\label{neutron_candidates.tbl}
}
\tablewidth{0pt}
\tabletypesize{\footnotesize}
%\rotate
\tablehead{
\colhead{CXOGNC J} & \colhead{Label} & \colhead{$\alpha$ (J2000.0)} & \colhead{$\delta$ (J2000.0)} & \colhead{PosErr}  & \colhead{NC}& \colhead{$E_{med}$}  \\  
                 &                 & \colhead{(\arcdeg)}          &  \colhead{(\arcdeg)}         &\colhead{(\arcsec)} & \colhead{(cts)}   & \colhead{(keV)}\\
\numberthecolumn & \numberthecolumn & \numberthecolumn & \numberthecolumn & \numberthecolumn & \numberthecolumn & \numberthecolumn  
\setcounter{column_number}{1}
}
\startdata
104104.34-594342.6  &   SB2\_500 &  160.268105 & -59.728505 & 0.76  &  11.3 & 0.96 \\
104218.32-594234.2  &   C1\_232  &  160.576366 & -59.709523 & 0.54  &  16.4 & 0.90 \\
104357.36-594046.0  &   C1\_961  &  160.989040 & -59.679461 & 0.45  &  22.6 & 0.80 \\
104431.25-595132.0  &   C3\_33   &  161.130245 & -59.858915 & 0.50  &  19.6 & 0.86 \\
104608.71-594306.4  &   E1\_85   &  161.536330 & -59.718466 & 0.10  & 522.0 & 0.99 \\
104731.02-593904.4  &   E1\_723  &  161.879256 & -59.651237 & 0.50  &  28.7 & 0.97 \\
104744.13-593804.8  &   E4\_12   &  161.933914 & -59.634681 & 0.64  &  19.9 & 0.99  
\enddata
\tablecomments{
Col.\ (1): IAU designation 
\\Col.\ (2): Source name used within the CCCP project 
\\Col.\ (3,4): Position 
\\Col.\ (5): Position error
\\Col.\ (6): Net X-ray counts extracted in the total band, 0.5--8~keV
\\Col.\ (7): Median X-ray energy in the total band
}
\end{deluxetable}
%-----------------------------------------------------------------------------

We have used R-band and H$\alpha$ images of Carina (Gruendl et al., in prep.) obtained with the MOSAIC~II camera on the Cerro Tololo Inter-American Observatory Blanco 4m Telescope \citep{Muller98} to search for counterparts to these neutron star candidates.  We found no counterparts for these sources to a 3$\sigma$ limit of R=21.5~mag.  They have no matches in VizieR.   We recommend further follow-up observations of these sources to establish their nature; if any of them are isolated neutron stars, this will constitute an important clue to Carina's past star formation activity.

\clearpage

%=============================================================================
\section{SELECTED CCCP SCIENCE GROUP RESULTS \label{sec:results}}

This section presents sample CCCP science results, arranged according to the science groups that form the management structure for the Project, although the CCCP science results transcend such artificial group divisions.  Subsequent papers in this {\em Special Issue} detail these results and explore many others.

\subsection{Data Products \label{sec:data}} 

\subsubsection{Survey Statistics \label{sec:xlf}}

The full CCCP catalog of 14,369 ACIS X-ray sources is given in \citet{Broos11a}.  Most sources are very weak---only 3\% have more than 100 net X-ray events extracted and 60\% have fewer than 10 net events.  The lowest-level calibrated photometric quantity that is available for all sources is apparent {\em photon} flux \citep[\S7.4 in][]{Broos10}.  Rough, model-independent estimates for apparent {\em energy} flux are also provided for all sources \citep[Equation~2 in][]{Broos11a}.  Model-dependent estimates for absorption and intrinsic energy flux \citep{Getman10} are provided for sources with sufficient counts that are likely to be low-mass young stars in the Carina complex \citep[Table~8 in][]{Broos11b}.  %Spectral models are presented for $<$100 massive stars that have intrinsic luminosities (0.5--10~keV) in the range  \tbr{$10^{?}$--$10^{?}$~erg~s$^{-1}$} \citep{Naze11}.

The survey exhibits strong spatial variation in point source detections, due mostly to the factor of ${\sim}$6 ($10^{0.8}$) difference in sensitivity between on-axis and far off-axis regions in each ACIS pointing (the so-called ``egg crate effect'' discussed earlier).
\citet[][Table~6]{Broos11a} estimate detection completeness limits for several off-axis angle ranges, measured in terms of photon flux and then converted to intrinsic luminosity assuming a plasma typical for low-mass young stars seen through an absorbing column typical for Carina.  They find that nearly all such canonical stars should be detected on-axis down to a total-band (0.5--8~keV) intrinsic luminosity of $10^{29.9}$~erg~s$^{-1}$ and should be detected far off-axis down to $10^{30.7}$~erg~s$^{-1}$.  That result is a characterization of source detection, however many science investigations impose additional requirements on their source samples, e.g., the availability of counterpart information or the availability of intrinsic luminosity estimates.  Such selections can result in higher completeness limits for the resulting sample.  
%For example, the source samples used for XLF analyses of Tr15 and Tr16, which require individual intrinsic luminosity estimates, were found to be complete to \tbr{${\sim}10^{30.3}$~erg~s$^{-1}$} \citep{Wang11} and \tbr{${\sim}10^{30.7}$~erg~s$^{-1}$} \citep{Wolk11} in the total-band (0.5--8~keV).
%***Check these numbers when we get Scott's draft -- Tr16 should be complete to a fainter limit than Tr15!***

Due to the short integration times of the individual ObsID's that make up the bulk of the CCCP survey, very little can be said about the important topic of X-ray flaring on pre-MS stars.  Only 660 sources exhibit significant variability in our data \citep[][Table~1]{Broos11a} and only rarely do we capture the full duration of a pre-MS star flare.
%\footnote
%{
%Here, significant variability is defined as ${\rm ProbKS\_single} < 0.005$, where ProbKS\_single is the p-value for the one-sample Kolmogorov-Smirnov statistic under the no-variability null hypothesis, published by \citep[][\tbr{Table~1}]{Broos11a}.
%}  

\subsubsection{Counterparts to X-ray Point Sources}

Assocations for $>$9800 CCCP sources have been identified in one or more of 5 visual and IR catalogs that cover the field and 13 smaller-scale catalogs targeting the richest clusters \citep{Broos11a}.  The multiwavelength data provided by these counterpart catalogs are invaluable to many CCCP studies (see Section~\ref{sec:multi}).  Identification of X-ray sources in other wavebands also lends support to the validity of weak X-ray detections, as demonstrated by the good agreement found between the CCCP catalog and a very high resolution near-IR observation of the Tr14 core \citep{Sana10}, presented by Broos et al.\ in their Figure~6.

\subsubsection{The Point Source Classifier \label{sec:classifier}}

\begin{figure}[htb] 
\begin{center}  
\includegraphics[width=1.0\textwidth]{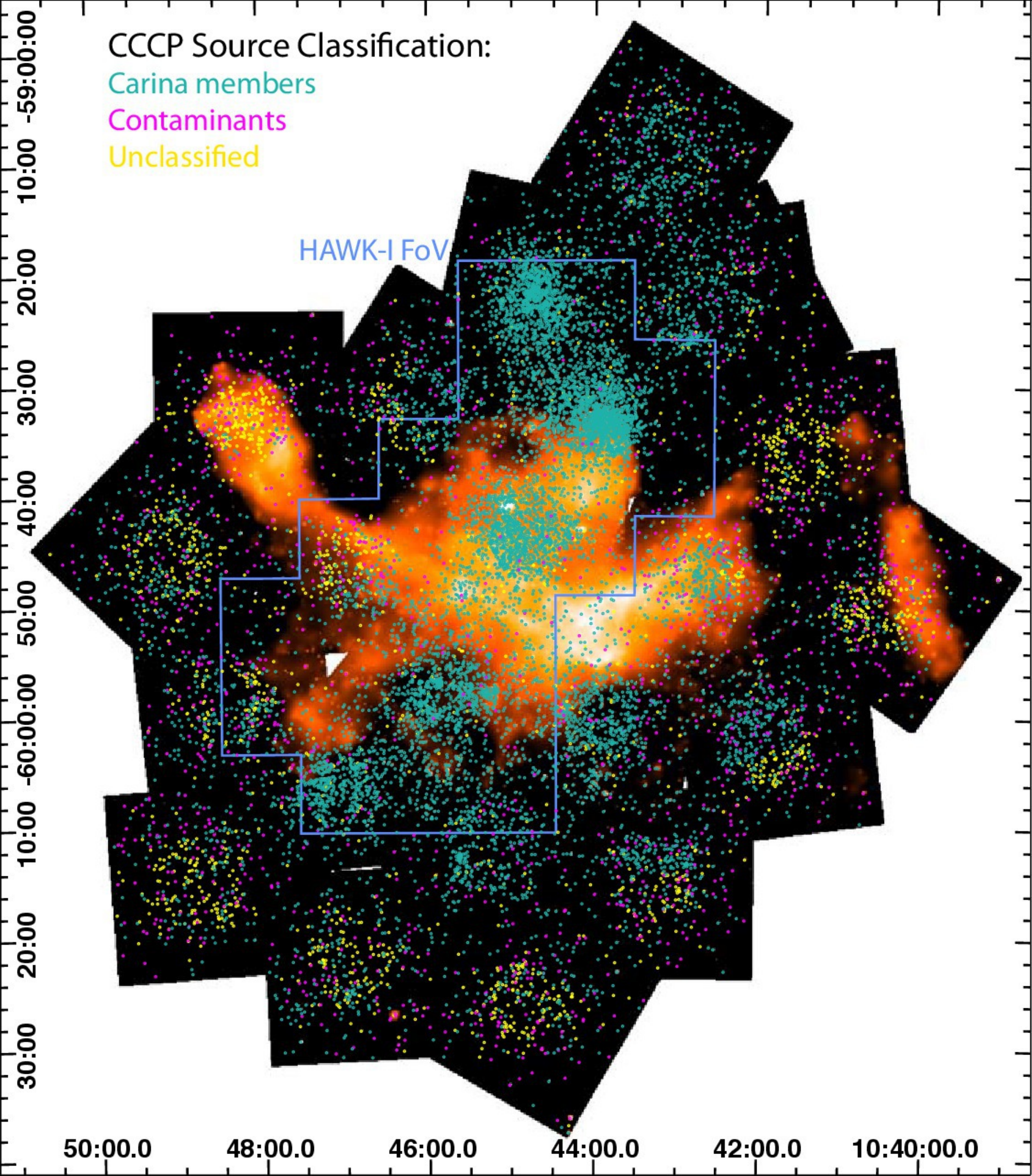}
\caption{
Classifying CCCP X-ray sources:  10,714 likely Carina members (cyan), 1609 likely contaminants (foreground stars, background stars, and AGN) (magenta), and 2045 unclassified sources (yellow).  The HAWK-I JHK study of Carina's massive clusters and their surroundings \citep{Preibisch11b} is outlined in blue.  Note again that part of the spatial distribution of point sources is due to the ``egg crate effect.''  These point source classifications are overlaid on the soft-band smoothed image from Figure~\ref{fig:cccpfull}, now scaled to show only the brighter regions of soft diffuse emission.
} 
\label{fig:classification}
\end{center}
\end{figure}

We developed a Naive Bayes source classifier to identify likely members of the Carina complex \citep{Broos11b}.  Using source location, X-ray properties, and visual/IR properties, each CCCP source is assigned probabilities for membership in each of four classes:  foreground stars, Carina members, background stars, and extragalactic objects.  Careful simulations were performed to characterize the observed properties expected for these contaminating populations \citep{Getman11}.  The classifier assigns 10,714 sources (75\%) as probable members of the Carina complex, 1609 sources as likely contaminants, and finds that 2045 have too little information to classify.  Figure~\ref{fig:classification} maps these three groups of sources.

%-------------------------------------------------------------------------
%-------------------------------------------------------------------------

The deep near-IR data from the HAWK-I survey \citep{Preibisch11b} were very helpful for classifying CCCP sources.  Within the HAWK-I field, 87\% of CCCP sources have counterparts in some other catalog \citep{Broos11a}.  Outside of the HAWK-I field, only 44\% of CCCP sources have counterparts.  This counterpart information is vital for source classification, as can be seen in Figure~\ref{fig:classification}; most unclassified sources lie outside the HAWK-I FoV.  The major problem here is that the 2MASS survey does not go deep enough to detect most pre-MS stars at the distance of Carina.  Deep wide-field visual (e.g., H$\alpha$) and near-IR data are required to improve our source classifications for CCCP X-ray sources.

\clearpage

%=============================================================================
\subsection{Massive Stars \label{sec:massive}}

The Carina star-forming complex is famous for its large number of OB stars \citep[e.g.,][]{Feinstein95, Walborn95, Smith06b} and for some of the nearest examples of the most massive O stars; the O3 spectral type was discovered here \citep{Walborn71}.  These massive stars, along with 3 well-known WR stars and $\eta$~Car, are widely distributed among Carina's many clusters and in the intercluster regions of the complex; a few of the most famous massive stars were marked in Figure~\ref{fig:cccpfull}.  The CCCP detects 68/70 known O stars in Carina plus 61/130 known early-B stars \citep{Naze11}.  The possible X-ray emission mechanisms for these massive stars are reviewed and explored by \citet{Gagne11}.  The bright, variable, massive multiple system QZ~Car is studied separately \citep{Parkin11b}.   
%Give details on the 2 O stars we miss, if Yael or Marc doesn't.
%Something we don't detect -- SBW1, the B1.5 Iab supergiant at 7 kpc (Smith, Bally, \& Walawender 2007).  This is just like Sher 25 in NGC 3603.

In a new type of study combining \Chandra data with 2MASS and {\em Spitzer} photometry, \citet{Povich11b} use IR SED-fitting on an X-ray-selected sample to establish a list of 94 candidate massive stars that have been overlooked in historical studies of Carina or missed due to obscuration.  Clearly follow-up visual/IR spectroscopy is needed to establish spectral types for these sources; if confirmed as massive stars, they will alter our understanding of Carina's energy budget and suggest that the complex is even more affected by massive star feedback than we have appreciated.

\subsubsection{A Remarkably Variable Candidate OB Star \label{sec:superflare}}

The most exceptional of the few variable sources seen in the CCCP is source C3\_225 (CXOGNC J104457.51-595429.5).  It was seen by {\em ROSAT} and {\em XMM} \citep{Albacete03,Antokhin08} as a fairly faint and unremarkable X-ray source, and is a bright visual \citep[V=11.7~mag,][]{Massey93} and 2MASS (K=8.9~mag) source.  This source has been identified as a candidate OB star reddened by $A_{V}=3.7$~mag \citep[][OBc~\#41 in their Table~2]{Povich11b} by fitting stellar photosphere models to SEDs covering the near- and mid-IR bands.

\begin{figure}[htb] 
\begin{center}  
\includegraphics[width=0.44\textwidth]{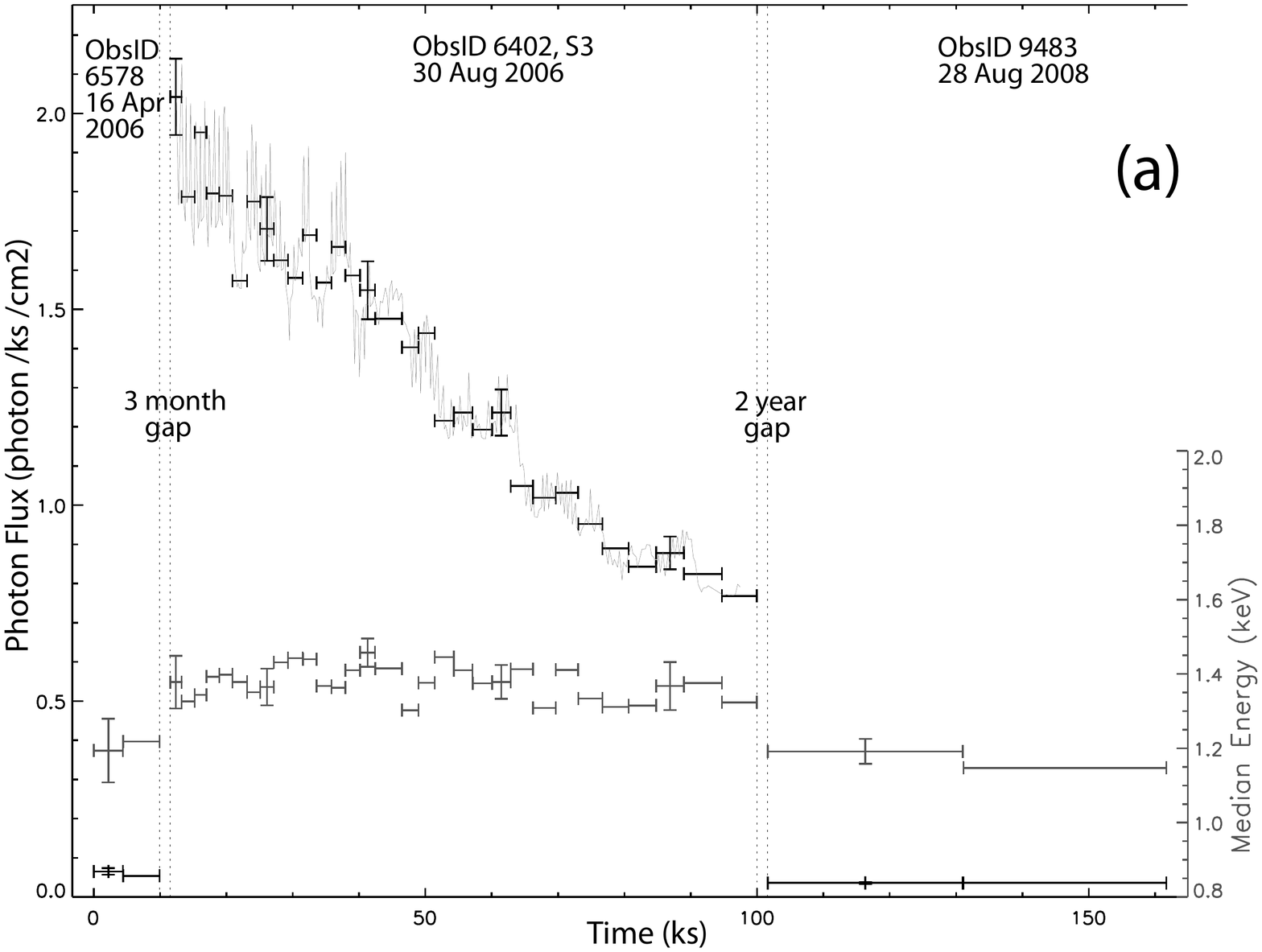}
\hspace*{0.2in}
\includegraphics[width=0.45\textwidth]{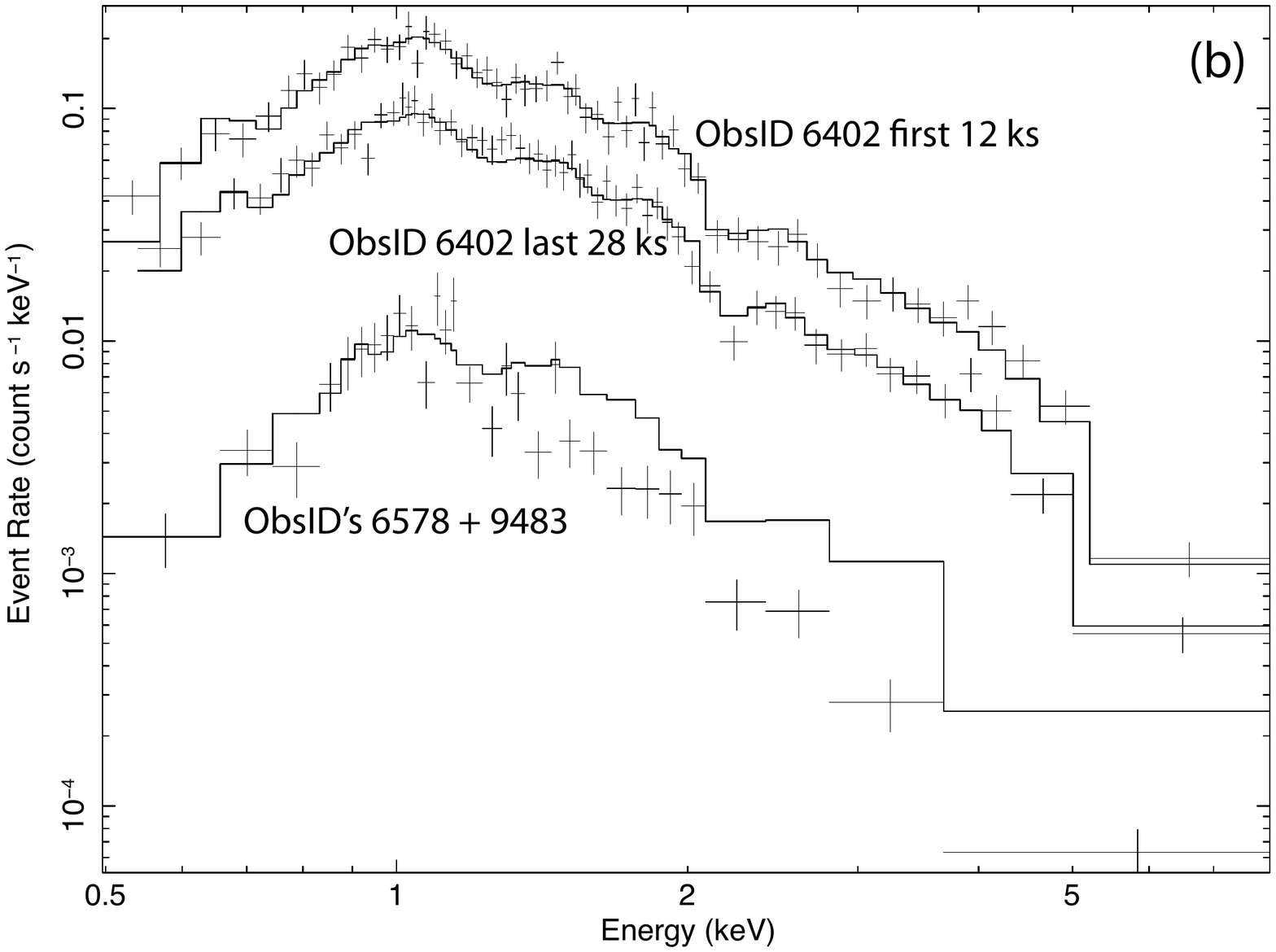}
\caption{The remarkable variable OB candidate CXOGNC~J104457.51-595429.5.
(a)  A multi-ObsID \Chandra lightcurve \citep{Broos10}.  Photon flux is shown in black (left ordinate axis) and median energy in medium gray (right ordinate axis).  For clarity, error bars are shown for only a few points.   An adaptively-smoothed lightcurve is overlaid in light gray on the photon flux data for ObsID~6402; 1000-sec modulations are due to the telescope dither period.
(b) Spectra (points with error bars) for 3 time samples from this source, with the same spectral model overplotted (histogram curves) on all 3 spectra.  The first (upper spectrum) and last (middle spectrum) segments of ObsID~6402 (when the source showed strong variability) are well-fitted by the same model, while the quiescent epochs (lower spectrum) are significantly softer than the model shown.
} 
\label{fig:variable}
\end{center}
\end{figure}

This source also serves to illustrate a very useful feature of {\it ACIS Extract}, multi-ObsID lightcurves (Figure~\ref{fig:variable}a).  C3\_225 is not bright or variable in its primary CCCP observation, Clusters Pointing 3 (ObsID 9483) but, by examining the ACIS S-array data from the Tr16 pointing (ObsID 6402) obtained 2 years earlier, we discovered an exceptionally bright decaying source (Figure~\ref{fig:variable}) consistent with the position of CCCP source C3\_225 and well-separated from all other CCCP sources.  
% cd /bulk/cochise1/targets/carina/data/extract_Sarray/point_sources.noindex/104457.51-595429.5
%  ds9 6402_S3/neighborhood.evt -region 6402_S3/extract.reg -region /bulk/cochise1/targets/carina/data/xray_positions.reg &
An even earlier ACIS GTO observation of the CCCP C3 pointing (ObsID~6578, 10~ks on the Treasure Chest Cluster) showed a low photon flux and median energy consistent with the CCCP C3 observation.  In ObsID~6402 this source sits right at the edge of the ACIS-S array, moving on and off the S3 CCD as the telescope dithers and introducing a 1000-sec modulation into its lightcurve consistent with the dither period.  Aside from this unfortunate geometry, the source clearly shows a decaying lightcurve throughout the long (87-ks) Tr16 observation; at the beginning of the observation it is so bright that it would have suffered substantial photon pile-up if it had not been imaged so far off-axis. 
The ``quiescent'' photon flux averaged over ObsID's 6578 and 9483 was $3.83 \times 10^{-5}$~photon~cm$^{-2}$~s$^{-1}$;
% See extract/point_sources.noindex/104457.51-595429.5/photometry/104457.51-595429.5.sequenced_lc.ps and
%     extract/point_sources.noindex/104457.51-595429.5/photometry/source.photometry.
at the onset of ObsID 6402 the source was brighter by a factor of 56.
%, at $2.13 \times 10^{-3}$~photon~cm$^{-2}$~s$^{-1}$.
% See extract_Sarray/point_sources.noindex/104457.51-595429.5/EPOCH_6402_S3with456bkgcnts/104457.51-595429.5.sequenced_lc.ps and
%     extract_Sarray/point_sources.noindex/104457.51-595429.5/EPOCH_6402_S3with456bkgcnts/source.lc

%-------------------------------------------------------------------------
%-------------------------------------------------------------------------

The time evolution of the median X-ray energy statistic (right ordinate axis and lower curve in Figure~\ref{fig:variable}a) suggests that the shape of the spectrum was constant during the X-ray decay ($E_{\rm median} \simeq 1.38$~keV) in ObsID~6402 and that it was softer during the quiescent epochs ($E_{\rm median} \simeq 1.18$~keV).  Indeed, spectra extracted from the first 12~ks and the last 28~ks of ObsID~6402 (upper and middle spectra in Figure~\ref{fig:variable}b) were found to be well-fitted by a single model; the {\em XSPEC} \citep{Arnaud96} model used was {\it TBabs}({\it apec} + {\it apec} + {\it apec}) with solor abundances.  The absorption estimate from SED-fitting ($A_{V}=3.7$~mag) was adopted by freezing $N_H$ to $0.59 \times 10^{22}$ cm$^{-2}$ in the {\em XSPEC} fit, and plasma temperatures of 0.2, 1.3, and 4.5~keV were derived from the fit.
This model is shown for all 3 spectra in Figure~\ref{fig:variable}, scaled independently for the 3 samples; the quiescent spectrum is significantly softer than this model (lower spectrum in Figure~\ref{fig:variable}b).
% /bulk/cochise1/targets/carina/data/extract_Sarray/point_sources.noindex/104457.51-595429.5/three_epochs.xcm
Assuming that this object is at Carina's distance (2.3~kpc), the inferred absorption-corrected (intrinsic) X-ray luminosity (0.5--8~keV)
%, averaged over the first 12~ks of ObsID 6402, is $L_{t,c} = 10^{33.8}$~erg~s$^{-1}$.
% distance      = 2300.
% cm_per_parsec = 3.086D18
% dscale        = 4D*!PI*(distance * cm_per_parsec)^2
% alog10(dscale * 3.7661e-12)
% alog10(1.0505e-11 * dscale)
%We estimate the intrinsic luminosity 
at the onset of ObsID~6402 was $10^{33.9}$~erg~s$^{-1}$; this is a lower bound on the peak of the source brightness, which must have occurred before ObsID~6402 began, since we see only a decaying lightcurve for the duration of this observation.
% Scale the epoch average by 0.00213/0.00179.
Similar extreme O-star spectral and temporal behavior has been seen in other young clusters, e.g., the central O4-O4 binary in M17 and W51~IRS2E \citep{Townsley05}.  It may be an effect of stellar interactions in multiple systems or of fossil magnetic fields \citep{Gagne11}.  An important first step towards understanding this source would be spectroscopy to establish its spectral type.

\subsubsection{Near Neighbors Around Massive Stars \label{sec:groupies}}

The CCCP has many examples of known massive stars surrounded by other nearby X-ray sources.  Although we cannot say for certain that these surrounding sources are pre-MS companions physically associated with the massive star, there seem to be too many of these examples, often with multiple companions, than would be expected from chance superpositions of unrelated sources.

%***Interest in this is picking up in the literature -- see recent preprint by Maiz-Apellaniz on ``lucky'' imaging.***

\begin{figure}[htb] 
\begin{center}
\includegraphics[width=0.45\textwidth]{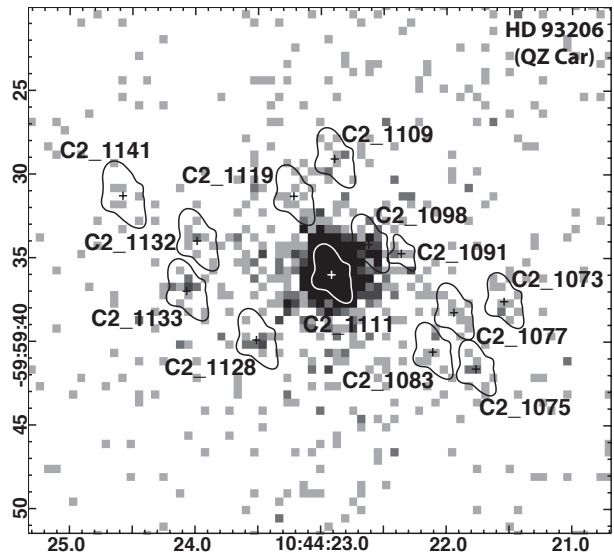}
\caption{
X-ray point sources grouped around QZ~Car.  The plusses show the source centers; surrounding polygons are the extraction regions.  CCCP source labels are shown.  The underlying image shows the binned ACIS event data.
} 
\normalsize 
\label{fig:massive}
\end{center}
\end{figure}

The most striking example of this is the bright massive multiple system QZ~Car (Figure~\ref{fig:massive}); X-rays from QZ~Car itself are studied in \citet{Parkin11b}.  Within a $\sim$15$\arcsec$ radius of QZ~Car our machinery finds 12 more unobscured, much fainter sources, forming a tight group.  Such close distributions of lower-mass stars around a massive star are difficult to detect in other wavebands due to the brightness of the massive star,
%, but massive stars have different X-ray emission mechanisms than low-mass stars, leading to different correlations between X-ray and bolometric luminosity.  For massive stars, this correlation is $L_{X} \propto 10^{-7}$~erg~s$^{-1}$ *REF*, whereas for pre-MS the correlation is $L_{X} \propto 10^{-4}$~erg~s$^{-1}$ *REF -- Norbert claims the relation is -3, not -4*.  
but they are occasionally seen in \Chandra surveys; another good example is the O5 star HD~46150 in the Rosette Nebula \citep{Wang08}.  We are quite confident that these sources are legitimate, even though they will be difficult to follow up in visual or IR studies; CCCP sources brighter than QZ~Car do not show such source groupings so this is not an artifact of photon pile-up or image reconstruction.
 
%-------------------------------------------------------------------------
%-------------------------------------------------------------------------

\begin{figure}[htb] 
\begin{center}  
\includegraphics[width=0.32\textwidth]{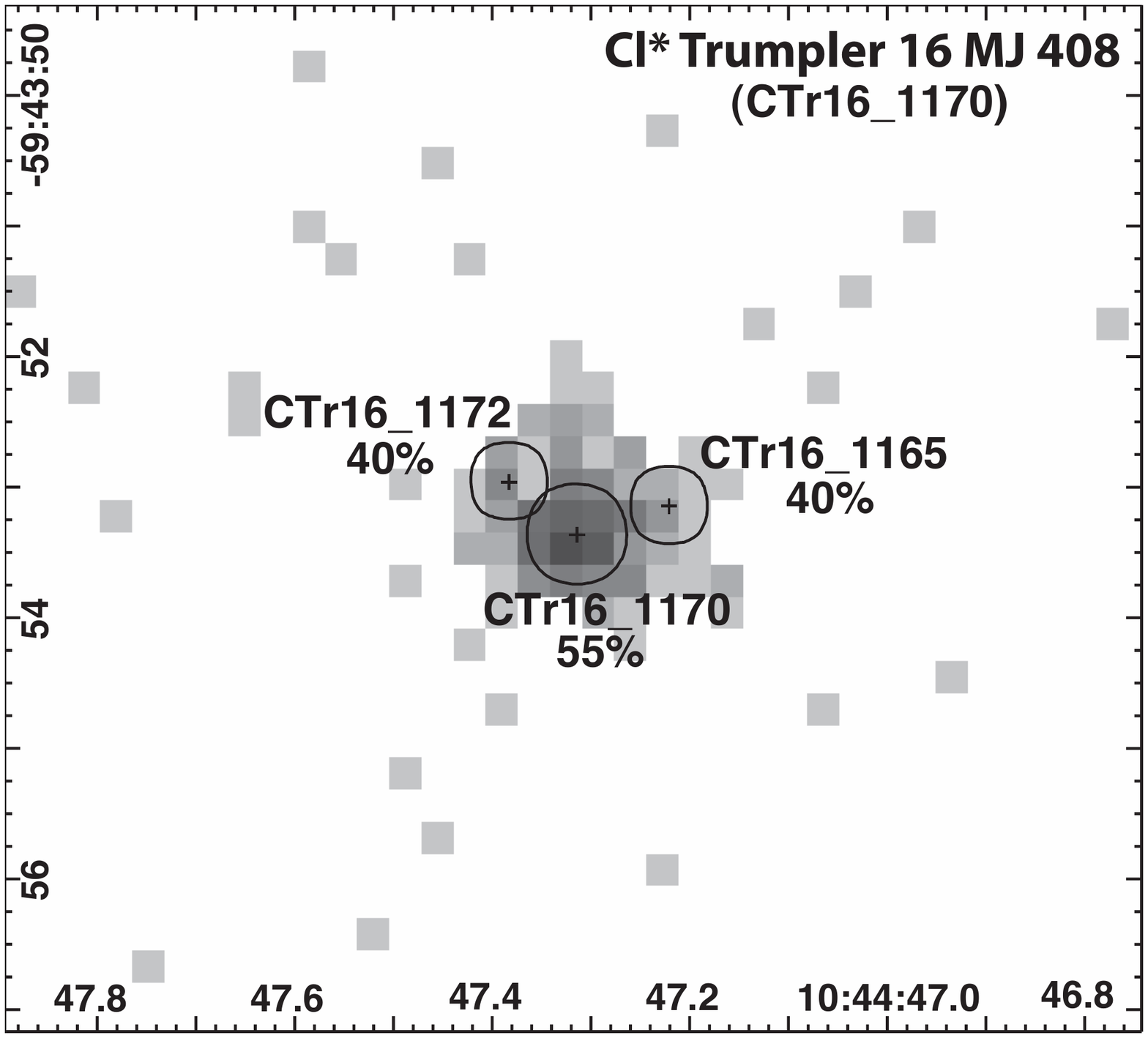}
\includegraphics[width=0.32\textwidth]{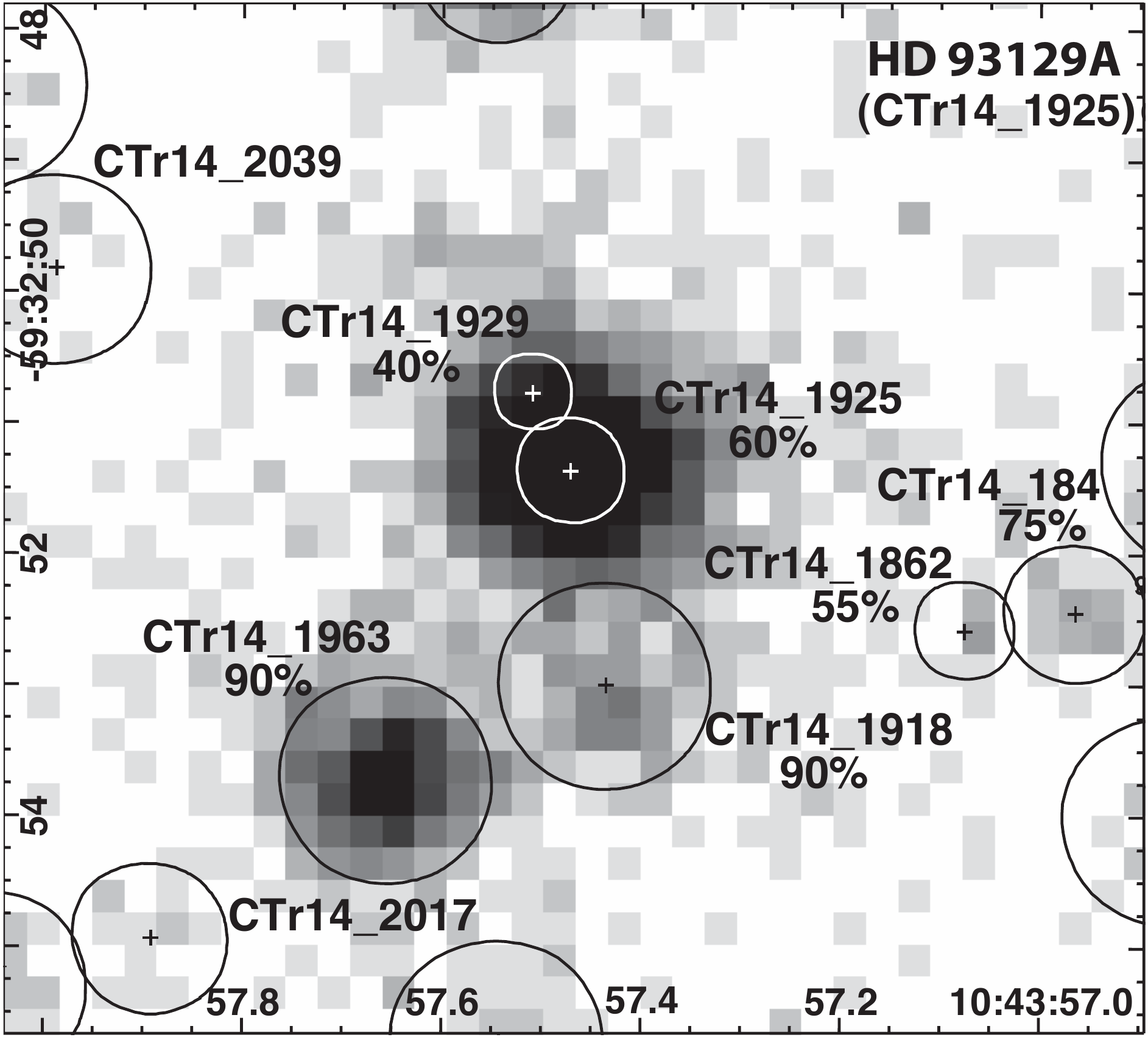}
\includegraphics[width=0.32\textwidth]{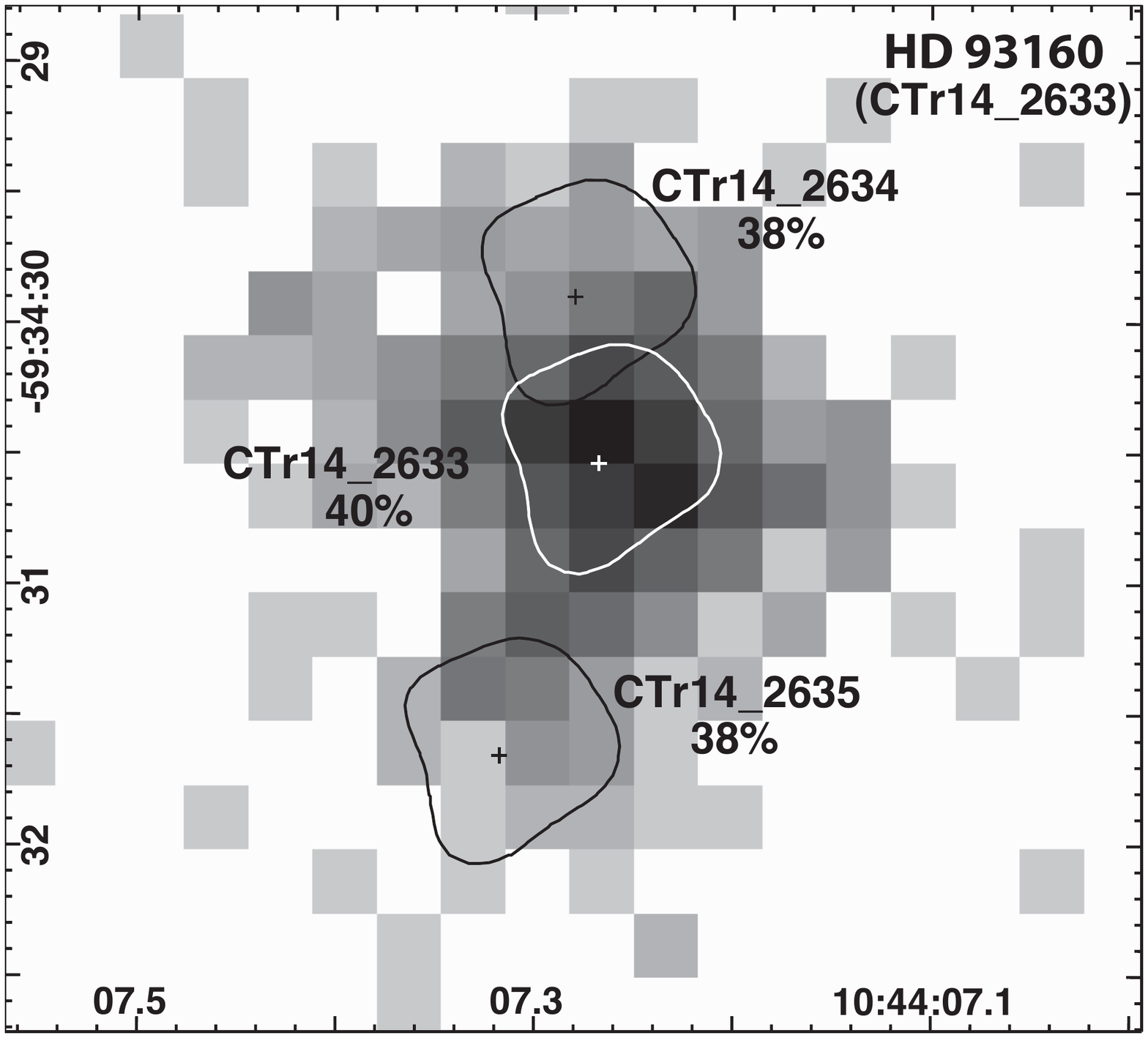} \\
\includegraphics[width=0.32\textwidth]{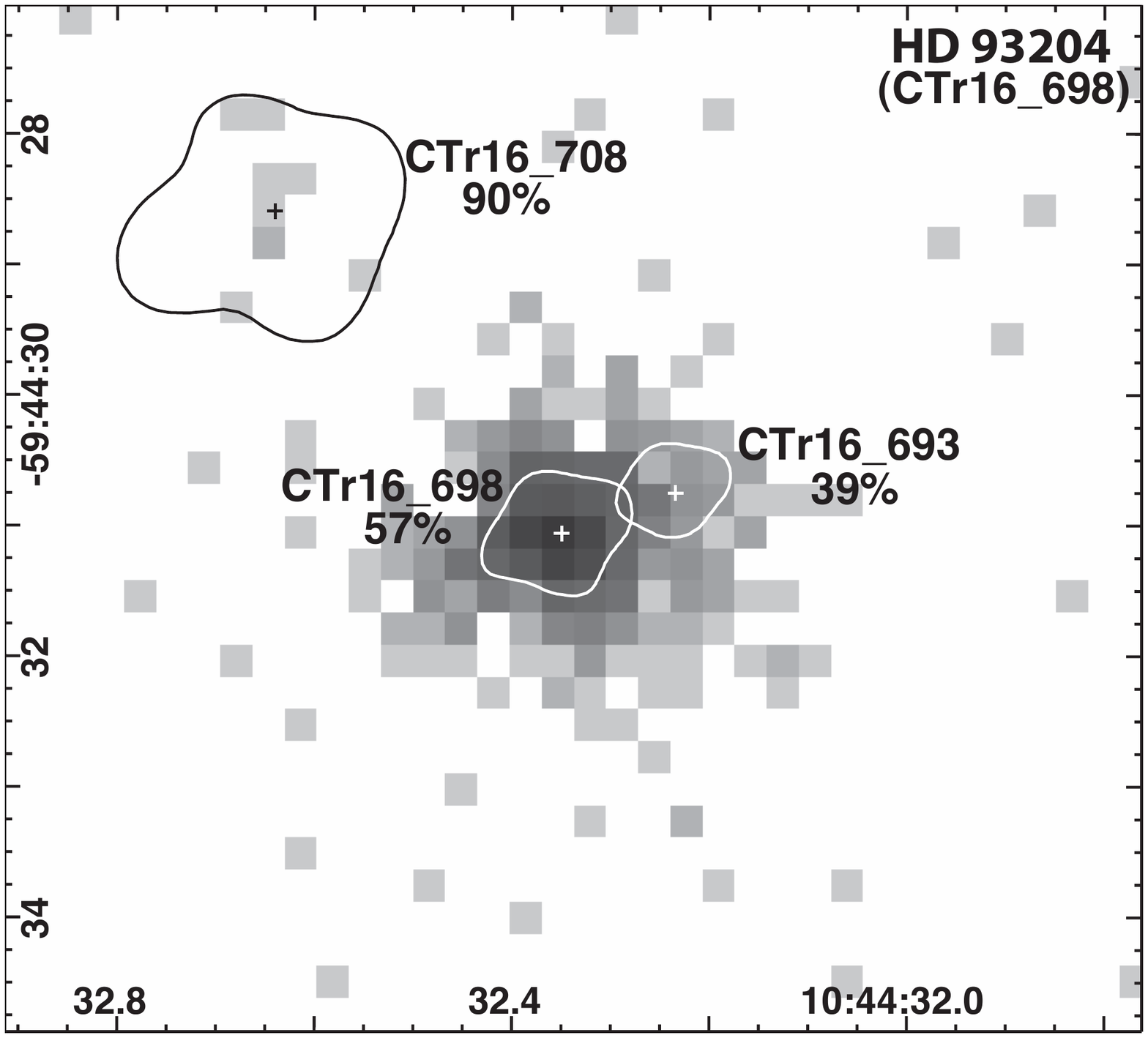}
\includegraphics[width=0.32\textwidth]{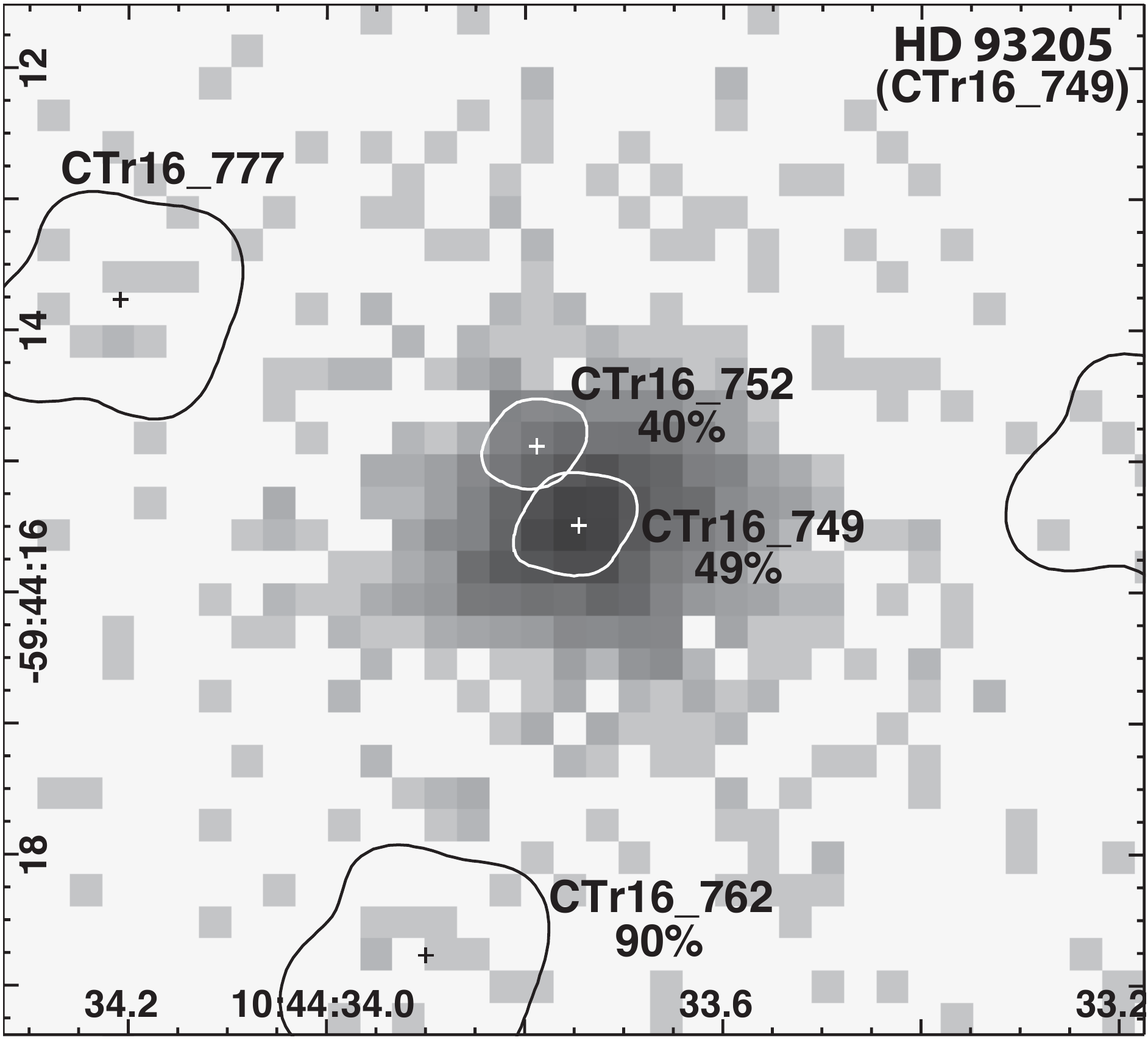}
\includegraphics[width=0.32\textwidth]{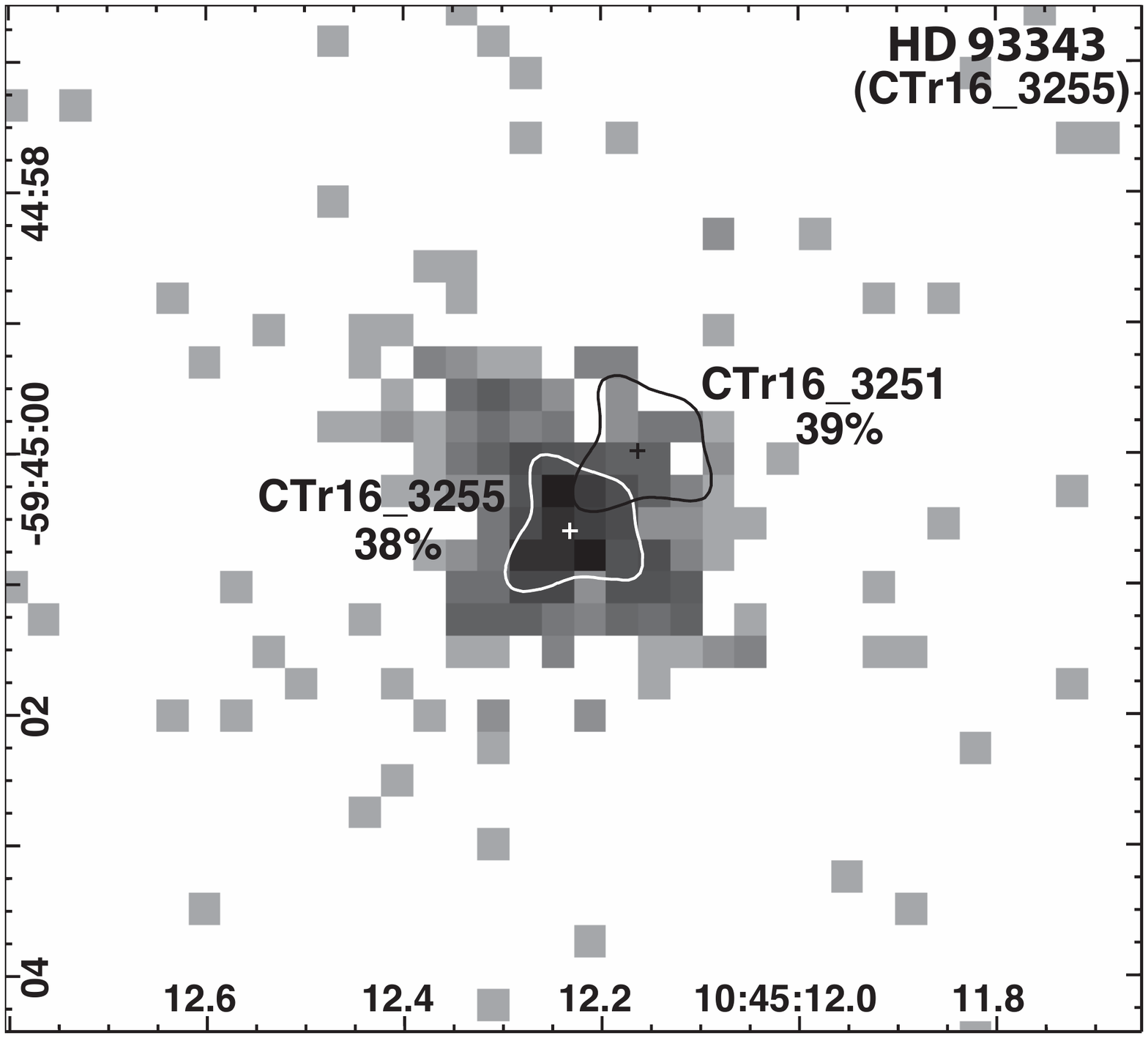} \\
\includegraphics[width=0.32\textwidth]{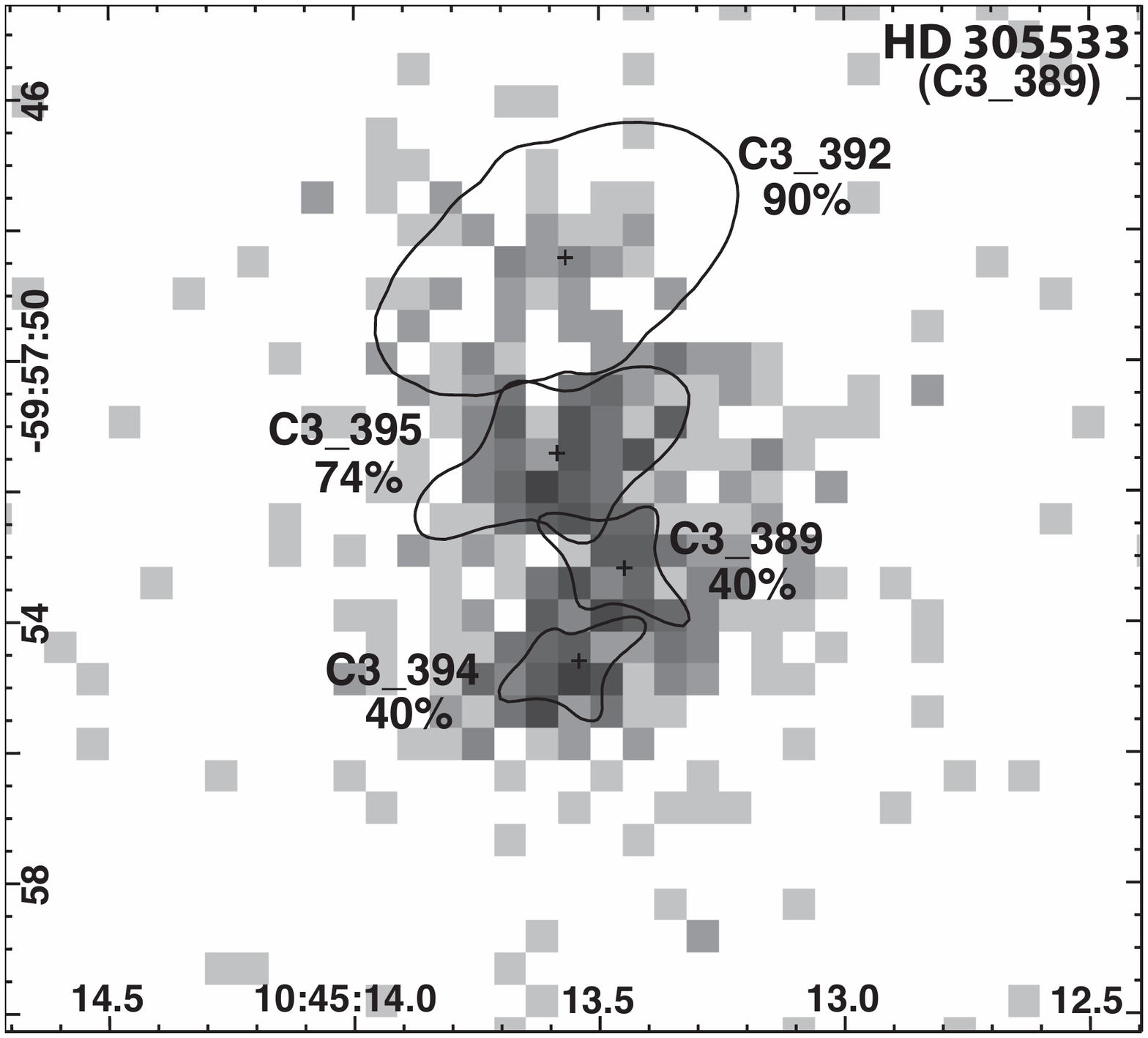}
\includegraphics[width=0.32\textwidth]{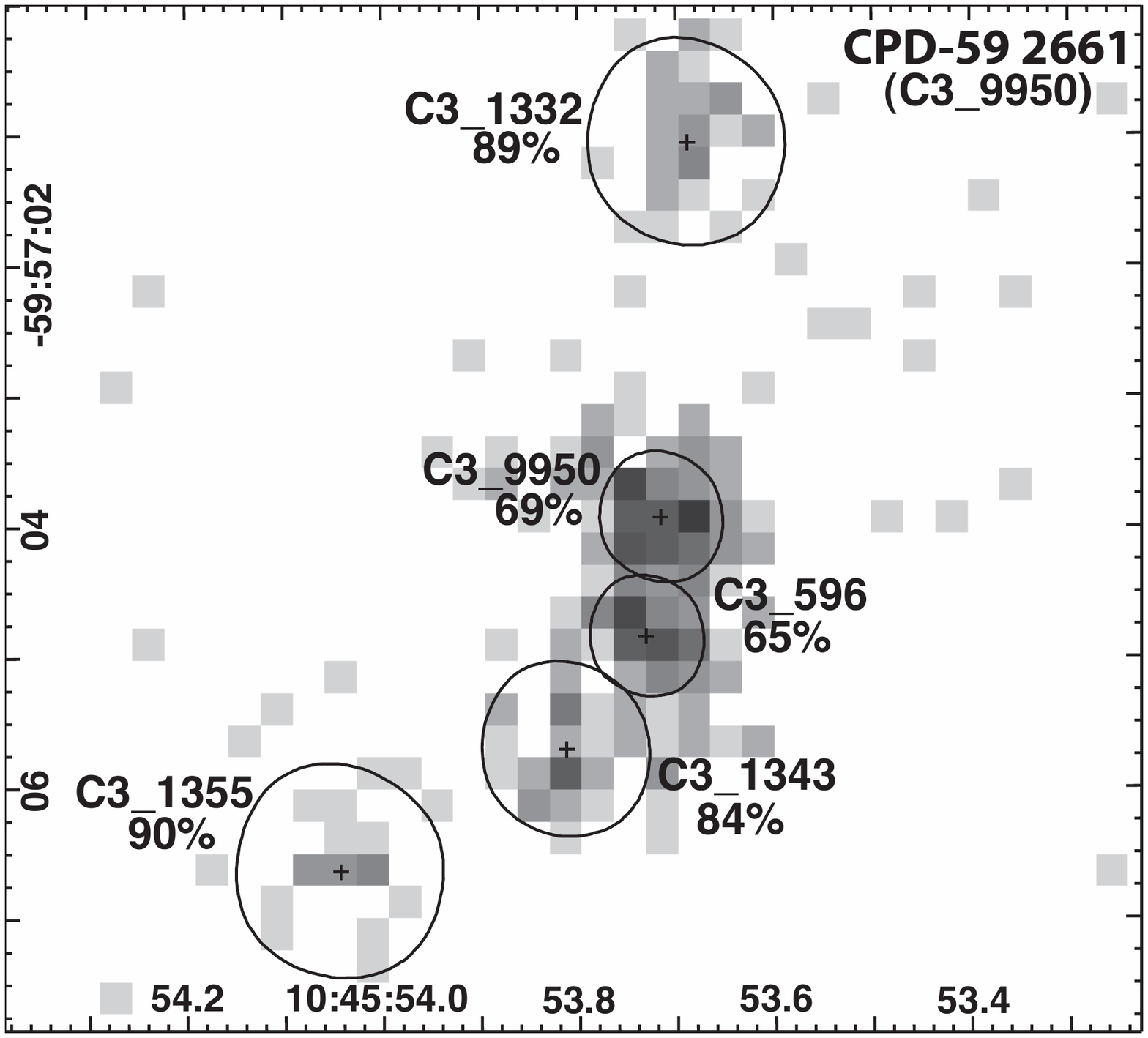}
\includegraphics[width=0.32\textwidth]{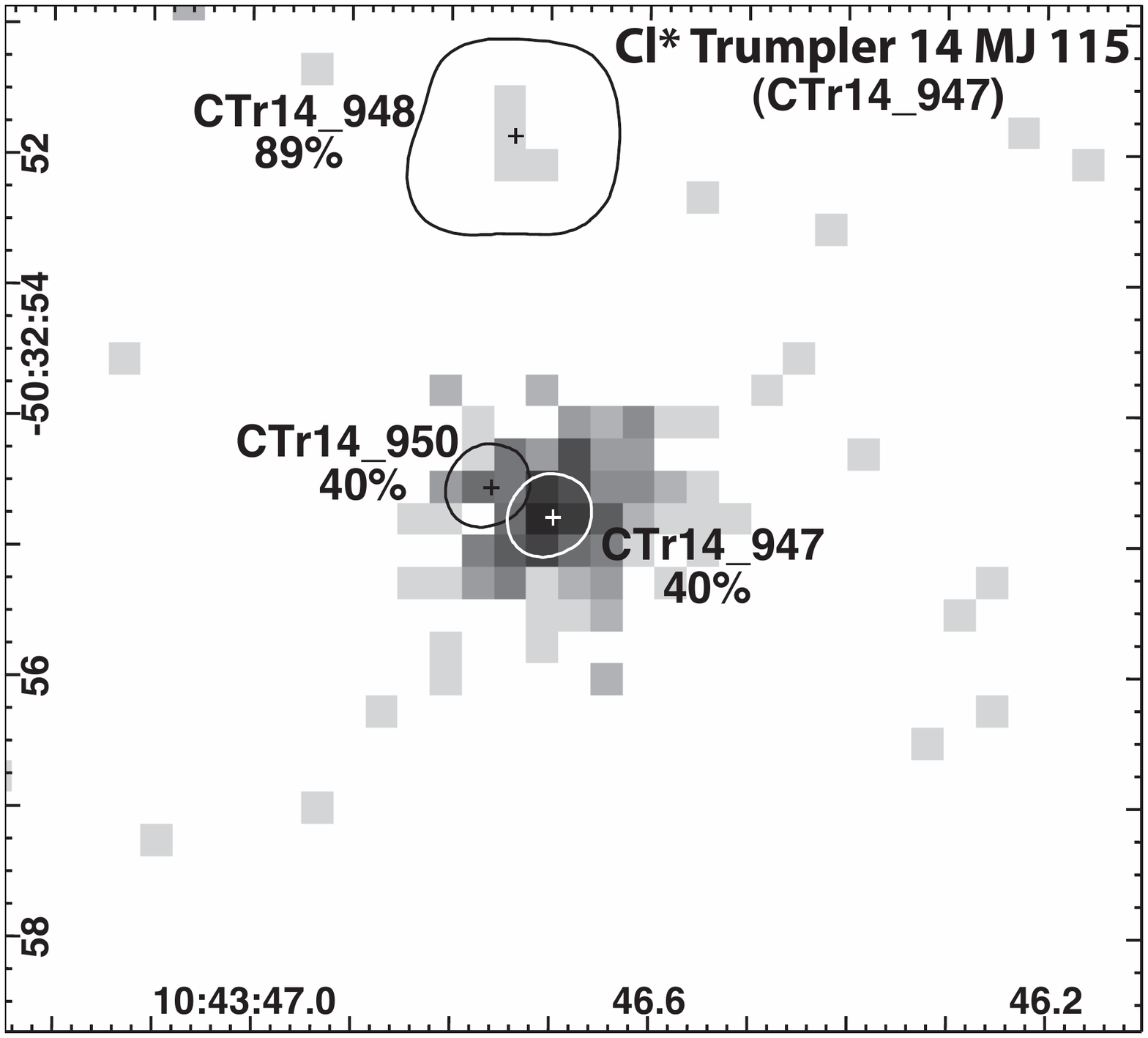} \\
\includegraphics[width=0.32\textwidth]{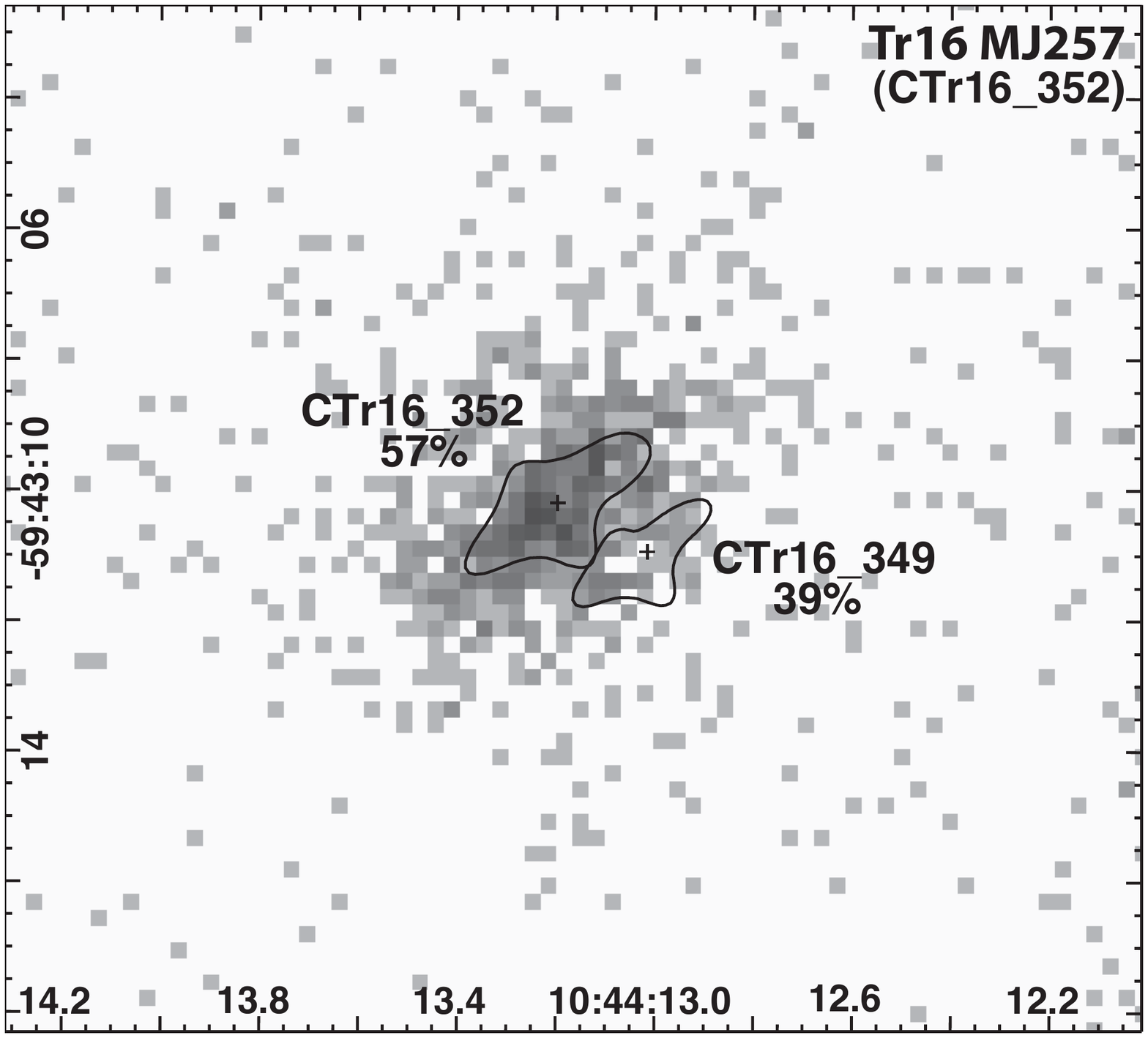}
\includegraphics[width=0.32\textwidth]{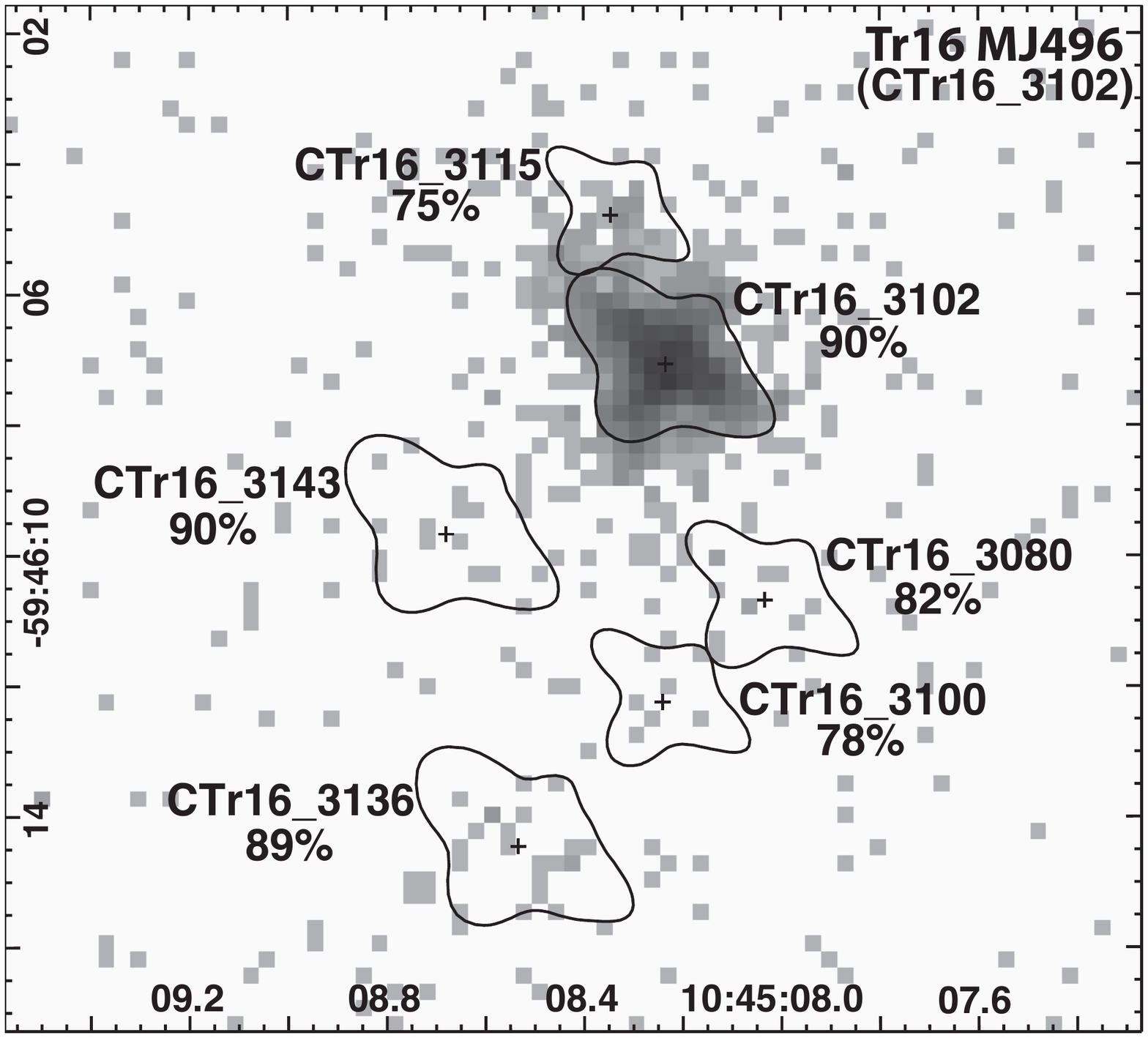}
\includegraphics[width=0.32\textwidth]{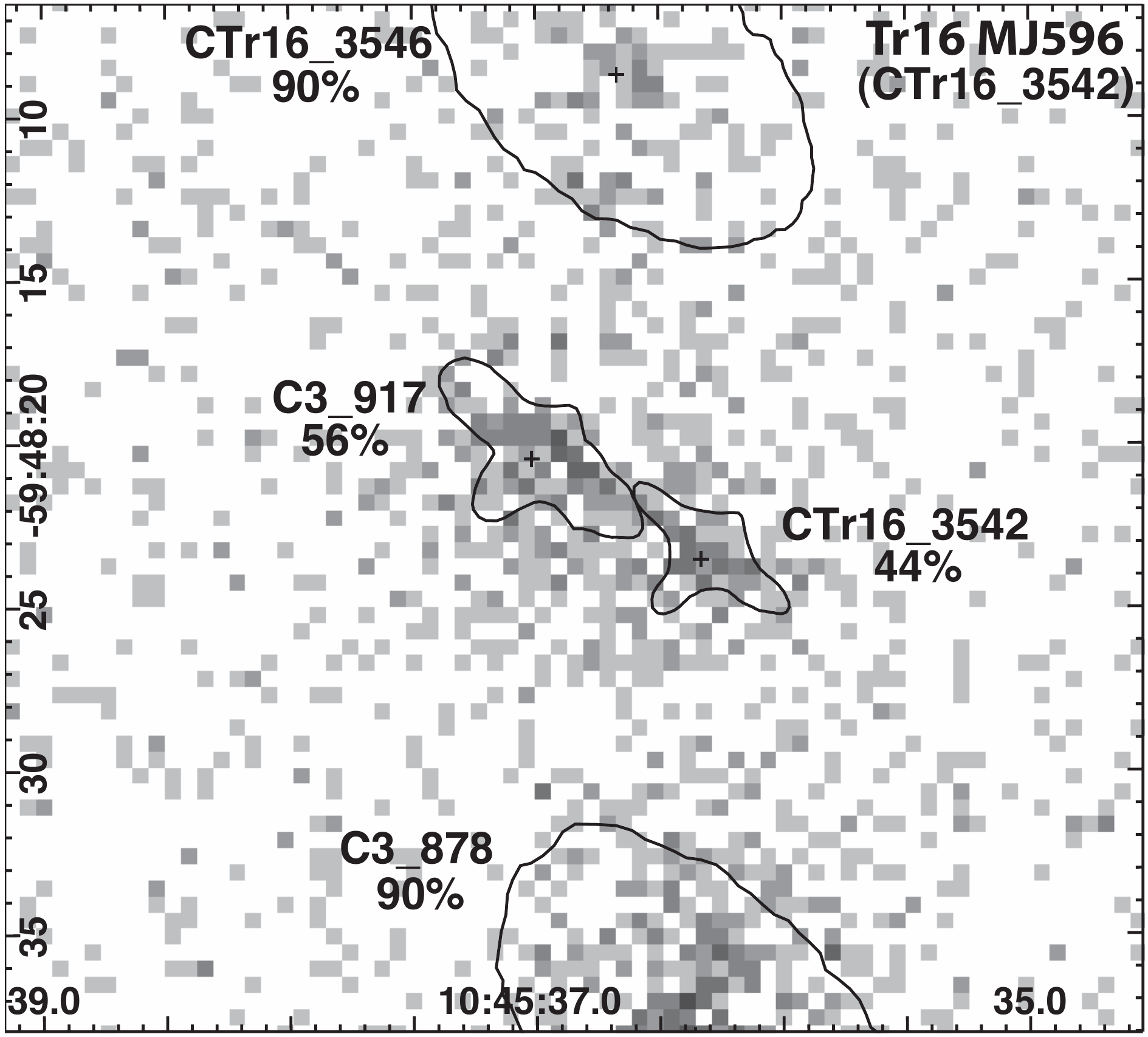}
\caption{Massive stars with close companions that reduce their extraction regions in standard processing.    Polygons show the extraction regions (for fields with multiple ACIS observations, only the largest extraction region is shown for each source); PSF fractions contained within each extraction region are noted, as are CCCP source labels.  Each panel is labeled in the upper right corner by the massive star name and its ACIS source label.  The underlying images show the binned ACIS event data.
} 
\label{fig:multiples}
\end{center}
\end{figure}

As mentioned above, no other massive star in the CCCP shows such a rich localized grouping of possible pre-MS companions as QZ~Car, but there are several other examples of O stars that require reduced extraction apertures (below the standard value of 90\% of the PSF extent) to minimize the influence of a nearby fainter source.  Figure~\ref{fig:multiples} shows these sources.  Again it is not clear whether the companion source is physically related to the massive star or just a chance superposition.  In either case, though, these reduced-aperture massive stars are important to note, because crowded sources can lead to mismatches when comparing to counterpart catalogs and reduced extraction apertures may not yield the optimal spectrum for the massive star (i.e., if the companion is faint enough, it might be better to ignore it and suffer its mild corruption of the massive star's spectrum rather than to accept a reduced number of counts for the massive star).

%***Massive stars people please heed this -- we may want to re-extract some massive stars, ignoring faint, close companions!***

%-------------------------------------------------------------------------
%-------------------------------------------------------------------------

\clearpage

\subsubsection{The Wolf-Rayet Stars \label{sec:wr}}

Carina has 3 Wolf-Rayet (WR) stars, all of which appear to be young (hydrogen-burning) stars rather than classical (core helium-burning) WR stars \citep{SmithConti08}.  Two of them (WR~22 and WR~25) are known binaries \citep{Gosset09,Gamen06} and the other one (WR~24) is a suspected binary \citep{Skinner10}.  No variability was seen, though, in the CCCP data for any of these sources.  Detailed analysis of these complicated sources is beyond the scope of this paper, but they are briefly introduced below.  We emphasize some important CCCP data pathologies here because they can lead to misinterpretation of the WR source properties.

\paragraph{WR~25}

\begin{figure}[htb] 
\begin{center}
\includegraphics[width=0.32\textwidth]{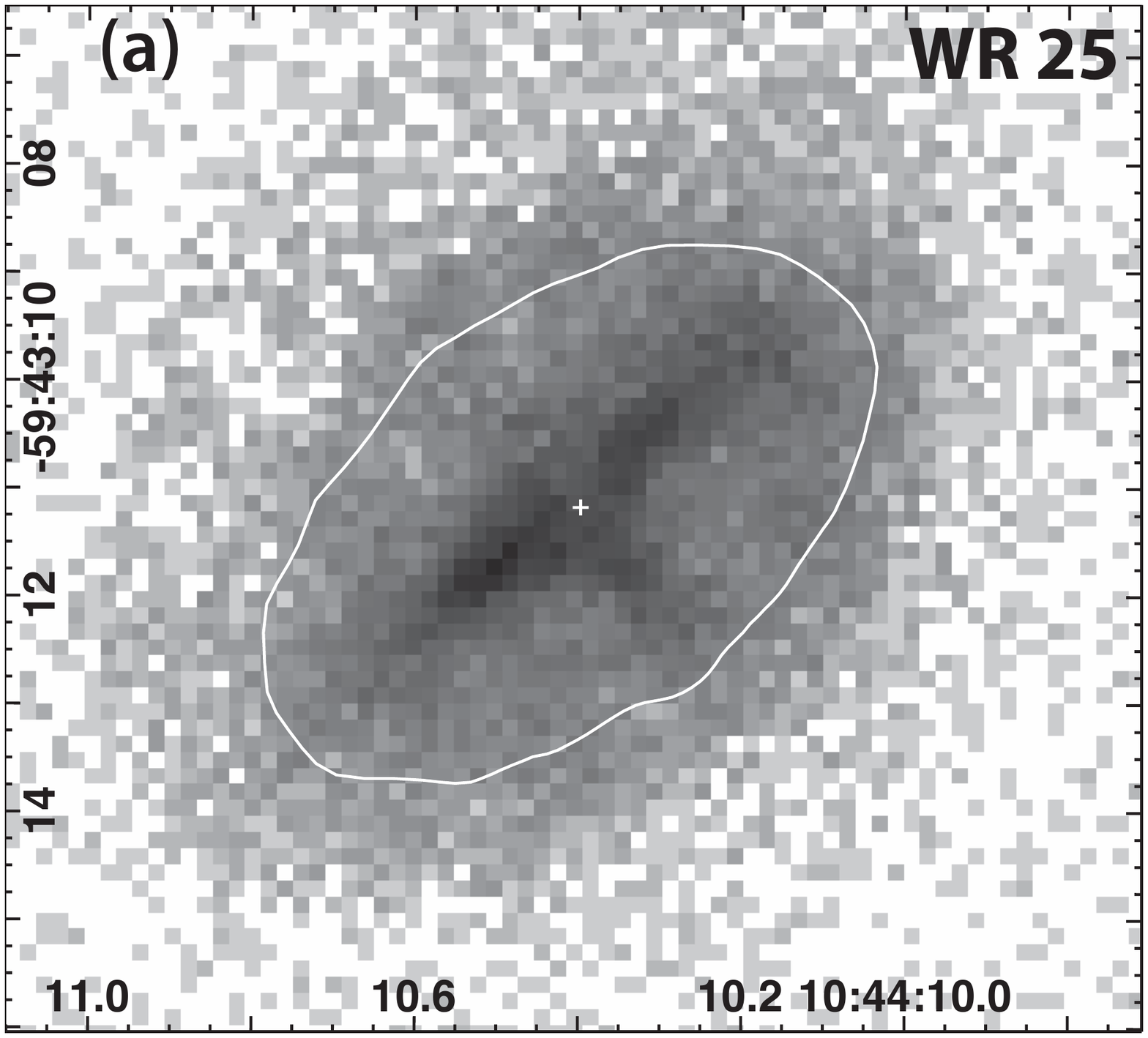}
\includegraphics[width=0.32\textwidth]{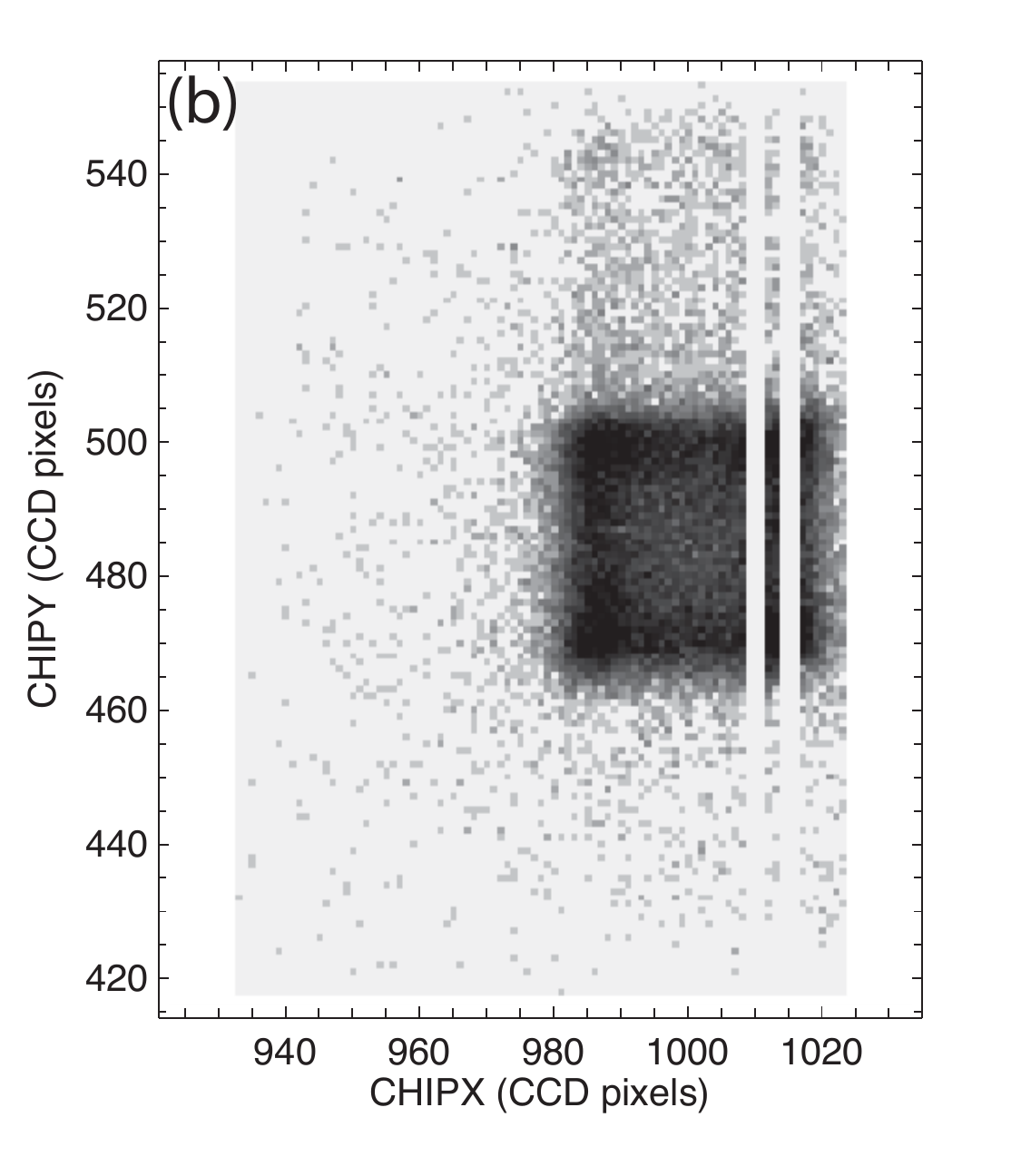}
\includegraphics[width=0.328\textwidth]{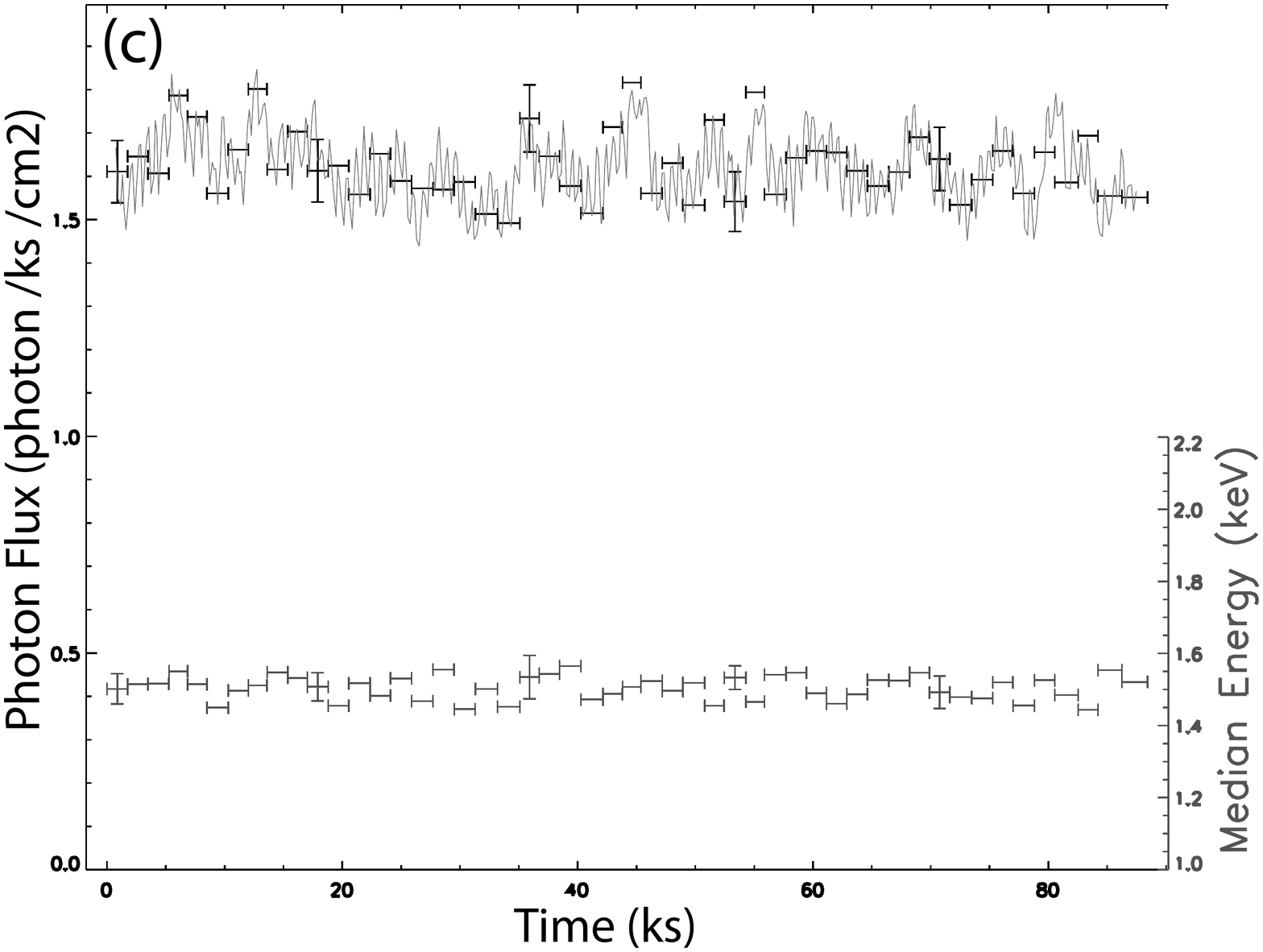}
\caption{WR~25 in the CCCP data.  Note that this source is significantly piled up.
(a)  An image of WR~25 finely-binned to show the multiple cusps in the PSF this far off-axis (4.8$\arcmin$).  The light-grey polygon shows the extraction region for this source.  
(b)  An image of WR~25 in CHIP coordinates, showing the box traced out by the telescope dither, truncated by the edge of the CCD at CHIPX = 1023.  A fainter box centered at roughly (1005, 525) pixels comes from another source.  The vertical bars with no events are due to bad CCD columns.
(c)  Lightcurve from ObsID~6402.  High-frequency modulations are due to the telescope dither interacting with these bad columns and the chip gap.
} 
\normalsize 
\label{fig:wr25}
\end{center}
\end{figure}

WR~25 (HD~93162; Figure~\ref{fig:wr25}) is a bright, variable X-ray source \citep{Pollock06}, a WR+O eccentric binary with a period of 208 days \citep{Gamen06}.  It was observed just once in the CCCP dataset, in the long exposure of Tr16 (see Table~\ref{tbl:obslog}).  WR~25 was so bright in that observation that it suffers substantial photon pile-up; we attempt to deal with this problem using simulations to reconstruct the source spectrum without pile-up, as described in \citet[][Appendix A]{Broos11a}.  Our reconstructed spectrum is well-fit by models derived from {\em XMM} data \citep{Pollock06,Antokhin08} and shows a soft thermal plasma component (kT$\sim$0.6~keV) plus a harder component (kT$\sim$2.5~keV) with many line-like residuals that are not well-fit by simple models.  More sophisticated spectral fitting will be the subject of future work.

%-------------------------------------------------------------------------
%-------------------------------------------------------------------------

WR~25 makes a clear readout streak in CCCP full-field images.  This source sits on the edge of an ACIS-I CCD, dithering in and out of a chip gap so the readout streak is narrower than normal and runs right along the edge of the CCD.  The source also dithers across bad columns, contributing to modulations in the lightcurve.  This instrumentally complex dataset must be treated with extreme caution.

\paragraph{WR~22}

\begin{figure}[htb] 
\begin{center}
\mbox{
\includegraphics[width=0.6\textwidth]{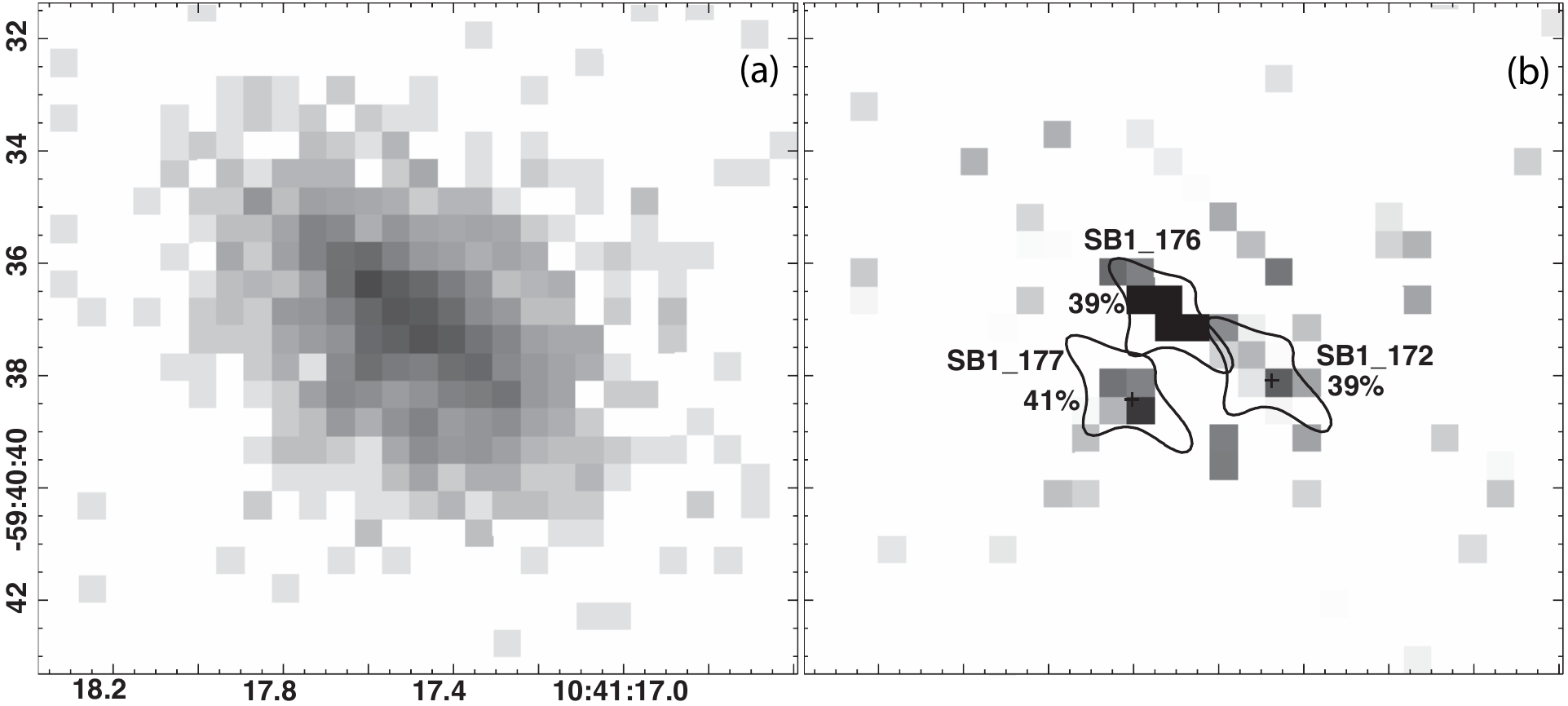}
\hspace*{0.1in}
\includegraphics[width=0.325\textwidth]{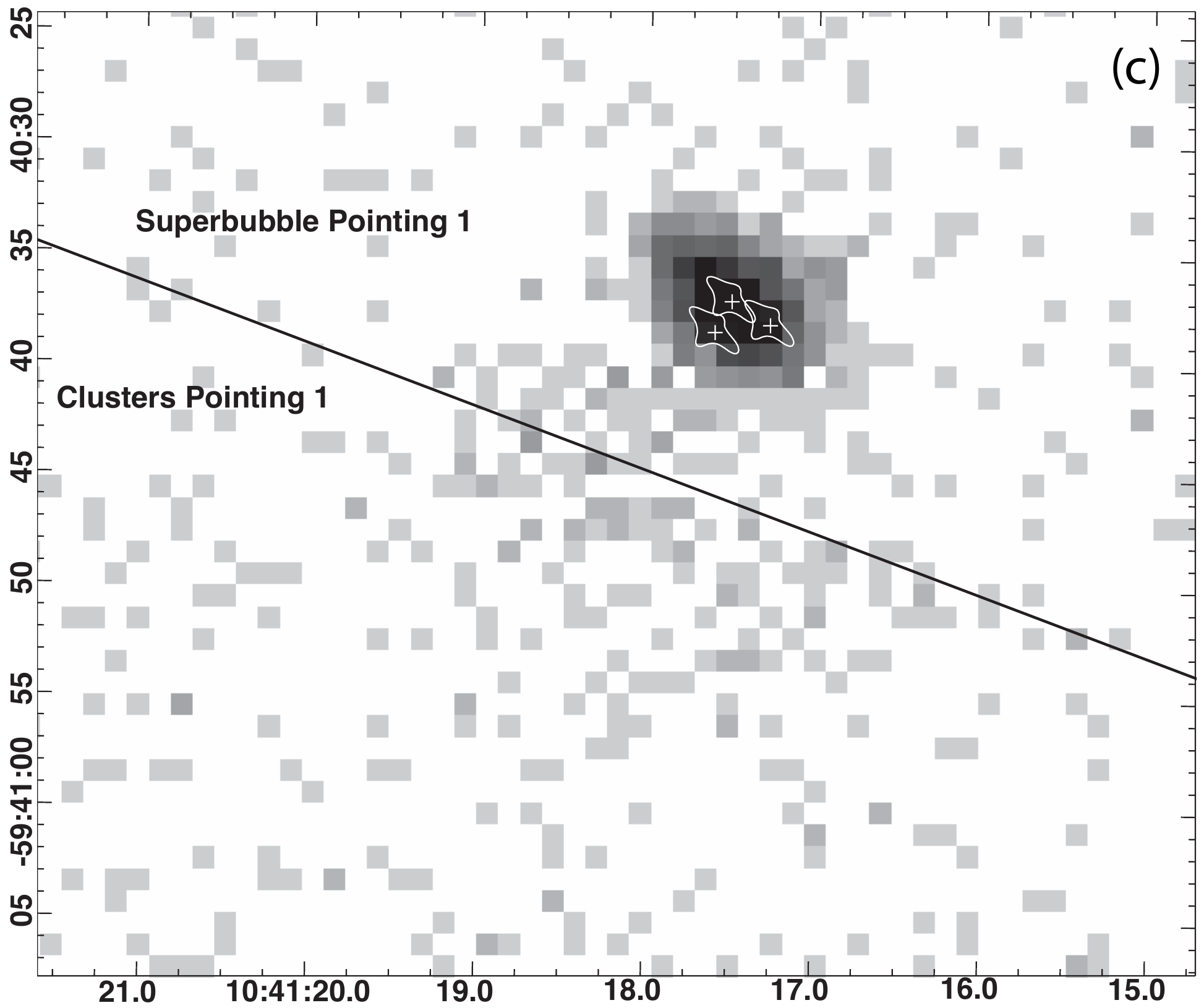}}
\caption{
(a) The original CCCP image of the WR~22 region, binned by 1 sky pixel.  
(b) The reconstructed image of WR~22 and its surroundings.  CCCP source labels and extraction regions are shown; percentages show the fraction of the total PSF extracted (reduced from the nominal 90\% to minimize overlap).  Coordinates are the same as in (a).
(c) A wider view of the CCCP field around WR~22, binned by 2 sky pixels.  The diagonal black line roughly indicates the edge of Clusters Pointing 1 (note that the telescope dithers across a $16\arcsec \times 16\arcsec$ area so the actual sky location of the edge of a pointing varies with time).  The fan-shaped distribution of events southeast of WR~22 is an observing artifact, not diffuse emission.
} 
\normalsize 
\label{fig:wr22}
\end{center}
\end{figure}

WR~22 is a WR+O binary with an 80-day orbit that has been studied in detail with several {\em XMM} observations \citep{Gosset09}.  The CCCP data were obtained in 2 ObsID's taken 1 day apart in May 2008 (see Table~\ref{tbl:obslog}).  The spectrum is consistent with the results of Gosset et al., well-fit by a two-temperature thermal plasma with components at about 0.6 and 4~keV.  The CCCP shows interesting new spatial information, though:  two other X-ray sources lie within 2$\arcsec$ of WR~22 (Figure~\ref{fig:wr22}).  Assuming that the X-ray emission from these two sources is constant, they would contribute $\sim$17\% of the {\em XMM} flux from WR~22 at the phase of the CCCP observations, since they would not be resolved by {\em XMM}.  They have no known counterparts in other wavebands, due in part to the brightness of WR~22.  If these are pre-MS stars, they are likely variable in X-rays, so it is impossible to know how much they might have contributed to the X-ray emission from WR~22 in the {\em XMM} observations.

%-------------------------------------------------------------------------
%-------------------------------------------------------------------------
 
The CCCP mosaic shows an unusual fan-shaped spray of events southeast of WR~22 (Figure~\ref{fig:wr22}c).  This is due to scattered light in the Clusters Pointing 1 observation; while it was cleanly imaged in Superbubble Pointing 1, WR~22 was located just off the edge of the field in Clusters Pointing 1 and was thus not detected in this pointing, but it contributed photons to the observation due to scattered light.  The telescope dither smears out that scattered light and contributes to its large spatial extent.  It is important not to attribute these scattered light events to ``diffuse emission'' or to any other astrophysical origin.  This odd data anomaly is not seen around any other source in the CCCP.

\paragraph{WR~24}

WR~24 has not yet been observed by {\em XMM}, but \citet{Skinner10} analyzed one of the CCCP observations of this source (ObsID 9482, 56~ks) and found a good spectral fit with a two-temperature thermal plasma with components at 0.7 and 3.3~keV.  They assumed a distance of 3.24~kpc for this source, which would place it behind the Carina complex; we consider this to be unlikely because this rare type of star is almost exclusively found to be associated with giant \hii regions.  Reducing its distance to 2.3~kpc to be commensurate with our assumption for the Carina complex would reduce its X-ray luminosity as reported by Skinner et al.\ by a factor of two.

We combined 6 CCCP ObsID's (including custom extractions for 2 ObsID's where WR~24 was observed far off-axis on the S2 CCD) into a composite spectrum with a total of 170~ks integration time.  Fitting this with a two-temperature thermal plasma with variable abundances (but with the abundances of the soft and hard components linked together), we can recover a fit similar to that of Skinner et al., with temperatures of 0.7 and 5.8~keV and supersolar abundances for O, Mg, Si, S, and Ar.  Applying this fit to each ObsID's spectrum separately, it appears that the soft plasma component is fairly constant over time but the hard component may vary somewhat (although it is not well-constrained in the shorter observations).  A better fit to the composite spectrum, however, is obtained with a different family of models.  These have soft components and elemental abundances similar to the above model, but the hard component is much less hard (kT$\sim$1~keV) and is highly obscured ($N_{H} \sim 5 \times 10^{22}$~cm$^{-2}$).  Again, this target and this large dataset merits a much more detailed future treatment to arrive at a physically meaningful model.

%\clearpage

%=============================================================================
\subsection{Revealed Stellar Populations \label{sec:revealed}}

\begin{figure}[htb] 
\begin{center}
\includegraphics[height=0.8\textheight]{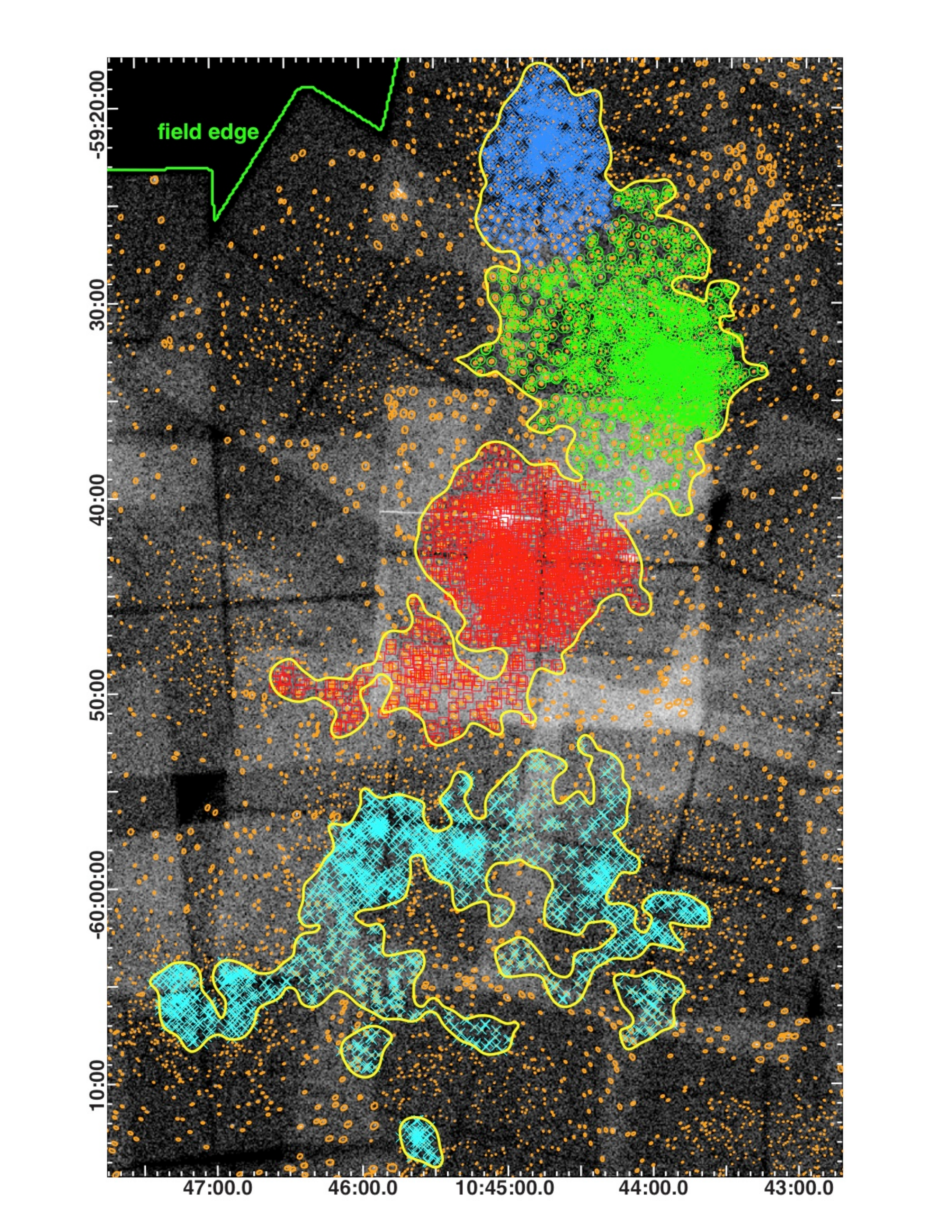}
\caption{
Carina's ``backbone'' of stellar clusters, as defined by source density contours (in yellow) from \citet{Feigelson11}; again 1~pc $\simeq 1.49\arcmin$.  This region is an expanded view of the central part of the CCCP mosaic, where the source density is highest.  The underlying image shows the binned event data; chip gaps, readout streaks, missing coverage, and overlapping fields are apparent.  ACIS sources are depicted here via their extraction polygons (shown in orange); superposed on these are symbols showing sources that will be further examined in detailed studies of Carina's historically-recognized constituent clusters:  blue diamonds show the Tr15 region \citep{Wang11}, green circles show the Tr14 region (Townsley et al.\ in prep.), red boxes show the Tr16 region \citep{Wolk11}, and cyan X's show the southern region, including Collinder~228, Bochum~11, and the Treasure Chest (Stassun et al.\ in prep.).
} 
\normalsize 
\label{fig:clusters}
\end{center}
\end{figure}

It is clear from the median energies presented in Figure~\ref{fig:sourcepops} that most CCCP X-ray sources are ``revealed,'' i.e., not highly-obscured, since intervening obscuration preferentially removes soft X-rays, leading to higher median energies.  Our point source classifier (Section~\ref{sec:classifier} and Figure~\ref{fig:classification}) further shows that most of our point sources are young stars in the Carina complex.  Thus several important observations regarding Carina's revealed stellar population can be made just by casually examining the CCCP X-ray point source locations.  More detailed inferences can be made by studying the clustering of CCCP sources classified as Carina members; this analysis is performed in \citet{Feigelson11} and a simple illustration is shown in Figure~\ref{fig:clusters}.

%-------------------------------------------------------------------------
%-------------------------------------------------------------------------

The region between Tr14 (to its southeast) and Tr16 (to its northwest) has a very long exposure ($\sim$144~ks) yet it is comparatively sparsely populated with X-ray sources.Ê This is not due to heavy obscuration.  This clearly indicates that Tr14 and Tr16 are separate clusters and that they are not triggering substantial star formation between them.  Similarly, the small clumps and clusters distributed across the field south of Tr16 do not clearly form an extension of the Tr16 cluster, as has been proposed in the literature \citep[reviewed by][]{Smith08}.  These papers suggest that the V-shaped dust lane south of Tr16 led to an artificial separation of \object{Collinder~228} into a distinct cluster in visual studies; now it is clear that Collinder~228 is neither an extension of Tr16 nor a well-developed, spherically-symmetric monolithic cluster like Tr14 or Tr15.  

Bochum~11, at RA$\sim10^{h}47^{m}$ and Dec$\sim-60^{\circ}06\arcmin$, is clearly a cluster in X-rays, but it is not very centrally-concentrated or spherically symmetric.Ê It appears similar to other small clumps and clusters distributed throughout the complex south of Tr16.  Examining the full survey images (Figures~\ref{fig:cccpfull} and \ref{fig:fluximage}), we note that few clusters are seen deep in the South Pillars.  A few obscured clumps of probably very young stars do appear in these southern regions; they will be discussed in the next section and in other CCCP papers \citep{Feigelson11,Povich11a}.  Bochum~10, at the northern edge of the CCCP survey, is not completely covered by our data, but no cluster is discernable for the part we do cover.

In this northern region and around the periphery of the entire survey a large population of young stars is found, far from any rich cluster or molecular cloud (see Figure~\ref{fig:classification} showing source classifications); these many thousands of stars trace a distributed population.Ê These may represent one or more older populations (perhaps 4--10~Myr) of young stars, all of which will be bright X-ray sources because enhanced X-ray emission due to magnetic activity continues for $\sim$100~Myr in young stars \citep{Preibisch05b}.Ê As described above, we expect fewer than 1 contaminant source (foreground star, background star, or extragalactic object) per square arcminute in the CCCP survey, so this distributed young stellar population can be studied without being overwhelmed by contaminants.  Several papers in the {\em Special Issue} discuss the distributed population.

The famous cluster Tr16 appears to be more an agglomeration of clumps than a single, centrally-concentrated cluster.  The X-ray properties of these clumps, including an important obscured clump southeast of Tr16 first noticed by \citet{Sanchawala07a}, are studied in detail by \citet{Wolk11}.  Late-B stars close to \etacar in Tr16 are identified; $>$30\% of them are found to have low-mass pre-MS companions \citep{Evans11}.  The slightly older monolithic cluster Tr15 has a robust population of X-ray sources, although it appears to be missing most of its massive stars (earlier than spectral type O9), perhaps due to supernova activity \citep{Wang11}.  Tr14 and the clumps and clusters in the South Pillars will be the subject of future work (Townsley et al.\ in prep., Stassun et al. in prep.).

\clearpage

%=============================================================================
\subsection{Obscured Stellar Populations \label{sec:obscured}}

\subsubsection{Clumps and Subclusters}

\begin{figure}[htb] 
\begin{center}
\includegraphics[width=1.0\textwidth]{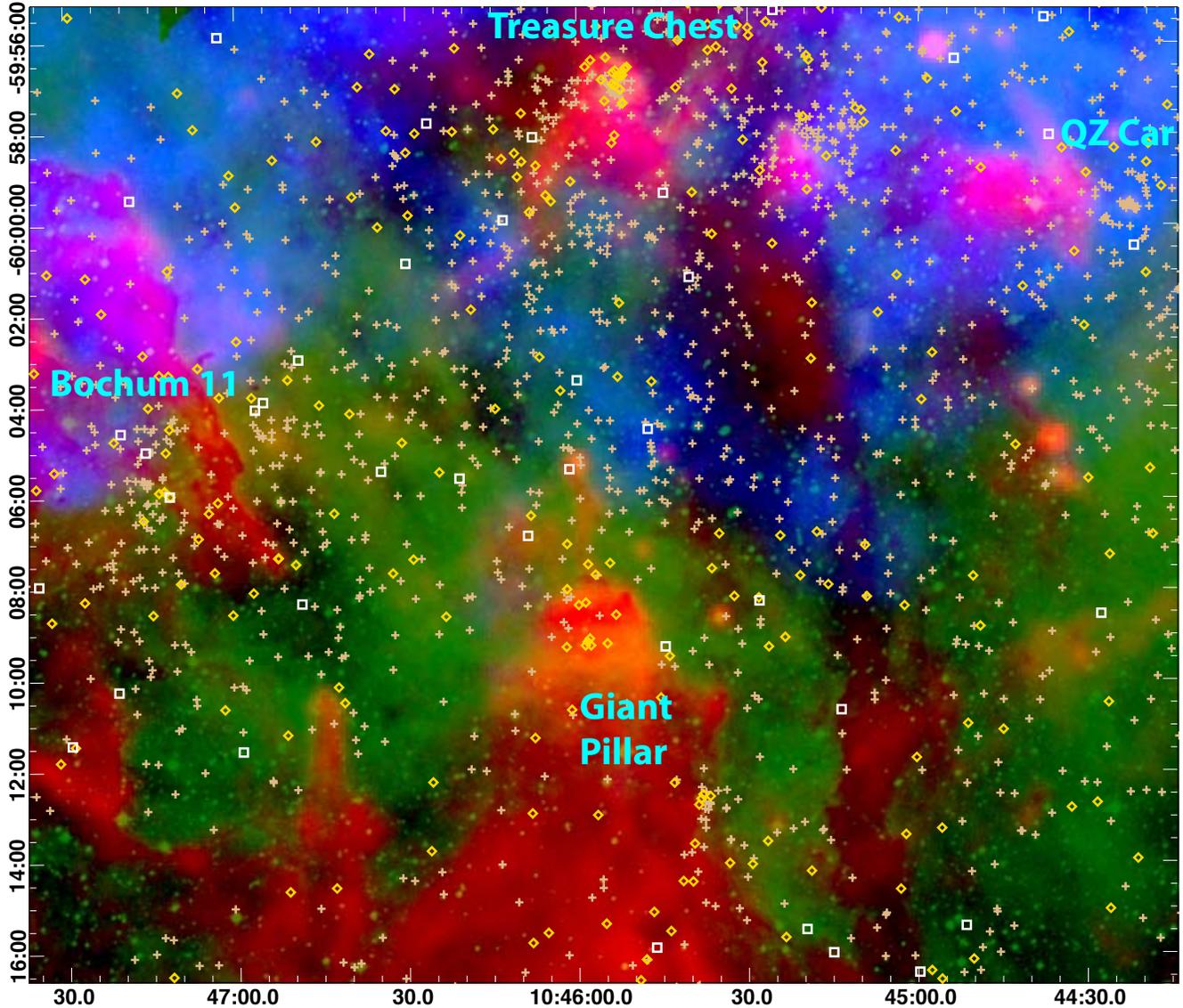}
\caption{
X-ray clumps and clusters in the South Pillars.  This image illustrates both revealed and obscured stellar populations in the CCCP by showing the positions of X-ray sources as median-energy-coded symbols from Figure~\ref{fig:sourcepops} (sources with 0.5~keV$< E_{med} <$2~keV as tan plusses, sources with 2~keV$< E_{med} <$4~keV as yellow diamonds, and sources with 4~keV$< E_{med} <$8~keV as white boxes).  These X-ray point sources are displayed on a multiwavelength image with {\em MSX} 8~$\mu$m data in red, Digitized Sky Survey data in green, and adaptively-smoothed diffuse X-ray emission, 0.5--2~keV, in blue, where X-ray point sources were removed before smoothing \citep{Broos10}.
Because of their strong dependence on the survey sensitivity, weak X-ray sources (net counts $<$5.0) are not shown.
} 
\normalsize 
\label{fig:populations}
\end{center}
\end{figure}

From Figure~\ref{fig:sourcepops}c it is clear that there are no major obscured young stellar clusters in the Carina complex; any embedded clusters with more than $\sim$50 members would be detected as a `Group' of CCCP sources in the spatial analysis of \citet{Feigelson11}.  These X-ray results confirm and extend the recent {\em Spitzer} South Pillars study \citep{Smith10b} that also found no new massive stellar clusters embedded in, or obscured by, Carina's remaining molecular material in the south.  Rather both the CCCP and IR studies show a variety of small clumps and subclusters in both the northern and southern parts of the Carina complex, with varying degrees of obscuration and varying fractions of young, disky IR-excess sources \citep{Smith10b,Feigelson11,Povich11a}.  A few of these clumps are shown in Figure~\ref{fig:populations}, which samples a region in the South Pillars that is particularly illustrative of the range of CCCP science.  

%-------------------------------------------------------------------------
%-------------------------------------------------------------------------

The underlying multiwavelength image in Figure~\ref{fig:populations} shows the complexity of Carina's ISM, with heated dust and PAH emission (shown in red) tracing pillars and other shreds of molecular material, dense ionized gas (shown in green) at the edges of the pillars, and hot plasma (shown in blue) suffusing the upper regions.  We have excised the many CCCP X-ray point sources from this image, then reintroduced them conceptually with symbols that code their median energies.  Unobscured X-ray sources (tan plusses) mix with obscured X-ray sources (yellow diamonds and white boxes), sometimes forming distinct clumps.  The Treasure Chest is a known embedded young cluster \citep{Smith05a}, but close to it are other clumps of X-ray sources that are on average less obscured.  A group of obscured sources resides at the top of the Giant Pillar, but a tight clump of X-ray sources also sits on its western flank.  The young cluster Bochum~11 has an X-ray population with a wide range of obscurations; perhaps some of these sources lie deeper within the molecular cloud and are just seen superposed on the cluster.  The group of sources around the massive system QZ~Car (Figure~\ref{fig:massive}) is unusually dense.

\clearpage

\subsubsection{A Mysterious Clump in the South Pillars \label{sec:hard_triplet}}

\begin{figure}[htb] 
\begin{center}  
\includegraphics[width=0.5\textwidth]{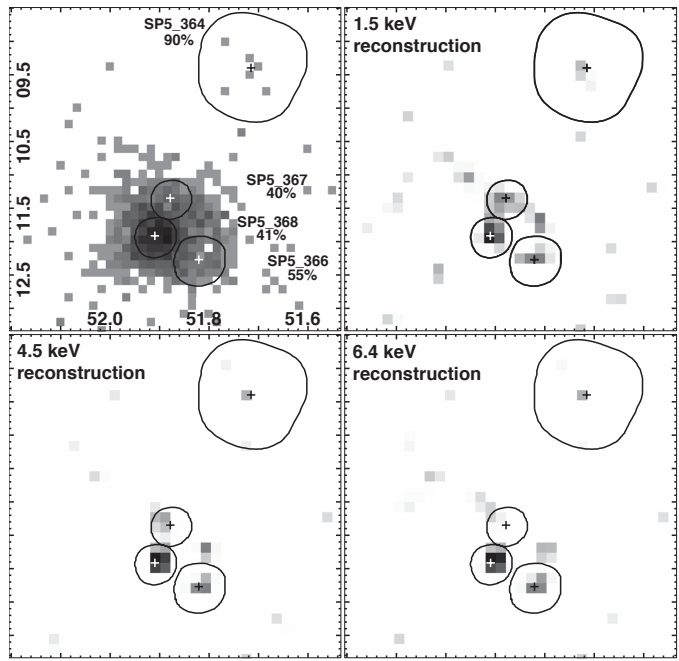}
\caption{
A small clump of highly obscured sources at the edge of the South Pillars; the brightest source (labeled SP5\_368) is CXOGNC~J104451.91-602511.9, with 377 net counts and a median energy of 5.25~keV.                                                   
(a) The ACIS event data binned by 0.5 sky pixels (0.25$\arcsec$).  CCCP source labels and PSF fractions contained within the extraction regions for each source are shown.  Extraction regions are reduced for the three close sources to minimize the effects of crowding in their spectra.
(b) Maximum likelihood reconstruction of the field shown in (a), using the 1.5~keV PSF that is the default PSF for all automated extractions in the CCCP project.
(c) A reconstruction of the same field using the 4.5~keV PSF.
(d) A reconstruction of the same field using the 6.4~keV PSF.  These harder PSF's are more appropriate for these obscured sources, which have median energy $>$5~keV.  
The extraction regions from (a) are reproduced in the reconstructed images for comparison.
} 
\label{fig:obscured}
\end{center}
\end{figure}

We have found a hard, highly-obscured clump of 4 sources near the western edge of the South Pillars, at RA$\sim10^{h}44^{m}51.91^{s}$, Dec$\sim-60^{\circ}25\arcmin11.9\arcsec$ (Figure~\ref{fig:obscured} and Table~\ref{tbl:mystery}).  Our image reconstruction and source detection process \citep{Broos11a} uses by default a 1.5~keV PSF to reconstruct the field and find point sources.  Since 3 of the sources in this obscured clump lie with 1$\arcsec$ of each other, we were suspicious of over-reconstruction, since the 1.5~keV PSF is sharper than higher-energy PSFs.  Thus we also reconstructed the field with PSFs at 4.5~keV and 6.4~keV (Figure~\ref{fig:obscured}c and d respectively).  All reconstructions recover the 3 closely-spaced sources plus another, isolated source $\sim 2.5\arcsec$ to the northwest; all of these sources have median energies $>$5~keV (Table~\ref{tbl:mystery}).  All reconstructions also show a likely fourth source in the bright clump (at RA$\sim10^{h}44^{m}51.81^{s}$, Dec$\sim-60^{\circ}25\arcmin11.8\arcsec$); this potential source failed our automatic source detection criteria (it is not quite bright enough in the reconstruction) so it was not extracted.

%-------------------------------------------------------------------------
%-------------------------------------------------------------------------

The 3 closely-spaced sources were bright enough to merit spectral fits.  We fit them in {\em XSPEC} \citep{Arnaud96} using a simple one-temperature plasma model with absorption, {\em TBabs*apec}.  X-ray source properties and spectral fit parameters are given in Table~\ref{tbl:mystery}.

%-----------------------------------------------------------------------------
\begin{deluxetable}{llllcrccrrr}
\tablecaption{Obscured Clump in the South Pillars \label{tbl:mystery}
}
\tablewidth{0pt}
\tabletypesize{\footnotesize}
%\rotate
\tablehead{
\colhead{CXOGNC J} &\colhead{Label} &\colhead{$\alpha$ (J2000.0)} &\colhead{$\delta$ (J2000.0)} &\colhead{PosErr}    &\colhead{NC} &\colhead{$E_{med}$} & \colhead{Var.} & \colhead{$N_{H}$} & \colhead{kT} & \colhead{$L_{tc}$}\\  
                   &                &\colhead{(\arcdeg)}          & \colhead{(\arcdeg)}         &\colhead{(\arcsec)} &\colhead{(cts)}     &\colhead{(keV)}           &                &\colhead{(cm$^{-2}$)}   &\colhead{(keV)}  &\colhead{(erg~s$^{-1}$)}              \\
\numberthecolumn &\numberthecolumn &\numberthecolumn &\numberthecolumn &\numberthecolumn &\numberthecolumn &\numberthecolumn &\numberthecolumn &\numberthecolumn &\numberthecolumn &\numberthecolumn
\setcounter{column_number}{1}
}
\startdata

104451.91-602511.9 & SP5\_368 &  161.216292 & -60.419976 & 0.01 & 377\phd\phn & 5.3 & c       & $41 \times 10^{22}$ & \{15\} & $2 \times 10^{33}$ \\
104451.82-602512.2 & SP5\_366 &  161.215922 & -60.420075 & 0.03 &  52\phd\phn & 5.5 & a       & $70 \times 10^{22}$ & \{15\} & $5 \times 10^{32}$ \\
104451.87-602511.3 & SP5\_367 &  161.216162 & -60.419819 & 0.02 &  26\phd\phn & 5.8 & a       & $70 \times 10^{22}$ & 6 & $8 \times 10^{32}$ \\
104451.71-602509.3 & SP5\_364 &  161.215478 & -60.419277 & 0.20  &  4.6       & 5.5 & \nodata & \nodata & \nodata & \nodata
\enddata
\tablecomments{
Col.\ (1): IAU designation 
\\Col.\ (2): Source name used within the CCCP project 
\\Col.\ (3,4): Position 
\\Col.\ (5): Position error
\\Col.\ (6): Net X-ray counts extracted in the total band, 0.5--8~keV
\\Col.\ (7): Median X-ray energy in the total band
\\Col.\ (8): Variability characterization based on K-S statistic (total band): (a) no evidence for variability ($0.05<P_{KS}$); (b) possibly variable ($0.005<P_{KS}<0.05$); (c) definitely variable ($P_{KS}<0.005$)
\\Col.\ (9): Absorbing column from spectral fit
\\Col.\ (10): Thermal plasma temperature from spectral fit; values in braces were frozen in the fit
\\Col.\ (11): Total-band absorption-corrected luminosity, assuming D=2.3~kpc
}
\end{deluxetable}
%-------------------------------------------------------------------------

\clearpage

\begin{figure}[htb] 
\begin{center}
\includegraphics[width=0.45\textwidth]{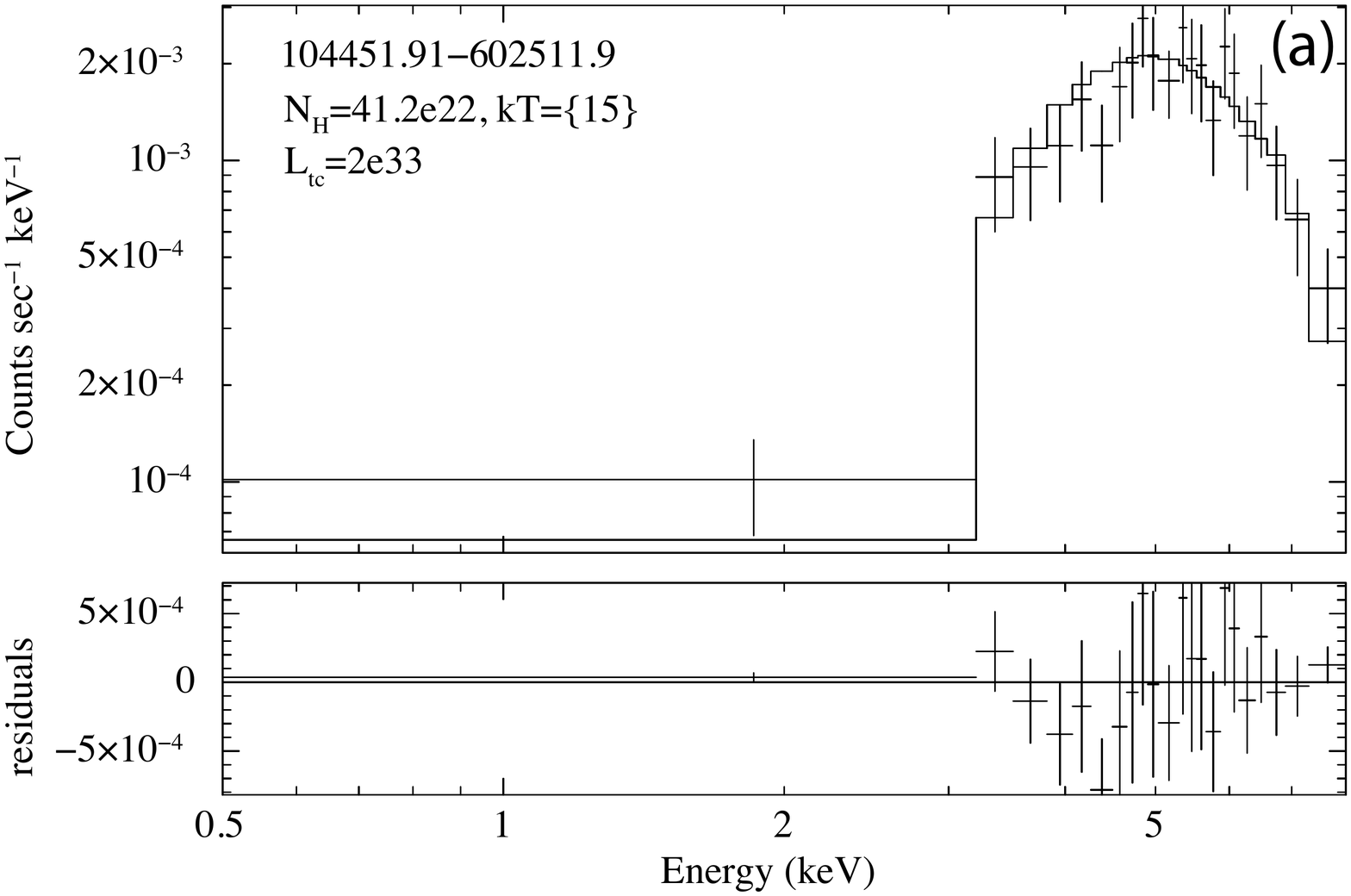}
%\hspace*{0.1in}
\includegraphics[width=0.45\textwidth]{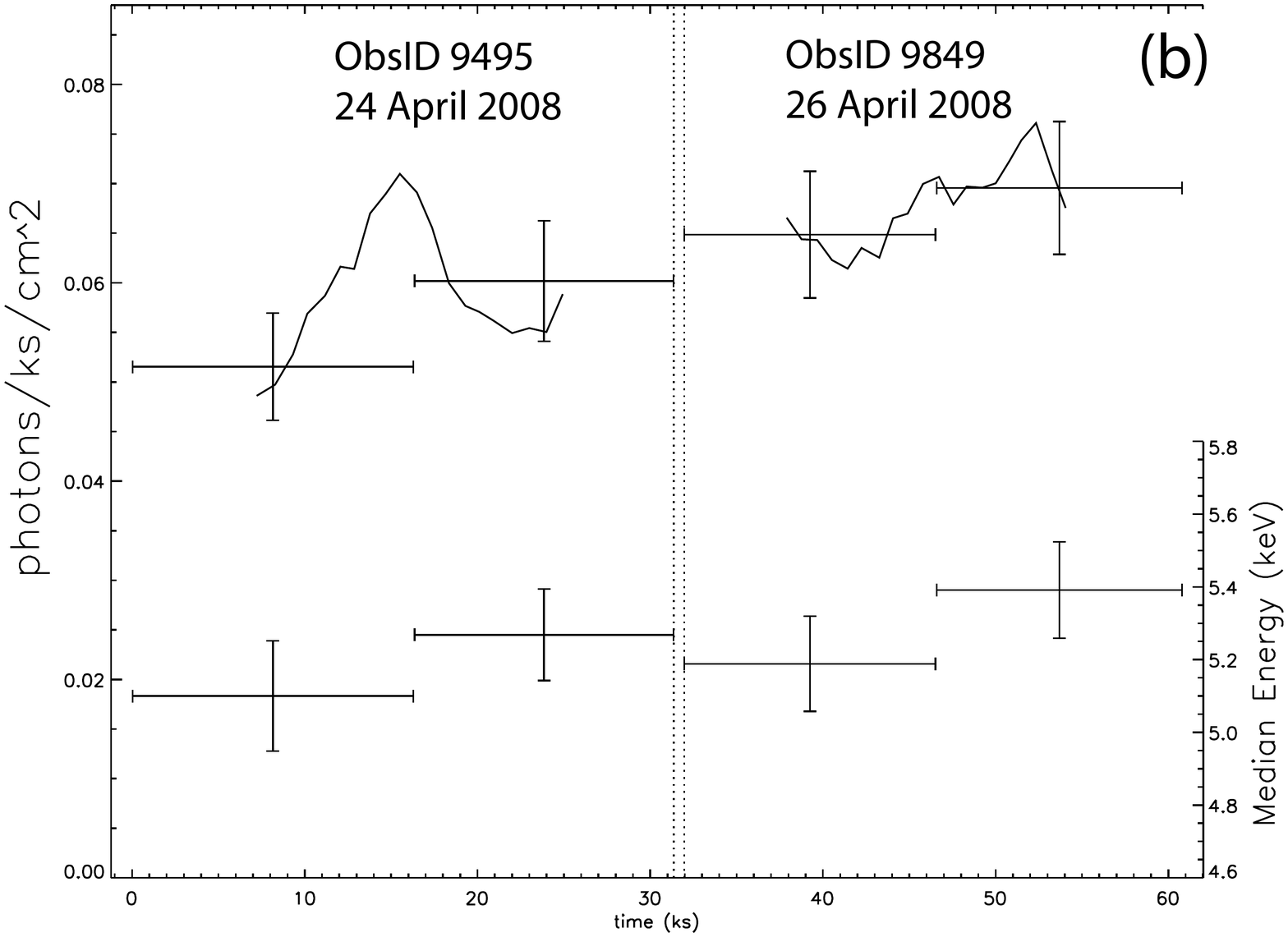}
\caption{
(a)  Spectrum of the bright, highly-obscured source CXOGNC~J104451.91-602511.9 (labeled SP5\_368 in Figure~\ref{fig:obscured}), one of a small group of highly-obscured sources in CCCP South Pillars Pointing 5.
(b)  Multi-observation lightcurve of the same source.  It is variable in the first ObsID.
%***P\_KS is 0.0048 for first ObsID, 0.89 for second.***
} 
\normalsize 
\label{fig:SP5_368}
\end{center}
\end{figure}

The ACIS spectrum and lightcurve of the brightest source in the clump (SP5\_368) are shown in Figure~\ref{fig:SP5_368}.  It shows a highly-obscured source with a spectrum so hard that ACIS cannot determine its temperature accurately (we froze kT at 15~keV, an arbitrary upper limit that we impose for very hard spectra).  A power-law fit yields a flat slope.  The source is clearly highly-obscured, with essentially no counts below 3~keV; using the \citet{Vuong03} gas-to-dust ratio $N_{H} = 1.6 \times 10^{21} * A_{V}$, our X-ray spectral fit yields an estimate of $A_{V} > 250$~mag of extinction.  If it is in fact a star at the distance of the Carina complex, its high X-ray luminosity requires it to be a massive star.  SP5\_368 is a variable X-ray source, according to the lightcurve from its first CCCP observation.  This lends support to the reconstruction results that there appears to be a group of point sources here; an extended source is unlikely to show such short-timescale variability.  Looking back at Table~\ref{tbl:mystery}, the other sources in this clump are similarly hard, highly-obscured, and intrinsically bright.  Such high obscuration could be masking a softer, even brighter spectral component; this would be expected if these are massive stars.

%-------------------------------------------------------------------------
%-------------------------------------------------------------------------

The nature of this highly-obscured tight clump of X-ray sources is most puzzling.  The short-term variability of SP5\_368, its very flat spectrum, and the fact that there appears to be a clump of similarly-obscured sources rather than a single source make an AGN interpretation implausible.  A 2MASS source is coincident with the clump; it is undetected at J but confidently seen at longer wavelengths (H=15, K=14).  \citet{Povich11a} performed SED-fitting to the near- and mid-IR data (including 24~$\mu$m) and find a fairly generic medium- to low-mass young stellar object (YSO); the IR source is inconsistent with an embedded massive YSO.  Perhaps the IR source is just a chance (and if so highly unfortunate) superposition on the X-ray clump.  Given the high obscuration inferred from the ACIS spectrum, the clump of X-ray sources is unlikely to be closer than the Carina complex; in fact a better explanation for the high absorbing column is that it is a background structure, perhaps a more distant unembedded massive stellar cluster in the Sagittarius-Carina spiral arm, highly obscured due to intervening spiral arm structures (including Carina's South Pillars).  These X-ray sources would then be even more intrinsically luminous; allowing substantially higher X-ray luminosities then suggests a globular cluster or nearby galaxy harboring several X-ray binaries.  Clearly deep, high-spatial-resolution near- and mid-IR imaging will be necessary to illuminate the nature of this mysterious clump of X-ray sources.

\clearpage

%=============================================================================
\subsection{Diffuse Emission \label{sec:diffuse}}

As described in Section~\ref{sec:intro}, Carina's diffuse X-ray emission has been recognized for over 30 years.  A major goal of the CCCP is to determine what fraction of the unresolved emission seen by {\em Einstein} and {\em ROSAT} is truly diffuse (rather than the integrated emission from tens of thousands of unresolved pre-MS stars) and to learn the origin of the diffuse emission by studying its spatial and spectral properties.  In the CCCP, our ability to study the diffuse emission is limited by our relatively shallow exposure; if we require a certain minimum quality in the diffuse spectra (say $>$1000 net counts for example), this limits the size of diffuse structures that we can probe.  Clearly, regions with higher apparent surface brightness can be divided into more cells for spectral fitting than fainter regions.  

\begin{figure}[htb] 
\begin{center}  
\includegraphics[width=1.0\textwidth]{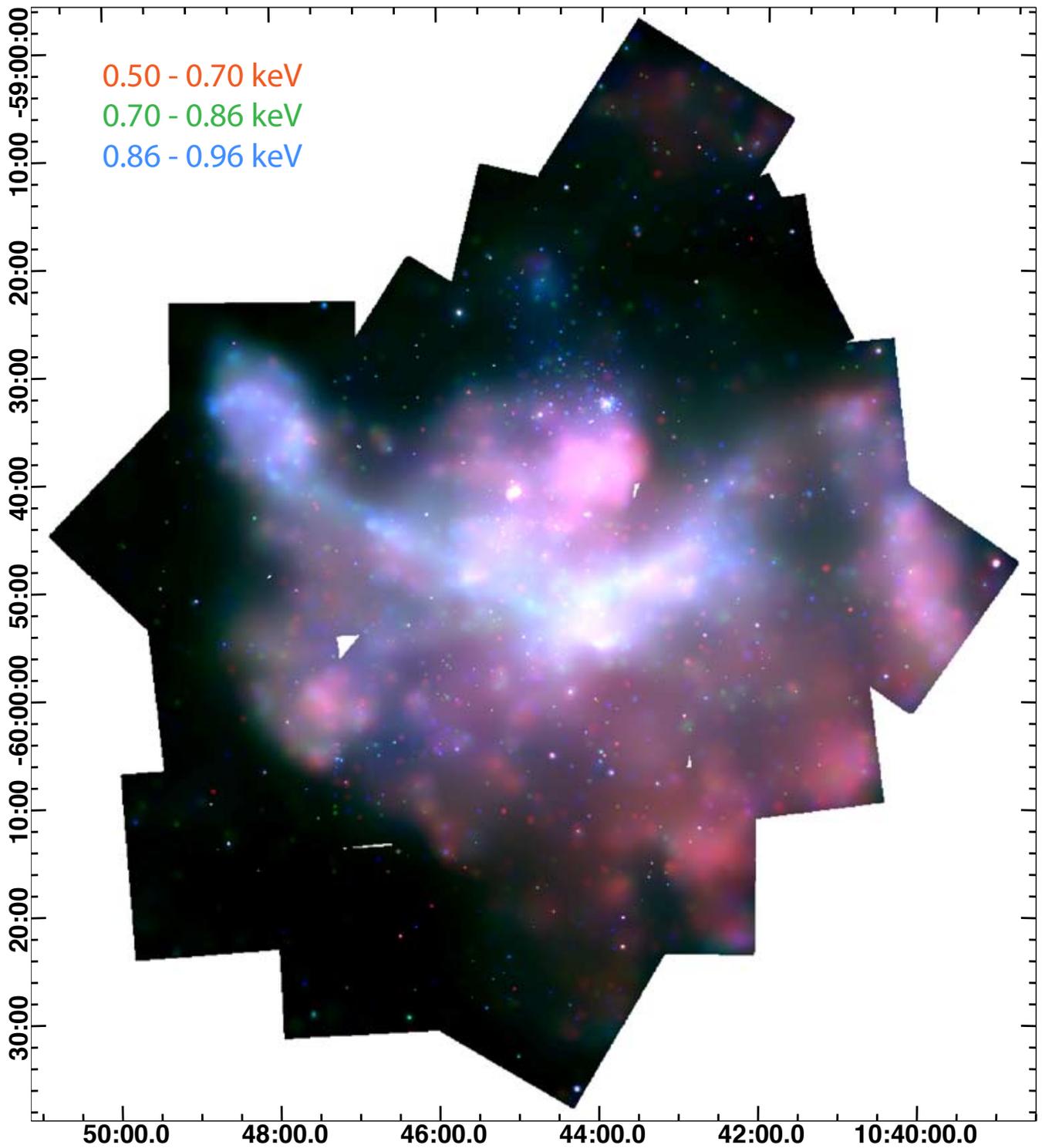}
\caption{
The \Chandra Carina data, smoothed with {\it csmooth}, now shown in 3 soft bands (0.50--0.70~keV in red, 0.70--0.86~keV in green, 0.86--0.96~keV in blue) to highlight the diffuse emission.} 
\label{fig:cccpdiffuse}
\end{center}
\end{figure}

Figure~\ref{fig:cccpdiffuse} illustrates the soft diffuse X-ray emission in Carina with 3 narrow-band images smoothed with {\it csmooth}, chosen from the global diffuse spectrum (see Section~\ref{sec:global} below) to isolate certain soft spectral features.  Note that these 3 narrow bands are contiguous in energy but sample a softer and narrower energy range (0.50--0.96~keV) than the band more usually defined as soft (0.5--2.0~keV).  

%-------------------------------------------------------------------------
%-------------------------------------------------------------------------

The soft diffuse structures revealed by \Chandra recover the large-scale morphology of the {\em Einstein} and {\em ROSAT} diffuse emission (see Figure~\ref{fig:intro}) but of course reveal more detail.  Although we caution readers not to interpret small-scale ``puffy'' structures too literally (they may reflect our choice of adaptive-kernel smoothing scales), overall the brightness and color variations in Figure~\ref{fig:cccpdiffuse} show that \Chandra gives us a much richer view of hot plasmas in \hii regions than previous missions.  The softest emission (0.50--0.70~keV) has wider spatial extent than the harder emission (0.70--0.96~keV), with apparent structure on 1--10~pc scales.  Sharp edges, especially around the periphery of the diffuse emission, imply that hot plasma is either confined by cavity edges or it is shadowed by foreground colder structures (or both).  

\begin{figure}[htb] 
\begin{center}
\includegraphics[width=1.0\textwidth]{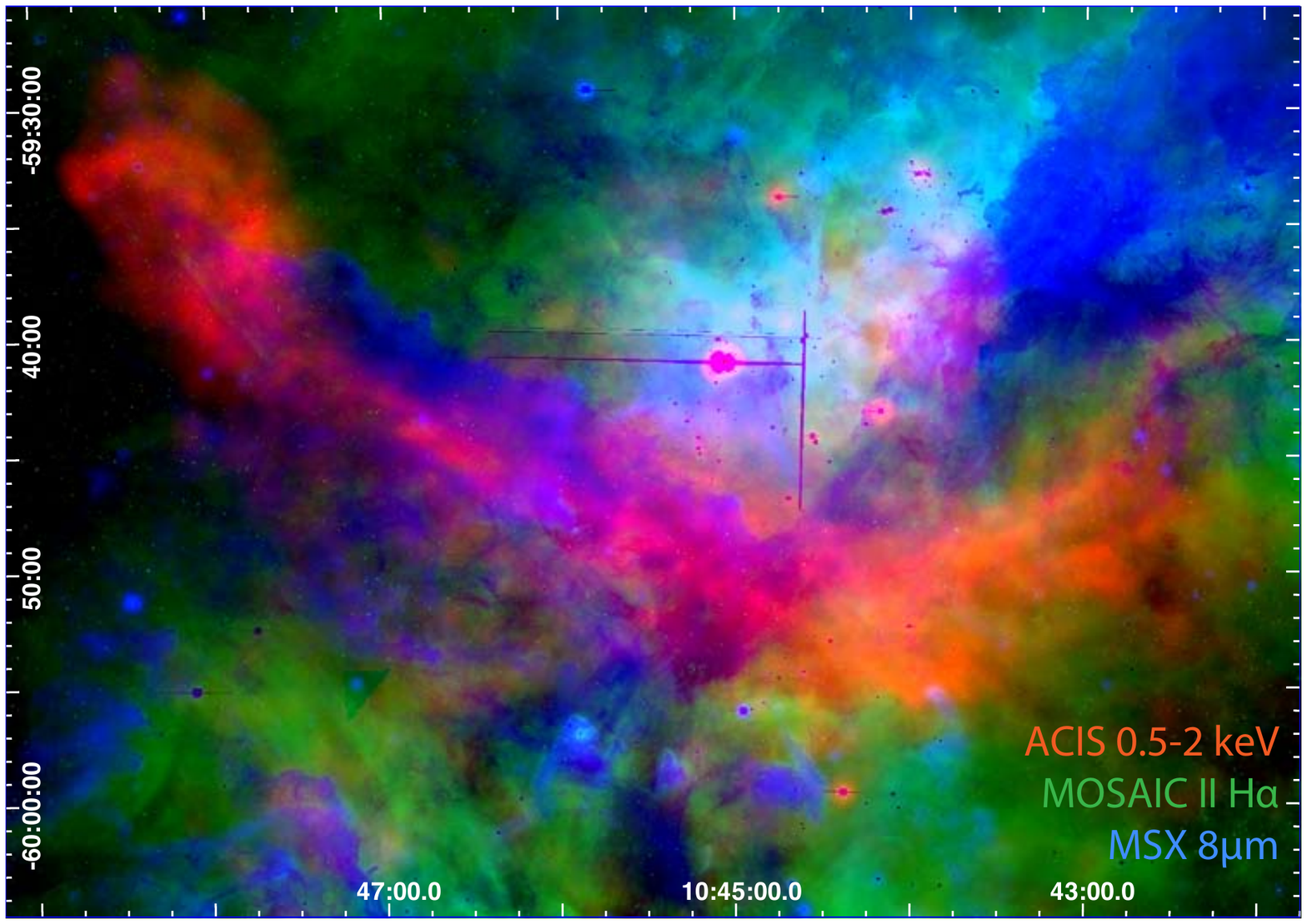}
\caption{The brightest diffuse X-ray emission (red) is displayed on a multiwavelength image with MOSAIC~II H$\alpha$ data in green and {\em MSX} 8~$\mu$m data in blue.  The soft (0.5--2~keV) X-ray emission is adaptively-smoothed with our custom software \citep{Broos10} that accepts an input image where X-ray point sources have been masked away, leaving only diffuse emission.  \etacar sits just north of the center of this image; it is strongly saturated in the MOSAIC~II and {\em MSX} data.
} 
\normalsize 
\label{fig:brightmulti}
\end{center}
\end{figure}

The brightest and hardest of the soft diffuse emission crosses the center of the Nebula with sharp linear structures only a few parsecs wide but many tens of parsecs long.  At the eastern edge of the survey, this linear feature stops abruptly and arcs to the north then east again, making a distinctive hook-shape (see also Figure~\ref{fig:brightmulti}).  Examining Figures~\ref{fig:intro}a and \ref{fig:fluximage}, we see that the eastern linear ``arm'' traces the edge of Carina's upper superbubble lobe \citep{Smith00}, encountering ever-increasing ISM densities as it approaches the Galactic midplane.  Perhaps the hook in the diffuse X-ray emission is caused by some encounter with dense material plowed up by Carina's upper superbubble lobe or other dense material near the Galactic Plane.  

To the west in Figure~\ref{fig:cccpdiffuse}, sharp bright linear features again stop in an arc-like structure running north-south and opening to the west, outlining the eastern edge of a void in the diffuse emission that is outlined on its western edge by a large linear swath of softer emission running north-south.  In Galactic coordinates (Figure~\ref{fig:fluximage}), this sharp western linear structure (the brightest diffuse X-ray emission in the field) is seen to run nearly parallel to the Galactic Plane at b$\sim-0.85^{\circ}$, situated just below the western arm of the V-shaped dust lane familiar from visual images of Carina.

The brightest diffuse X-ray emission is illustrated more clearly in Figure~\ref{fig:brightmulti}, now in context with visual and mid-IR data.  Its morphology is quite complex, likely due to a mix of shadowing by cooler gas and dust and confinement within the catacombs of Carina's ISM cavities.  For example, the linear swath of diffuse X-ray emission at RA$\sim10^{h}46^{m}15^{s}$ to $10^{h}47^{m}00^{s}$ and Dec$\sim-59^{\circ}45\arcmin$ fills a linear ``crevice'' in the {\em MSX} 8~$\mu$m emission; these linear structures follow the eastern arm of the V-shaped dust lane seen in the MOSAIC~II H$\alpha$ data (Gruendl et al.\ in prep).  The apex of this V shows a complicated mix of 8~$\mu$m emission and diffuse X-ray emission.  The brightest X-ray emission appears to be shaped in part by the ionized gas seen in the H$\alpha$ data, although again the morphology is not simple.  The western arm of the V is largely dominated by the Carina I molecular cloud; this structure could easily shadow any soft diffuse X-ray emission or displace any hot plasma.

%-------------------------------------------------------------------------
%-------------------------------------------------------------------------

South of Tr14 and west of Tr16 lies a bright, very soft diffuse structure covering $\sim$5~pc (centered at roughly RA$\sim10^{h}44^{m}$, Dec$\sim-59^{\circ}38\arcmin$; see Figure~\ref{fig:cccpdiffuse}).  To its west lies the Carina I molecular cloud that probably confines this soft plasma.  That molecular cloud may also shadow other parts of the diffuse emission.  Strong shadowing is seen in a 10$\arcmin$-long structure at RA$\sim10^{h}43^{m}$, Dec$\sim-60^{\circ}20\arcmin$.  This may be a cold cloud lying in front of the Carina complex, as it appears dark in visual and IR images as well.  Quantitative spectral fitting results for the CCCP's diffuse X-ray structures are presented in \citet{Townsley11a}.  

%The diffuse X-ray emission is compared to visual echelle spectra that detail the kinematics of the ionized gas in \citet{Gruendl11}.

\clearpage

%=============================================================================
\subsection{Multiwavelength Studies \label{sec:multi}}

%***SuperCOSMOS (http://www-wfau.roe.ac.uk/sss/halpha/index.html) has "short red" and H-alpha images of Carina!***

\begin{figure}[htb] 
\begin{center}  
\includegraphics[height=0.5\textheight]{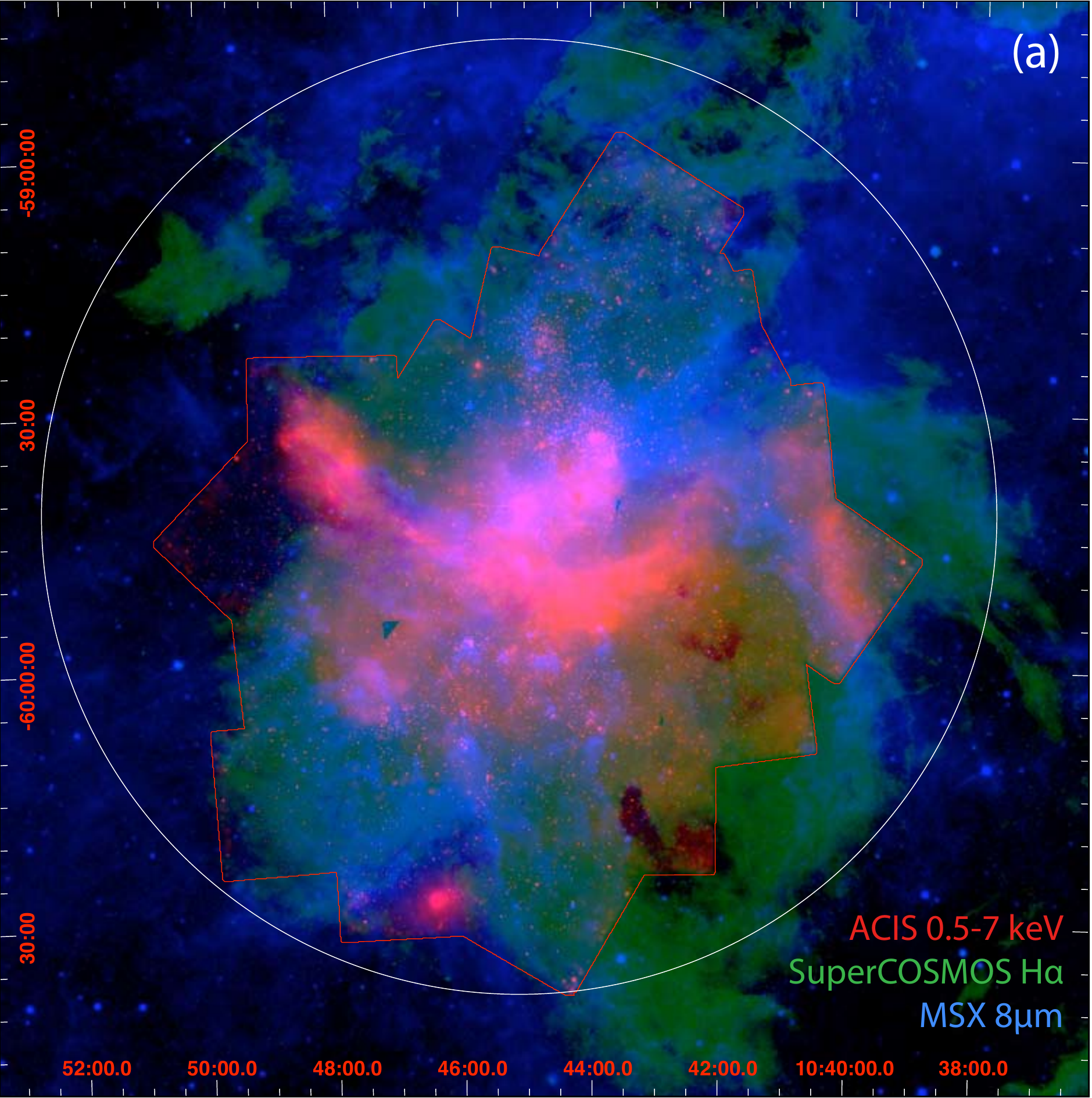}
\includegraphics[height=0.4\textheight]{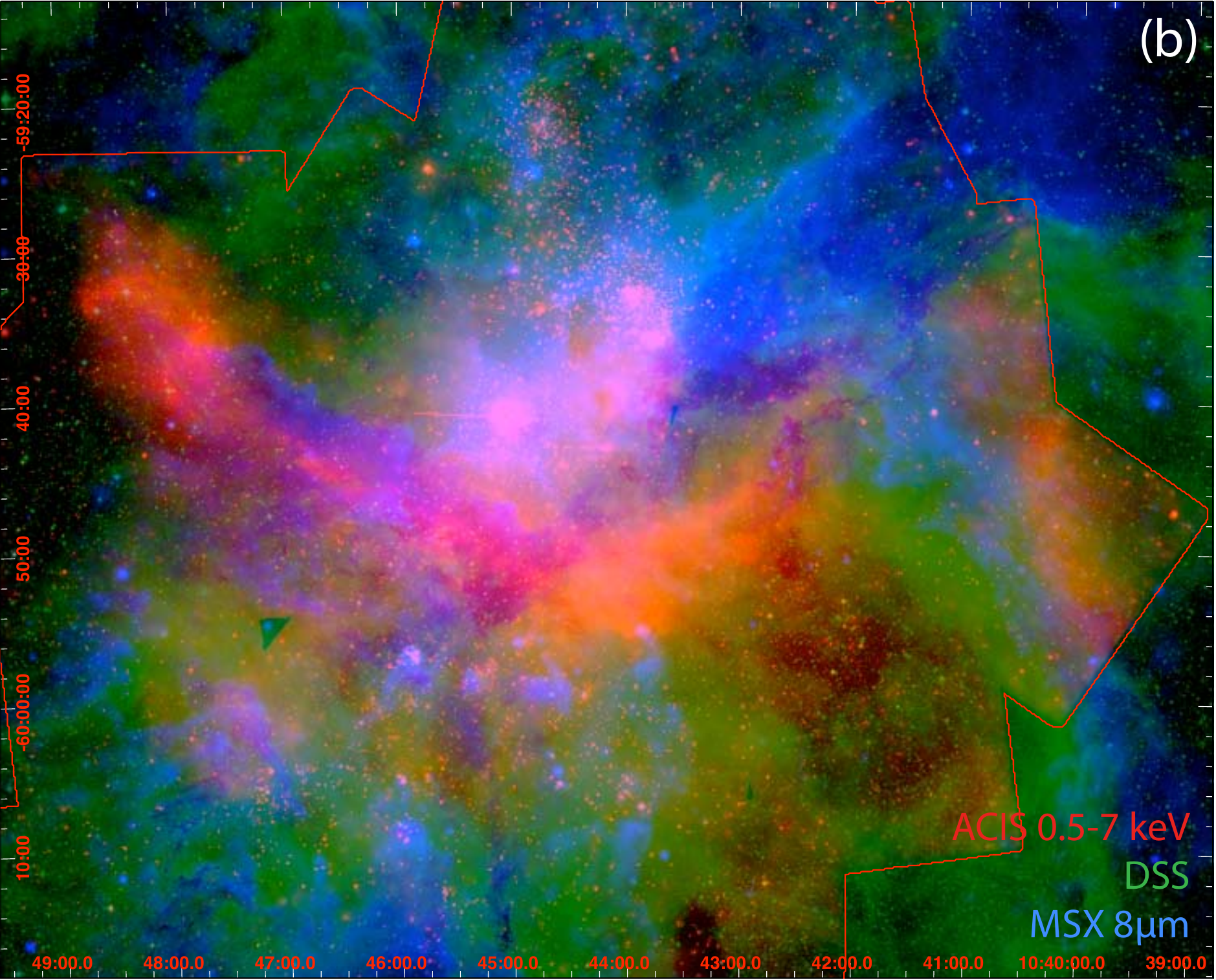}
\caption{Multiwavelength views of the CCCP field.
(a) ACIS total-band data smoothed with {\em csmooth} in red, the SuperCOSMOS H$\alpha$ data \citep{Parker05} in green, and the {\em MSX} 8~$\mu$m data in blue.  The CCCP survey region is outlined in red; the {\em ROSAT} field from Figure~\ref{fig:intro} is outlined by the large white circle.
(b) An expanded view of the CCCP field center, with ACIS and {\em MSX} data as in (a) and now showing the Digitized Sky Survey data in green.
} 
\label{fig:cccpmulti}
\end{center}
\end{figure}

The extensive X-ray point source populations and diffuse structures seen in the CCCP are best understood by putting them in a multiwavelength context, as several figures throughout this paper have already shown.  The last decade has seen much multiwavelength work in the Carina Nebula, including extensive surveys with {\em HST} \citep{Smith10a} and {\em Spitzer} \citep{Smith10b}.  Figure~\ref{fig:cccpmulti} shows some preliminary examples of multiwavelength comparisons in Carina, using {\em MSX} mid-IR data tracing PAH emission and heated dust (now shown in blue), visual SuperCOSMOS H$\alpha$ data \citep{Parker05} tracing dense ionized gas (shown in green), and soft X-ray emission tracing hot plasma (shown in red to make it easier to see).  

%-------------------------------------------------------------------------
%-------------------------------------------------------------------------

The PAH emission and dense ionized gas emission are often strikingly anticoincident.  The X-ray structures show more complex morphology in multiwavelength comparisons, sometimes superposed on cooler structures with little correlation (e.g., the eastern ``hook'').  In other cases, such as the cold pillar in the south that shadowed the X-rays as described in the last section, the visual and IR data are perfectly commensurate (this pillar appears dark at all three wavelengths studied here).  

\begin{figure}[htb] 
\begin{center}
\mbox{  
\includegraphics[width=0.4\textwidth]{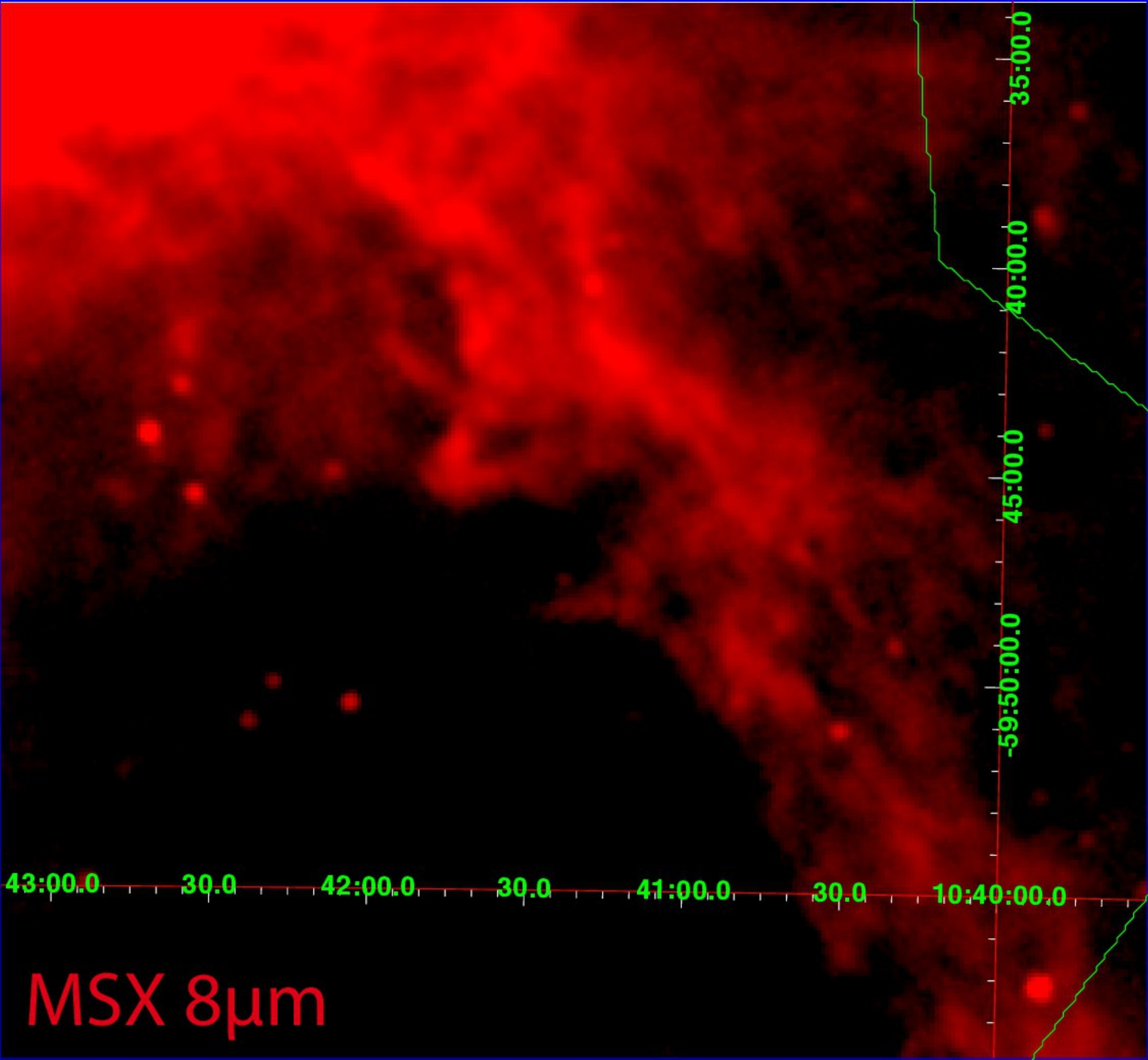}
\includegraphics[width=0.4\textwidth]{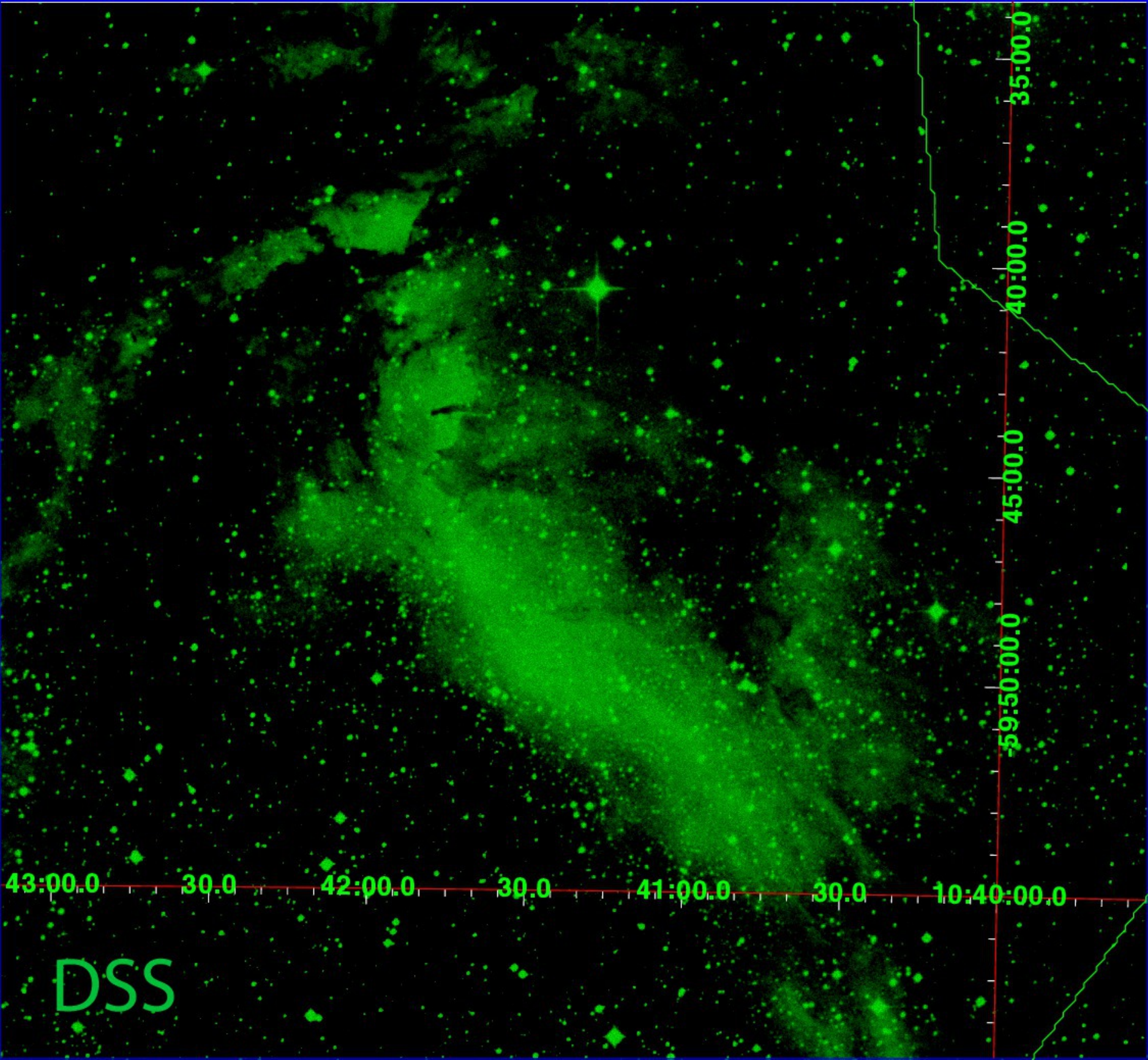}}
\mbox{ 
\includegraphics[width=0.4\textwidth]{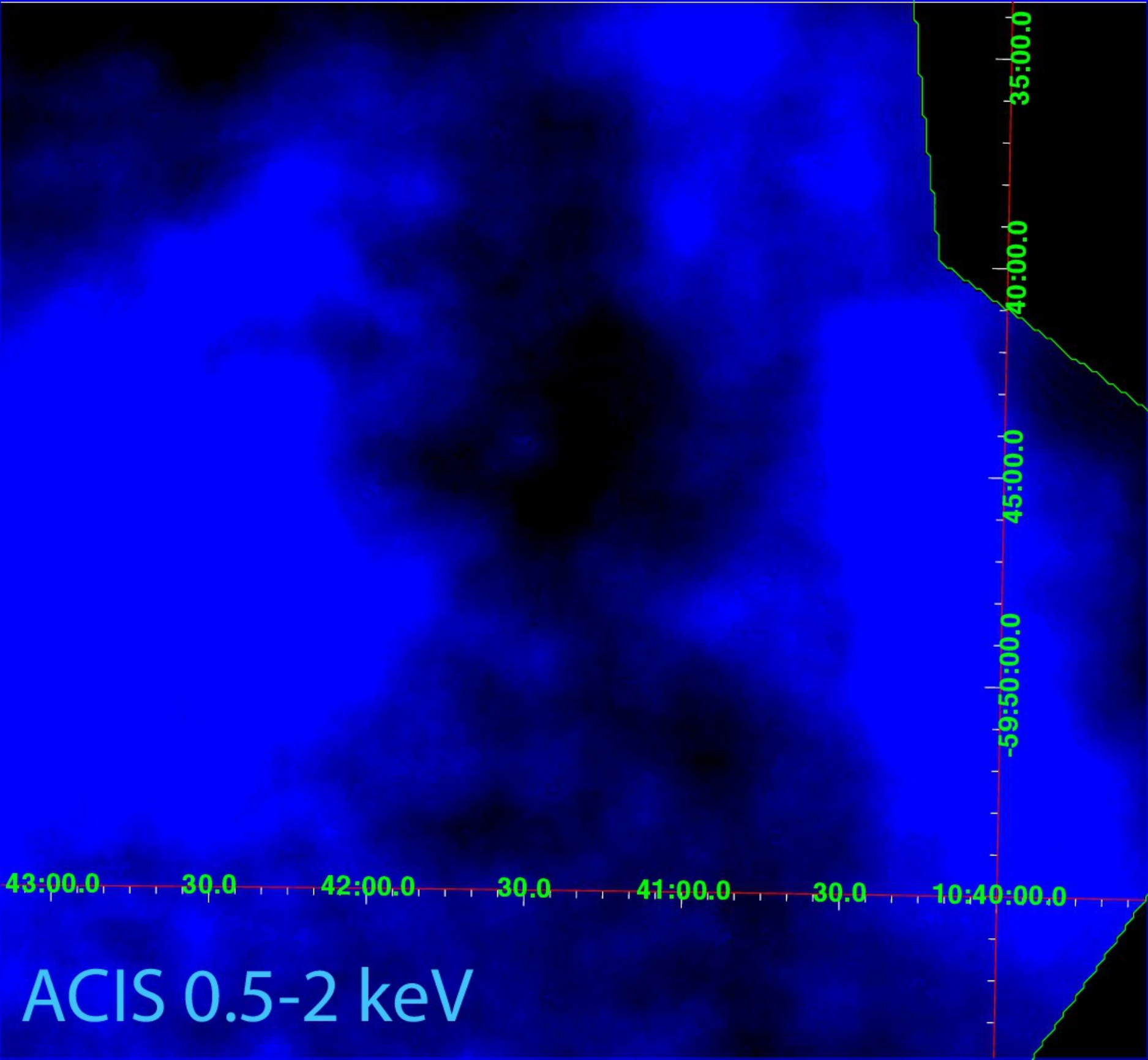}
\includegraphics[width=0.4\textwidth]{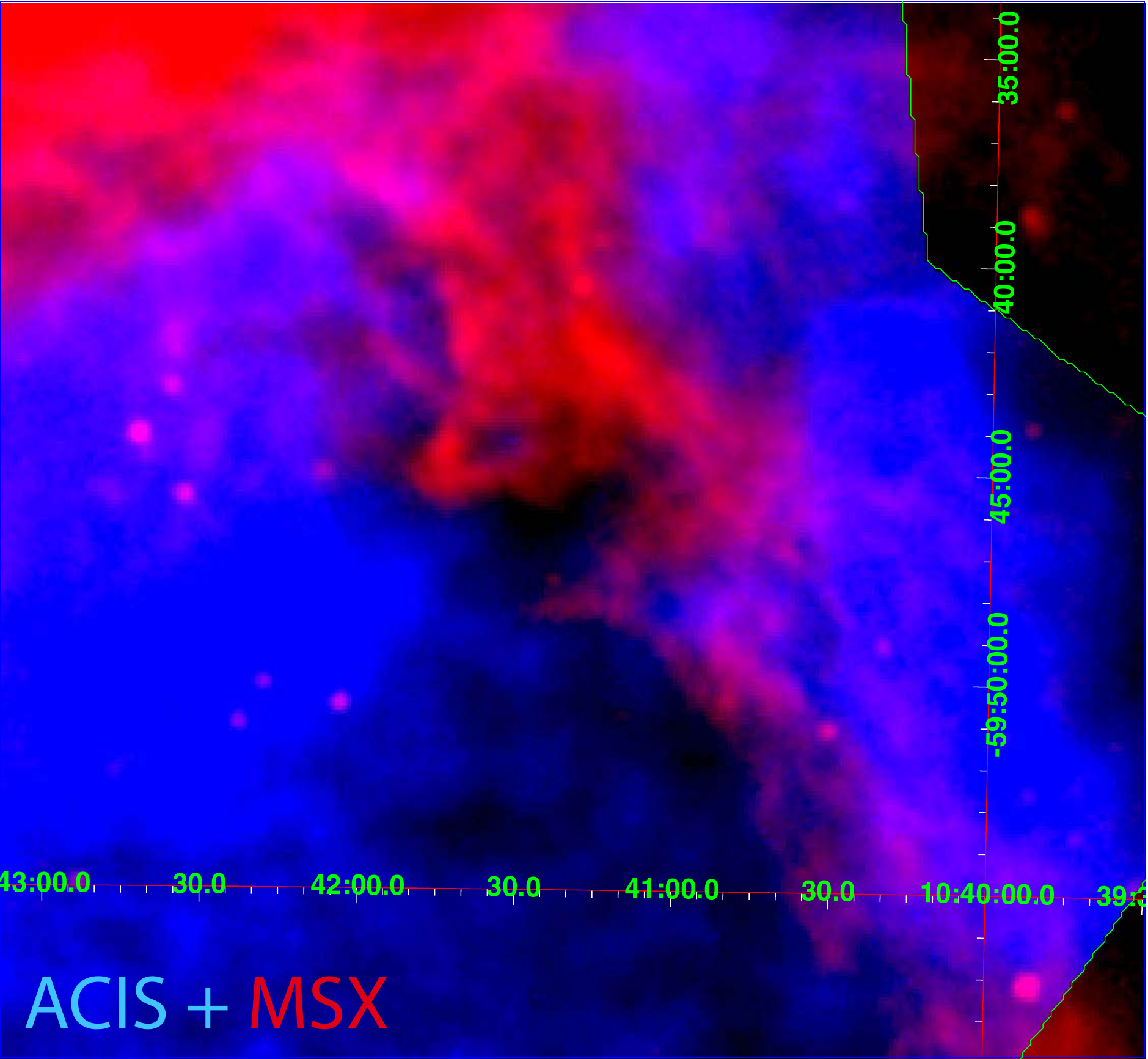}}
\mbox{
\includegraphics[width=0.4\textwidth]{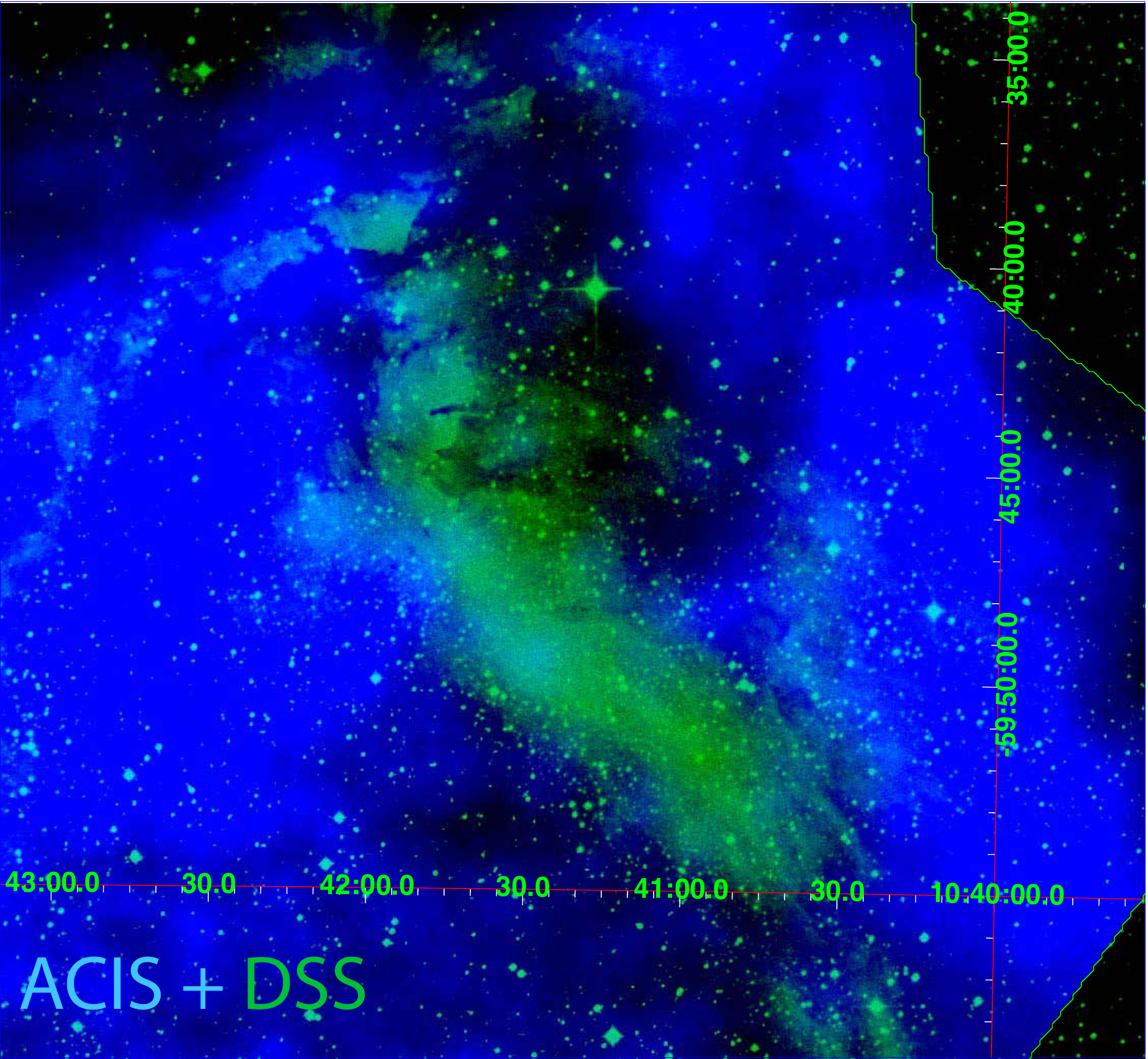}
\includegraphics[width=0.4\textwidth]{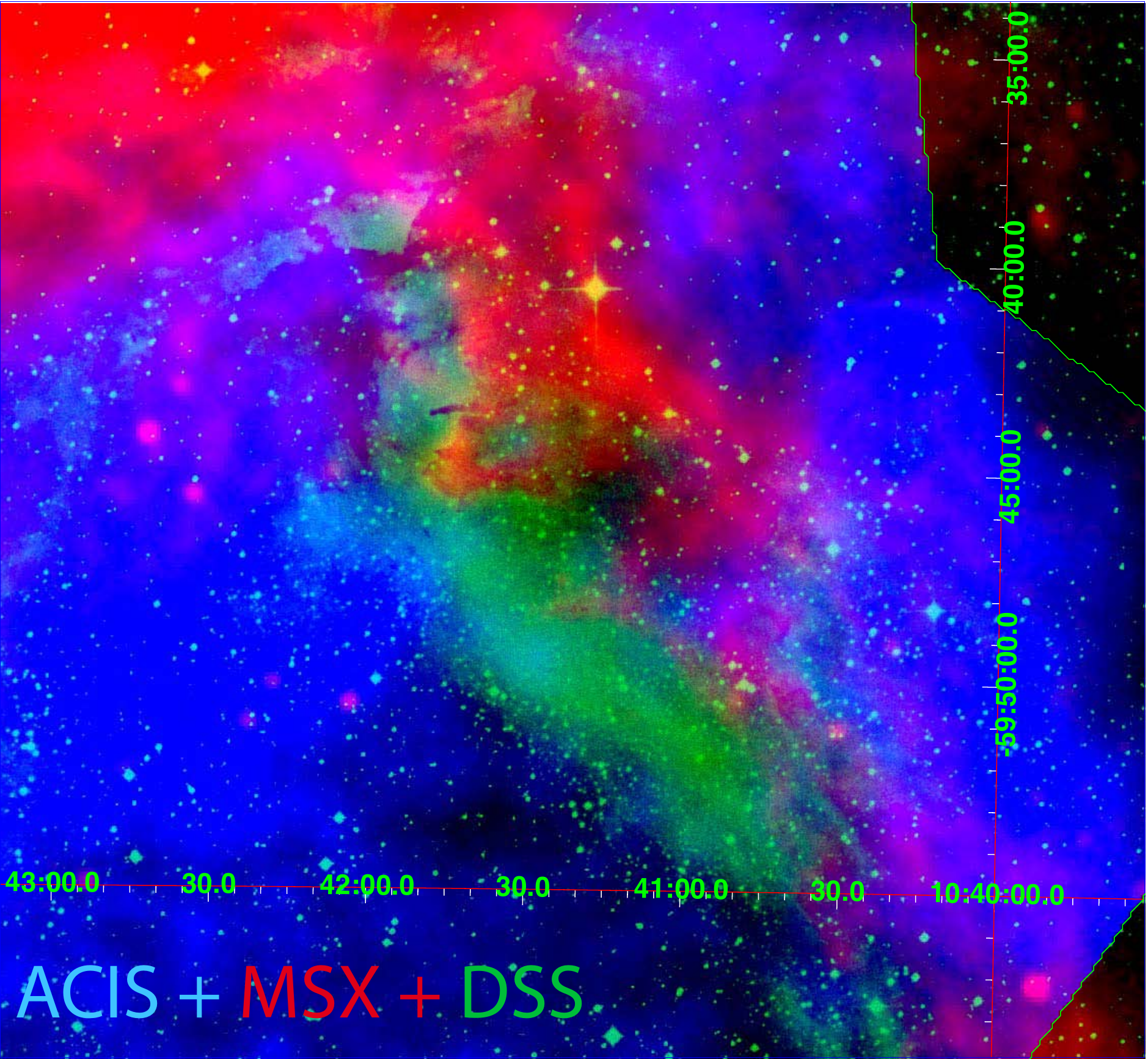}} 
\caption{X-ray shadowing by heated dust and ionized gas.  Note that the color-to-dataset mapping is different here than in Figure~\ref{fig:cccpmulti}.
(a)  {\em MSX} 8~$\mu$m image of the Moray Eel Filament.
(b)  DSS POSS2 red image of the same field.
(c)  ACIS 0.5--2~keV (soft band, point sources removed) adaptively-smoothed image of the same field.  Note the eel's eye (the faint blue dot of soft X-rays at RA$\sim10^{h}41^{m}30^{s}$, Dec$\sim-59^{\circ}45\arcmin$).
(d)  Combining the {\em MSX} image (a) with the ACIS image (c).
(e)  Combining the DSS image (b) with the ACIS image (c).
(f)  Combining the {\em MSX} image (a) with the DSS image (b) and the ACIS image (c).
} 
\label{fig:moray}
\end{center}
\end{figure}

The long dust filament running down the western side of the lower superbubble lobe appears to shadow the soft X-ray plasma filling that lobe; its distinctive shape inspired us to call it the ``Moray Eel Filament'' (Figure~\ref{fig:moray}).  The head of this moray eel (seen in profile facing east, with his mouth open, taking a bite out of the diffuse X-rays) is X-ray-dark except for a small region corresponding to the eel's ``eye,'' where soft X-rays shine through.  A bright swath of X-ray emission runs along the western edge of the filament.

%-------------------------------------------------------------------------
%-------------------------------------------------------------------------

This filament was included in a recent {\em Spitzer} study by \citet{Smith10b}; these authors conclude that the Moray Eel Filament is not a site of active star formation.  This evidence, coupled with the X-ray shadowing seen in Figure~\ref{fig:moray}'s multiwavelength comparison images, lead us to conclude that the filament is a dusty pillar with a prominent ionization front on its eastern flank, lying on the near surface of the lower superbubble lobe, and that this lobe is at least partially filled with hot X-ray-emitting plasma.  The X-ray shadowing it causes is a clear indication that the X-ray-emitting plasma is part of Carina, not a foreground source.  The bright swath of diffuse X-ray emission on the western side of the Moray Eel Filament, near the edge of the CCCP FoV, shows that hot plasma fills the lower superbubble lobe all the way to its western edge (seen most clearly in the SuperCOSMOS H$\alpha$ image in Figure~\ref{fig:intro}a).

%***Eric says to compare the Moray Eel filament to Eagle pillars, the Horsehead Nebula, and W5 ``mountains'' -- which of these structures is comparable in size (in pcs)?***

These simple multiwavelength images merely hint at the extensive ISM and bubble/superbubble studies that can be performed by comparing the CCCP maps of hot plasma with images of lower-energy phenomena.  More extensive multiwavelength comparisons, made with more state-of-the-art datasets, are underway (Smith et al.\ in prep.).  

\clearpage

%=============================================================================
\subsection{Global Synthesis \label{sec:global}}

It is important to ascertain how Carina compares to other famous sites of massive star formation in the Milky Way and beyond.  By considering the CCCP in these broader contexts, we can hope to bring a more measured understanding of Carina's place in the spectrum of star formation processes at work in the Local Group.  The last paper in this series, \citet{Townsley11b}, attempts to address some of these global considerations by comparing Carina's diffuse X-ray emission to that seen in other massive star-forming regions.  As a first step towards that goal, here we simply attempt to compare the CCCP with another recent large \Chandra survey of a nearby, well-studied star-forming region, the Orion Nebula Cluster (ONC).

In the CCCP field of view, the absorption-corrected, total-band (0.5--7~keV) X-ray luminosity of  Carina's diffuse emission is $L_{X} = 3 \times 10^{35}$~erg~s$^{-1}$, with an average surface brightness (SB) of $\log$SB = 32.2~erg~s$^{-1}$~pc$^{-2}$ \citep{Townsley11a}.  As mentioned above, diffuse X-ray emission from the part of the complex that lies outside of the CCCP field could increase this total luminosity estimate by a factor of 2.  Townsley et al.\ model Carina's diffuse X-ray emission with multiple thermal plasma components, with temperatures $kT \sim 0.3$ and 0.6~keV.

Five years before the CCCP, \Chandra observed the ONC in a nearly-contiguous set of exposures for $\sim$13 days \citep{Feigelson05}, in a study called the \Chandra Orion Ultradeep Project (COUP, PI E.\ Feigelson).  The shallow, wide-field CCCP is very much complementary to the deep, narrow-field COUP:  CCCP covers 1.42 square degrees on the sky while COUP covers 0.08 square degrees; CCCP has a nominal exposure time of 60~ks in each of 22 pointings while COUP stared at a single sky location for 838~ks \citep{Getman05}; the ONC is at a distance of $\sim$415~pc \citep{Menten07} while Carina is 5.5 times farther away, at $\sim$2.3~kpc \citep{Smith06a}.  Since survey sensitivity goes as the integration time divided by the square of the distance, COUP is $\sim$430 times more sensitive than the CCCP (a difference of 2.6 in the log).  Thus the CCCP is roughly equivalent to a 2~ks ACIS-I observation of the ONC.  Yet this paltry sensitivity resulted in a census of almost 11,000 X-ray-emitting young stars in Carina \citep{Broos11b}.  

COUP was deep enough to detect roughly half of the stellar population of the ONC, with $\sim 1400$ X-ray-detected members compared to an underlying population of $\sim$2800 stars (Hillenbrand \& Hartmann 1998). If the ONC were placed in the Carina Nebula, it would subtend a $3\arcmin \times 3\arcmin$ region in the CCCP mosaic.  If it was located at the aimpoint of one of the ACIS-I pointings, was not in a region of bright diffuse X-ray emission, did not suffer substantial extinction, and was not confused with another Carina cluster, we might imagine that we could detect pre-MS ONC members with as few as 3 net counts \citep{Broos11a}.  Scaling the net counts of COUP sources down by a factor of 430 and accounting for the $\sim$9 contaminant sources that we would expect in a $3\arcmin \times 3\arcmin$ CCCP region, we would then expect to detect $\sim$360 X-ray sources in our imaginary Carina ONC.  

The Carina ONC would be the fourth richest cluster in the complex after Tr14, Tr15, and Tr16 \citep[see Table~1 of][]{Feigelson11}.  It would have about the same spatial extent and massive star population as Bochum~11 (see Figures~\ref{fig:cccpfull} and \ref{fig:populations}) with about twice the Bochum~11 pre-MS population; the Carina ONC would have less than half of the detected X-ray point source sample of Tr15 \citep{Wang11}.  About 18\% of the COUP population, and 9\% of the total ONC stellar population, would be detected.  Estimates of the total number of pre-MS stars in Carina, sometimes based on scalings to the ONC, are given in other {\em Special Issue} papers \citep{Feigelson11,Povich11a}; they number into the many tens of thousands.  The unresolved X-ray emission in Carina must include a contribution from these unresolved pre-MS stars; \citet{Townsley11a} quantify this contribution and show that most of Carina's unresolved X-ray emission is truly diffuse.

Diffuse X-ray emission from O-star winds in the ONC was detected by {\em XMM-Newton} \citep{Gudel08}.  These authors estimate the intrinsic luminosity of this emission to be $L_{X} = 5.5 \times 10^{31}$~erg~s$^{-1}$ in the broad passband 0.1--10~keV and a surface brightness (SB) in the same band of roughly logSB = 31.5~erg~s$^{-1}$~pc$^{-2}$.  The plasma is soft, with $kT < 0.2$~keV.  In the narrower {\em Chandra}/ACIS-I bandpass of 0.5--7~keV, the intrinsic luminosity would be $2.6 \times 10^{31}$~erg~s$^{-1}$ with a surface brightness of $\log$SB = 31.2~erg~s$^{-1}$~pc$^{-2}$.

Compared to Carina's diffuse X-ray emission \citep{Townsley11a}, the ONC's diffuse X-ray emission is slightly softer ($<$0.2~keV), with an order of magnitude fainter surface brightness \citep{Gudel08}.  The total luminosity of Carina's diffuse emission, considering only the area covered by the CCCP, is $10^{4}$ times brighter than that seen by {\em XMM} in the ONC.  The ONC has one O7 star, one O9.5 star, and seven early-B stars (B0--B3), all with measured X-ray emission \citep{Stelzer05}.  The CCCP field contains at least 65 O stars \citep{Smith06b}, three WNH stars, and \etacar, so perhaps a factor of $10^{2}$ or even $10^{3}$ higher diffuse X-ray luminosity in Carina might be plausibly explained by massive star winds, but the factor of $10^{4}$ that we see might imply that other physical processes are at work.  These issues are explored in more detail in \citet{Townsley11a} and \citet{Townsley11b}.

%\clearpage

%=============================================================================

\section{SUMMARY \label{sec:summary}}

Our 1.42 square degree {\em Chandra}/ACIS-I survey of the Great Nebula in Carina reveals a rich population of $>$10,000 young stars.  Many reside in the well-known massive clusters Tr14, Tr15, and Tr16 that constitute the major recent star formation events in this complex.  The smaller clusters Bochum~11 and the Treasure Chest are also clearly seen as tight groupings of X-ray sources; conversely, the region containing the part of the cluster Bochum~10 that our survey covers shows a general enhancement in source density, but not an obvious cluster.  Since our survey is relatively shallow, we have probably captured only the brightest 10--20\% of the young stellar populations in Carina; put another way, the complex may well contain 50,000--100,000 young stars.  

Nearly all spectroscopically-confirmed O stars and many early-B stars are detected, as well as all three WR stars and of course $\eta$~Car.  Because of its extreme photon pile-up in the CCCP data, \etacar is not included in any of our analyses or in our source list.

The historically-studied clusters that make up this ``cluster of clusters'' do not merge in the X-ray sample; the X-ray data show no evidence that Collinder~228 is an extension of Tr16 or that Tr14 and Tr16 are part of the same cluster.  Tr15 is a well-populated, centrally-concentrated massive cluster with hundreds of members; it has been neglected in studies of young Milky Way star clusters mainly because of its more extreme neighbors, Tr14 and Tr16.  Many small clumps of X-ray sources with a range of obscurations are found throughout the CCCP field, but none of them constitutes a new major cluster.  %***Where do the OB stars, both old and new, fall wrt these clumps?***

There is evidence for a 5--10~Myr old population, more distributed than today's young clusters, that either formed unclustered or formed as one or more clusters but that have now drifted apart.  Perhaps cavity supernovae from this older population have helped to evacuate dust and gas from these clusters, hastening their demise through ``infant mortality'' \citep{Lada03}.  Carina's recently-noticed neutron star may well come from that population; in this paper we suggest six more candidate neutron stars that should be studied in more detail to rule out this interpretation. 

Diffuse X-ray emission persists in the \Chandra data, despite resolving out $>$14,000 X-ray point sources.  Because this diffuse emission is spectrally distinct from the composite point source emission and does not trace the spatial distribution of point sources, we believe it to be truly diffuse.  Visual and IR images of Carina's ionization fronts and dusty pillars show clear anticoincidence with Carina's diffuse X-ray emission, indicating that this diffuse emission is a part of the Carina complex and not a foreground structure.  Although part of it must be coming from the winds of massive stars, such \hii region wind emission has been found by \Chandra and {\em XMM} to be typically X-ray-faint, so winds are unlikely to account for the bulk of the diffuse X-ray emission.  

The X-ray luminosity of the diffuse emission, its complex morphology, and the recent discovery of a neutron star in the Carina complex all suggest that one or more cavity supernovae have exploded inside Carina's wind-blown cavities.  This means that \etacar will not be the first star to pollute the environment around Carina's disk-bearing young stars with supernova ejecta and that we can no longer consider Carina to be a ``pristine'' environment undergoing its first epoch of star formation.  This changes Carina's historical place in star formation science and forces us to think harder about the effects of multiple epochs of massive star and cluster formation and evolution when trying to understand the major sites of Galactic star formation that we see today.

The papers that follow in this {\em Special Issue} explore many of these issues in more depth and detail.  We hope that this ensemble work, centered around the CCCP dataset but enhanced by new multiwavelength studies, will help to bring about a more coherent understanding of star formation, massive stellar cluster evolution, and feedback in the Great Nebula in Carina and beyond.

%==========================================================================

\acknowledgments
We very much appreciate the time and effort donated by our anonymous referee to improve this paper.  This work is supported by \Chandra X-ray Observatory grant GO8-9131X (PI:  L.\ Townsley) and by the ACIS Instrument Team contract SV4-74018 (PI:  G.\ Garmire), issued by the \Chandra X-ray Center, which is operated by the Smithsonian Astrophysical Observatory for and on behalf of NASA under contract NAS8-03060.  S.J.W. is also supported by NASA contract NAS8-03060 ({\em Chandra}).  M.S.P. is supported by an NSF Astronomy and Astrophysics Postdoctoral Fellowship under award AST-0901646.  The Digitized Sky Surveys were produced at the Space Telescope Science Institute under U.S.\ Government grant NAG W-2166.  Near-IR observations were collected with the HAWK-I instrument on the VLT at Paranal Observatory, Chile, under ESO program 60.A-9284(K).  This research made use of data products from the Midcourse Space Experiment, the NASA/IPAC Infrared Science Archive, NASA's Astrophysics Data System, and the VizieR catalogue access tool, operated at CDS, Strasbourg, France.

{\em Facilities:} \facility{CXO (ACIS)}, \facility{ROSAT (PSPC)}, \facility{MSX ()}.

% =================START BIBLIOGRAPHY================ 

\end{document}